\documentclass[acmsmall]{acmart}

\acmJournal{PACMPL}
\acmVolume{1}
\acmNumber{CONF} 
\acmArticle{1}
\acmYear{2018}
\acmMonth{1}
\acmDOI{} 
\startPage{1}

\setcopyright{none}

\usepackage{booktabs}   
\usepackage{subcaption} 

\usepackage[all,british]{foreign}

\definecolor{MyBlue}{HTML}{0019D4}
\newcommand{\apl}[1]{{\textcolor{MyBlue}{\texttt{#1}}}}

\usepackage{mathpartir}

\usepackage{todonotes}

\usepackage{varwidth}

\usepackage{agda}
\usepackage{newunicodechar}
\usepackage[final]{listings}

\usepackage{microtype}

\makeatletter
\newcommand{\centersymbol}[2]{%
  \mathrel{\vphantom{#1#2}\mathpalette\center@symbol{{#1}{#2}}}%
}
\newcommand{\center@symbol}[2]{%
  \center@@symbol{#1}#2%
}
\newcommand{\center@@symbol}[3]{%
  \ooalign{\hss$#1\m@th#2$\hss\cr\hss$#1\m@th#3$\hss\cr}%
}
\makeatother

\newcommand{\APLrot}{\centersymbol{|}{\circ}}
\newcommand{\APLtrans}{\centersymbol{\backslash}{\circ}}
\newcommand{\APLredfirst}{\centersymbol{/}{-}}

\newunicodechar{∷}{::}
\newunicodechar{→}{\ensuremath{\to}}
\newunicodechar{ω}{\ensuremath{\omega}}
\newunicodechar{⊎}{\ensuremath{\uplus}}
\newunicodechar{≔}{\ensuremath{\coloneqq}}
\newunicodechar{∎}{\ensuremath{\blacksquare}}
\newunicodechar{≟}{\ensuremath{\stackrel{?}{=}}}
\newunicodechar{ᵣ}{\ensuremath{_r}}
\newunicodechar{ₗ}{\ensuremath{_l}}
\newunicodechar{▴}{\ensuremath{\blacktriangle}}
\newunicodechar{▾}{\ensuremath{\blacktriangledown}}
\newunicodechar{▹}{\ensuremath{\triangleright}}
\newunicodechar{∸}{\ensuremath{\dot{-}}}
\newunicodechar{⊛}{\ensuremath{\circledast}}
\newunicodechar{ᵀ}{\ensuremath{{}^{\text{T}}}}
\newunicodechar{⌿}{\ensuremath{\APLredfirst}}
\newunicodechar{ℕ}{\ensuremath{\mathbb{N}}}
\newunicodechar{∀}{\ensuremath{\forall}}
\newunicodechar{≡}{\ensuremath{\equiv}}
\newunicodechar{ℓ}{\ensuremath{\ell}}
\newunicodechar{⊤}{\ensuremath{\top}}
\newunicodechar{⋯}{\ensuremath{\cdots}}
\newunicodechar{₅}{\ensuremath{_5}}
\newunicodechar{₂}{\ensuremath{_2}}
\newunicodechar{₁}{\ensuremath{_1}}
\newunicodechar{₀}{\ensuremath{_0}}
\newunicodechar{₈}{\ensuremath{_8}}
\newunicodechar{₃}{\ensuremath{_3}}
\newunicodechar{₄}{\ensuremath{_4}}
\newunicodechar{₆}{\ensuremath{_6}}
\newunicodechar{₇}{\ensuremath{_7}}
\newunicodechar{′}{{'}}
\newunicodechar{ᵇ}{\ensuremath{^b}}
\newunicodechar{⇒}{\ensuremath{\implies}}
\newunicodechar{∃}{\ensuremath{\exists}}
\newunicodechar{λ}{\ensuremath{\lambda}}
\newunicodechar{≤}{\ensuremath{\leq}}
\newunicodechar{≢}{\ensuremath{\nequiv}}
\newunicodechar{∘}{\ensuremath{\circ}}
\newunicodechar{⍨}{\ensuremath{\stackrel{..}{\sim}}}
\newunicodechar{⌽}{\ensuremath{\APLrot}}
\newunicodechar{⊕}{\ensuremath{\oplus}}
\newunicodechar{⍵}{\ensuremath{\omega}}
\newunicodechar{⍺}{\ensuremath{\alpha}}
\newunicodechar{α}{\ensuremath{\alpha}}
\newunicodechar{⍤}{\ensuremath{\stackrel{..}{\circ}}}
\newunicodechar{⍴}{\ensuremath{\rho}}
\newunicodechar{⋄}{\ensuremath{\diamond}}
\newunicodechar{⊢}{\ensuremath{\vdash}}
\newunicodechar{⍉}{\ensuremath{\APLtrans}}
\newunicodechar{ι}{\ensuremath{\iota}}

\newcommand{\AD}[1]{\AgdaDatatype{#1}}
\newcommand{\AC}[1]{\AgdaInductiveConstructor{#1}}
\newcommand{\AF}[1]{\AgdaFunction{#1}}
\newcommand{\AB}[1]{\AgdaBound{#1}}
\newcommand{\AK}[1]{\AgdaKeyword{#1}}
\newcommand{\AR}[1]{\AgdaField{#1}}
\newcommand{\AM}[1]{\AgdaModule{#1}}
\newcommand{\AN}[1]{\AgdaNumber{#1}}
\newcommand{\AS}[1]{\AgdaString{#1}}

\newcommand\codeblock[1]{%
  {\begin{varwidth}{0.9\textwidth}#1\end{varwidth}}
}

\definecolor{xgray}{rgb}{0.5,0.5,0.5}
\definecolor{xblue}{rgb}{0.9,0.1,0.1}
\usepackage[final]{listings}

\newlength{\codeindent}
\lstdefinelanguage{sac}{morekeywords={with,fold,genarray,modarray,bool}}
 \newcommand{\sac}{\textsc{SaC}}

\begin{document}

\title[]{Choosing is Losing: How to combine the benefits of shallow and deep embeddings through reflection}         


\author[A. {\v{S}}inkarovs]{Artjoms {\v{S}}inkarovs}
\orcid{0000-0003-3292-2985}
\affiliation{%
  \institution{Heriot-Watt University}   
  \city{Edinburgh}
  \state{Scotland}
  \postcode{EH14 4AS}
  \country{UK}}
\email{a.sinkarovs@hw.ac.uk}             

\author[J. Cockx]{Jesper Cockx}
\affiliation{%
  \institution{TU Delft}
  \city{Delft}
  \postcode{2628 XE}
  \country{Netherlands}}
\email{j.g.h.cockx@tudelft.nl}           

\begin{abstract}
  Dependently-typed host languages empower users to verify a wide
  range of properties of embedded languages and
  programs written in them.  Designers of such embedded languages
  are faced with a difficult choice between using a shallow or a deep
  embedding.  The former is easier to use because the entire
  infrastructure of the host langauge is immediately
  available. Meanwhile, the latter gives full access to the
  structure of embedded programs, but is difficult to use in practice,
  especially when the embedded language is itself dependently typed.

  The main insight presented in this paper is that the choice between
  shallow and deep embedding can be eliminated by working in a host
  language with reflection capabilities: we start from a shallow
  embedding that can use all libraries and tools of the host language,
  and later use reflection to expose the deep structure of the
  embedded programs.
  Concretely, we apply this technique to embed three
  programming languages --- Kaleidoscope, SaC, and (a subset of) APL ---
  into the dependently typed theorem prover Agda, using dependent
  types to statically enforce several properties of interest.  We then
  use Agda's reflection capabilities to extract the embedded programs
  back into the original language, so that the existing toolchain can
  be leveraged.
  In this process, statically verified properties of the host language
  are mapped onto runtime checks in the target language, allowing
  extracted programs to interact safely with existing code.
  Finally, we demonstrate the feasibility of our approach with the
  implementation and extraction of a convolutional neural network in
  our embedding of APL.\@

%
\end{abstract}

%


\maketitle

\section{Introduction}

Dependently-typed systems such as Coq~\cite{10.5555/1965123},
Agda~\cite{10.5555/1813347.1813352}, Lean~\cite{conf/cade/MouraKADR15}
or Idris~\cite{JFP:9060502} make it possible to write down specifications
of programs as types and check that those specification are satisfied
during type checking. These verified programs can be either executed
directly, or compiled into a
general-purpose target language.  However, if we want to generate code
targeting a specific tool or language that is not supported by the
existing backends, we are stuck.

There are two common approaches to this problem: i) extending the
toolchain to produce code in the desired target; or ii) defining
a syntactic representation of the target language and implementing
the program in it.  The first approach gives the most freedom but it
may be technically challenging, as one has to modify the internals of
the theorem prover.  The second approach forces us to define
a deep embedding and express our specifications in it.  While this
seems easier up-front, the resulting encodings of the specifications
get impractically large, especially when the embedded language is
itself dependently typed.

In this paper we consider an approach that combines the benefits of
the two aforementioned solutions. We use a shallow embedding of the
target language and rely on \emph{compile-time
metaprogramming} to define custom extractors, making it possible to
write a verified program hand-in-hand with an executor for it --- all
within a single environment.  No need to touch internals of the host
language; no encoding artefacts.

Specifically, we use
\emph{reflection}~\cite{idris-refl, lean-refl,metacoq,agda-refl} (also
known as \emph{direct reflection}~\cite{nuprl-refl}), a form of
compile-time metaprogramming.\footnote{This use of the word
`reflection' is not to be confused with reflection in Java and similar
languages, which is a form of \emph{runtime metaprogramming}.}
That is, the host language has \texttt{quote}/\texttt{unquote}
primitives that translate expressions to their internal representation
and back.  The internal representation is regular data that can be
accessed and manipulated as usual.  The main goal of
reflection is to automate routine tasks of writing
boilerplate code, \eg{} tactics.  However,  APIs are
typically also complete enough for general
code generation tasks such as writing custom compiler backends.

In this paper, we are interested in writing verified embedded programs
that can be extracted to an existing language.  We approach this problem by
using an embed-typecheck-extract loop.  That is, we first embed a
language or a subset of it into a theorem prover, using dependent
types to encode properties of interest; next, we use the typechecker of
the host language to verify the properties; and finally, we use a
custom extractor to translate our implementation to the target
language.
Our approach is fine-grained: embeddings
(and consequently extractors) do not have to cover the entire
language, only the subset that is sufficient to specify the problem of
interest.  Also, instead of implementing the entire application in a
theorem prover, one might chose a particular fragment and integrate it
into an existing codebase.

Our proposed approach has several advantages:
\begin{itemize}
  \item Firstly, working with a shallow embedding is much easier than
  working with a deep one.  While deep embeddings provide full access
  to the program representation, they require us to encode the full
  syntax and typing rules of the target language,
  which soon becomes impractical for embedded languages with dependent
  types~\cite{10.1145/1863495.1863497}. This matters because we use
  dependent types in the embedding to encode the properties of
  interest.
  %
  In contrast, shallow embeddings do not require any additional
  encoding, since they are regular programs in the host language, and
  can make use of all available libraries and tools.

  \item Secondly, specifying embeddings and extractions in the same
  dependently-typed framework means we can modify the extractor
  without rebuilding the whole compiler of the host language.
  Extractors can be fine-tuned for the particular application, \eg{}
  treating a chosen function in a specific way.  Extractors can also make
  full use of dependent types, ensuring that the extraction process is
  sound.

  \item Thirdly, since the embedded program and the extractor are
  developed side-by-side, we can make use of features of the host
  language to customise the behaviour of the extractor
  further. Concretely, we use Agda's \emph{rewrite rules} as a
  lightweight method of applying verified domain-specific
  optimizations during extraction.
\end{itemize}

We implement extractors for Kaleidoskope~\cite{kaleidoscope} (Section~\ref{sec:basic-extr})
and SaC~\cite{sac} (Section~\ref{sec:array}) in Agda.
The former is a minimalist language that we use to demonstrate the basic
extraction principles.  The latter is a high-performance
array language that generates efficient code for various architectures, but
it has a restrictive type system.  Our SaC embedding guarantees
program termination, in-bound array indexing, and safety of arithmetic operations.
Finally, we also embed a small subset of APL~\cite{APL} into Agda (Section~\ref{sec:apl}), which
is sufficient to encode a Convolutional Neural
Network~\cite{cnninapl}.  Embedding APL is challenging because the
language is untyped, and its basic operators are heavily overloaded.
We define all the basic operators of APL in terms of the SaC embedding,
effectively obtaining a compiler from APL into executable binaries.

\paragraph{Contributions}
The key contributions of this paper are:
\begin{itemize}
    \item We introduce the concept of reflection-based extractors
            for shallowly-embedded languages that are embedded
            in metaprogramming-capable theorem provers.
    \item We extend the reflection API of
            Agda to make it better suited for defining
            custom extractors. Specifically, we extend the clause
            syntax with telescopes (Section~\ref{sec:normalisation}), we add new operations for
            selective reduction (Section~\ref{sec:controlling-reduction}), and we implement optional reconstruction
            of datatype parameters (Section~\ref{sec:embedded-array-lang}).
    \item We implement extractors for two languages: Kaleidoscope (Section~\ref{sec:basic-extr})
            and SaC (Section~\ref{sec:array}).  We use the latter as a basis for embedding
            a subset of APL (Section~\ref{sec:apl}).
    \item We demonstrate how the proposed embeddings can be used
            by encoding a real-world application and ensuring that
            it extracts correctly (Section~\ref{sec:cnn}).
\end{itemize}

\begin{code}[hide]%
\>[0]\AgdaKeyword{postulate}\<%
\\
\>[0][@{}l@{\AgdaIndent{0}}]%
\>[2]\AgdaPostulate{⋯}\AgdaSpace{}%
\AgdaSymbol{:}\AgdaSpace{}%
\AgdaSymbol{∀}\AgdaSpace{}%
\AgdaSymbol{\{}\AgdaBound{a}\AgdaSymbol{\}\{}\AgdaBound{A}\AgdaSpace{}%
\AgdaSymbol{:}\AgdaSpace{}%
\AgdaPrimitive{Set}\AgdaSpace{}%
\AgdaBound{a}\AgdaSymbol{\}}\AgdaSpace{}%
\AgdaSymbol{→}\AgdaSpace{}%
\AgdaBound{A}\<%
\\
\>[0]\AgdaKeyword{module}\AgdaSpace{}%
\AgdaModule{\AgdaUnderscore{}}\AgdaSpace{}%
\AgdaKeyword{where}\<%
\\
\>[0]\AgdaKeyword{module}\AgdaSpace{}%
\AgdaModule{Basics}\AgdaSpace{}%
\AgdaKeyword{where}\<%
\end{code}
\section{Background}
We start with a brief overview of key Agda constructions that
are used in this paper.  We also present relevant parts of the
reflection API.  For a more in-depth introduction to Agda refer
to the Agda user manual~\cite{agda}.

\subsection{Agda Basics}
Agda is an implementation of Martin-L{\"o}f's dependent type
theory~\cite{Martin-Lof-1972} extended with many convenience
constructions such as records, modules, do-notation, \etc{}
Types are defined using the following syntax:
\begin{mathpar}
\codeblock{
\begin{code}%
\>[0][@{}l@{\AgdaIndent{1}}]%
\>[2]\AgdaKeyword{data}\AgdaSpace{}%
\AgdaDatatype{ℕ}\AgdaSpace{}%
\AgdaSymbol{:}\AgdaSpace{}%
\AgdaPrimitive{Set}\AgdaSpace{}%
\AgdaKeyword{where}\<%
\\
\>[2][@{}l@{\AgdaIndent{0}}]%
\>[4]\AgdaInductiveConstructor{zero}\AgdaSpace{}%
\AgdaSymbol{:}\AgdaSpace{}%
\AgdaDatatype{ℕ}\<%
\\
\>[4]\AgdaInductiveConstructor{suc}%
\>[9]\AgdaSymbol{:}\AgdaSpace{}%
\AgdaDatatype{ℕ}\AgdaSpace{}%
\AgdaSymbol{→}\AgdaSpace{}%
\AgdaDatatype{ℕ}\<%
\end{code}
}
\and
\codeblock{
\begin{code}%
\>[2]\AgdaKeyword{data}\AgdaSpace{}%
\AgdaDatatype{Fin}\AgdaSpace{}%
\AgdaSymbol{:}\AgdaSpace{}%
\AgdaDatatype{ℕ}\AgdaSpace{}%
\AgdaSymbol{→}\AgdaSpace{}%
\AgdaPrimitive{Set}\AgdaSpace{}%
\AgdaKeyword{where}\<%
\\
\>[2][@{}l@{\AgdaIndent{0}}]%
\>[4]\AgdaInductiveConstructor{zero}\AgdaSpace{}%
\AgdaSymbol{:}\AgdaSpace{}%
\AgdaSymbol{∀}\AgdaSpace{}%
\AgdaSymbol{\{}\AgdaBound{n}\AgdaSymbol{\}}\AgdaSpace{}%
\AgdaSymbol{→}\AgdaSpace{}%
\AgdaDatatype{Fin}\AgdaSpace{}%
\AgdaSymbol{(}\AgdaInductiveConstructor{suc}\AgdaSpace{}%
\AgdaBound{n}\AgdaSymbol{)}\<%
\\
\>[4]\AgdaInductiveConstructor{suc}%
\>[9]\AgdaSymbol{:}\AgdaSpace{}%
\AgdaSymbol{∀}\AgdaSpace{}%
\AgdaSymbol{\{}\AgdaBound{n}\AgdaSymbol{\}}\AgdaSpace{}%
\AgdaSymbol{→}\AgdaSpace{}%
\AgdaDatatype{Fin}\AgdaSpace{}%
\AgdaBound{n}\AgdaSpace{}%
\AgdaSymbol{→}\AgdaSpace{}%
\AgdaDatatype{Fin}\AgdaSpace{}%
\AgdaSymbol{(}\AgdaInductiveConstructor{suc}\AgdaSpace{}%
\AgdaBound{n}\AgdaSymbol{)}\<%
\end{code}
}
\and
\codeblock{
\begin{code}%
\>[2]\AgdaKeyword{data}\AgdaSpace{}%
\AgdaOperator{\AgdaDatatype{\AgdaUnderscore{}≡\AgdaUnderscore{}}}\AgdaSpace{}%
\AgdaSymbol{\{}\AgdaBound{a}\AgdaSymbol{\}}\AgdaSpace{}%
\AgdaSymbol{\{}\AgdaBound{A}\AgdaSpace{}%
\AgdaSymbol{:}\AgdaSpace{}%
\AgdaPrimitive{Set}\AgdaSpace{}%
\AgdaBound{a}\AgdaSymbol{\}}\<%
\\
\>[2][@{}l@{\AgdaIndent{0}}]%
\>[6]\AgdaSymbol{(}\AgdaBound{x}\AgdaSpace{}%
\AgdaSymbol{:}\AgdaSpace{}%
\AgdaBound{A}\AgdaSymbol{)}\AgdaSpace{}%
\AgdaSymbol{:}\AgdaSpace{}%
\AgdaBound{A}\AgdaSpace{}%
\AgdaSymbol{→}\AgdaSpace{}%
\AgdaPrimitive{Set}\AgdaSpace{}%
\AgdaBound{a}\AgdaSpace{}%
\AgdaKeyword{where}\<%
\\
\>[2][@{}l@{\AgdaIndent{0}}]%
\>[4]\AgdaInductiveConstructor{refl}\AgdaSpace{}%
\AgdaSymbol{:}\AgdaSpace{}%
\AgdaBound{x}\AgdaSpace{}%
\AgdaOperator{\AgdaDatatype{≡}}\AgdaSpace{}%
\AgdaBound{x}\<%
\end{code}
}
\end{mathpar}
Unary natural numbers \AD{ℕ} is a type with two constructors:
\AC{zero} and \AC{suc}.  \AD{Fin} is an indexed type, where the index
is of type \AD{ℕ}.  Constructor names can be overloaded and are
disambiguated from the typing context, or can be prefixed with the
type name: \AC{ℕ.zero}, \AC{ℕ.suc}.
The \AD{Fin} \AB{n} type represents natural numbers that are bounded
by \AB{n}.  In the definition of \AD{Fin}, ∀ binds the variable
without needing to specify its type.  Curly braces indicate hidden
arguments, which can be left out at function applications: we have
\AC{suc} \AC{zero}~:~\AD{Fin} \AN{3}, assuming that Agda can infer a
(unique) value for the hidden argument.  Hidden arguments can be
passed explicitly using the syntax $\AC{zero}\ \{\AB{n} = \AB{x}\}$.
The propositional equality type \AF{\_≡\_} expresses equality of its
two arguments, and has a single constructor \AC{refl} stating that any
value \AB{x} is equal to itself.  It uses mixfix
syntax~\cite{10.1007/978-3-642-24452-0_5}: the
underscores in the name indicate placeholders for the arguments.
\AF{Set} is the name of the type of all small types.  Sets form a
predicative hierarchy, meaning that \AF{Set} \AB{i} is of type
\AF{Set} (\AF{ℓsuc} \AB{i}), and \AF{Set} is a synonym for \AF{Set}
\AF{ℓzero}.  The functions \AF{ℓsuc} and \AF{ℓzero} are used to construct elements of type \AD{Level}.

Functions are defined in a pattern-matching style:
\begin{code}[hide]%
\>[2]\AgdaKeyword{postulate}\AgdaSpace{}%
\AgdaOperator{\AgdaPostulate{\AgdaUnderscore{}+\AgdaUnderscore{}}}\AgdaSpace{}%
\AgdaSymbol{:}\AgdaSpace{}%
\AgdaDatatype{ℕ}\AgdaSpace{}%
\AgdaSymbol{→}\AgdaSpace{}%
\AgdaDatatype{ℕ}\AgdaSpace{}%
\AgdaSymbol{→}\AgdaSpace{}%
\AgdaDatatype{ℕ}\<%
\\
\>[0]\<%
\end{code}
\begin{mathpar}
\codeblock{
\begin{code}%
\>[0][@{}l@{\AgdaIndent{1}}]%
\>[2]\AgdaOperator{\AgdaFunction{\AgdaUnderscore{}*\AgdaUnderscore{}}}\AgdaSpace{}%
\AgdaSymbol{:}\AgdaSpace{}%
\AgdaDatatype{ℕ}\AgdaSpace{}%
\AgdaSymbol{→}\AgdaSpace{}%
\AgdaDatatype{ℕ}\AgdaSpace{}%
\AgdaSymbol{→}\AgdaSpace{}%
\AgdaDatatype{ℕ}\<%
\\
\>[2]\AgdaInductiveConstructor{zero}%
\>[11]\AgdaOperator{\AgdaFunction{*}}\AgdaSpace{}%
\AgdaBound{y}%
\>[16]\AgdaSymbol{=}\AgdaSpace{}%
\AgdaInductiveConstructor{zero}\<%
\\
\>[2]\AgdaSymbol{(}\AgdaInductiveConstructor{suc}\AgdaSpace{}%
\AgdaBound{x}\AgdaSymbol{)}%
\>[11]\AgdaOperator{\AgdaFunction{*}}\AgdaSpace{}%
\AgdaBound{y}%
\>[16]\AgdaSymbol{=}\AgdaSpace{}%
\AgdaBound{y}\AgdaSpace{}%
\AgdaOperator{\AgdaPostulate{+}}\AgdaSpace{}%
\AgdaSymbol{(}\AgdaBound{x}\AgdaSpace{}%
\AgdaOperator{\AgdaFunction{*}}\AgdaSpace{}%
\AgdaBound{y}\AgdaSymbol{)}\<%
\end{code}
}
\and
\codeblock{
\begin{code}%
\>[2]\AgdaFunction{abs}\AgdaSpace{}%
\AgdaSymbol{:}\AgdaSpace{}%
\AgdaDatatype{Fin}\AgdaSpace{}%
\AgdaInductiveConstructor{zero}\AgdaSpace{}%
\AgdaSymbol{→}\AgdaSpace{}%
\AgdaDatatype{ℕ}\<%
\\
\>[2]\AgdaFunction{abs}\AgdaSpace{}%
\AgdaSymbol{()}\<%
\end{code}
}
\and
\codeblock{
\begin{code}%
\>[2]\AgdaFunction{wth}\AgdaSpace{}%
\AgdaSymbol{:}\AgdaSpace{}%
\AgdaSymbol{(}\AgdaBound{a}\AgdaSpace{}%
\AgdaBound{b}\AgdaSpace{}%
\AgdaSymbol{:}\AgdaSpace{}%
\AgdaDatatype{ℕ}\AgdaSymbol{)}\AgdaSpace{}%
\AgdaSymbol{→}\AgdaSpace{}%
\AgdaDatatype{ℕ}\<%
\\
\>[2]\AgdaFunction{wth}\AgdaSpace{}%
\AgdaBound{a}\AgdaSpace{}%
\AgdaBound{b}\AgdaSpace{}%
\AgdaKeyword{with}\AgdaSpace{}%
\AgdaBound{a}\AgdaSpace{}%
\AgdaOperator{\AgdaFunction{*}}\AgdaSpace{}%
\AgdaBound{b}\<%
\\
\>[2]\AgdaSymbol{...}%
\>[10]\AgdaSymbol{|}\AgdaSpace{}%
\AgdaInductiveConstructor{zero}\AgdaSpace{}%
\AgdaSymbol{=}\AgdaSpace{}%
\AgdaInductiveConstructor{zero}\<%
\\
\>[2]\AgdaCatchallClause{\AgdaSymbol{...}}%
\>[10]\AgdaCatchallClause{\AgdaSymbol{|}}\AgdaSpace{}%
\AgdaCatchallClause{\AgdaSymbol{\AgdaUnderscore{}}}%
\>[17]\AgdaSymbol{=}\AgdaSpace{}%
\AgdaBound{b}\<%
\end{code}
}
\end{mathpar}
The definition of \AF{abs} uses the \emph{absurd pattern} (),
indicating an impossible case for the first argument, \ie{} there is
no constructor constructing a term of type \AD{Fin} \AC{zero}.
Clauses with absurd patterns do not have a body, as the type system
guarantees that they are never called at run-time.
In the definition of \AF{wth} we demonstrate the use of the \AK{with}
construction~\cite{10.1017/S0956796803004829} which makes it possible
to match on the result of an expression locally.

%

\subsection{Reflection}
Instead of explaining the full structure of all types that Agda uses
to encode reflected syntax, we consider a small but representative
sample: the function \AF{foo} (left) and its reflection (right).
\begin{code}[hide]%
\>[0]\AgdaKeyword{module}\AgdaSpace{}%
\AgdaModule{FunExample}\AgdaSpace{}%
\AgdaKeyword{where}\<%
\\
\>[0][@{}l@{\AgdaIndent{0}}]%
\>[2]\AgdaKeyword{open}\AgdaSpace{}%
\AgdaKeyword{import}\AgdaSpace{}%
\AgdaModule{Data.List}\<%
\\
\>[2]\AgdaKeyword{open}\AgdaSpace{}%
\AgdaKeyword{import}\AgdaSpace{}%
\AgdaModule{Data.Nat}\<%
\\
\>[2]\AgdaKeyword{open}\AgdaSpace{}%
\AgdaKeyword{import}\AgdaSpace{}%
\AgdaModule{Data.Fin}\AgdaSpace{}%
\AgdaKeyword{using}\AgdaSpace{}%
\AgdaSymbol{(}\AgdaDatatype{Fin}\AgdaSymbol{)}\<%
\\
\>[2]\AgdaKeyword{open}\AgdaSpace{}%
\AgdaKeyword{import}\AgdaSpace{}%
\AgdaModule{Data.Bool}\<%
\\
\>[2]\AgdaKeyword{open}\AgdaSpace{}%
\AgdaKeyword{import}\AgdaSpace{}%
\AgdaModule{Reflection}\<%
\\
\>[2]\AgdaKeyword{open}\AgdaSpace{}%
\AgdaKeyword{import}\AgdaSpace{}%
\AgdaModule{Data.Unit}\<%
\\
\>[2]\AgdaKeyword{open}\AgdaSpace{}%
\AgdaKeyword{import}\AgdaSpace{}%
\AgdaModule{Data.Product}\<%
\\
\>[2]\AgdaKeyword{open}\AgdaSpace{}%
\AgdaKeyword{import}\AgdaSpace{}%
\AgdaModule{Function}\<%
\\
\>[0]\<%
\end{code}
\begin{mathpar}
\codeblock{
\begin{code}%
\>[0][@{}l@{\AgdaIndent{1}}]%
\>[2]\AgdaFunction{foo}\AgdaSpace{}%
\AgdaSymbol{:}\AgdaSpace{}%
\AgdaDatatype{ℕ}\AgdaSpace{}%
\AgdaSymbol{→}\AgdaSpace{}%
\AgdaDatatype{ℕ}\<%
\\
\>[2]\AgdaFunction{foo}\AgdaSpace{}%
\AgdaNumber{0}%
\>[15]\AgdaSymbol{=}\AgdaSpace{}%
\AgdaInductiveConstructor{zero}\<%
\\
\>[2]\AgdaFunction{foo}\AgdaSpace{}%
\AgdaSymbol{(}\AgdaInductiveConstructor{suc}\AgdaSpace{}%
\AgdaBound{x}\AgdaSymbol{)}%
\>[15]\AgdaSymbol{=}\AgdaSpace{}%
\AgdaBound{x}\AgdaSpace{}%
\AgdaOperator{\AgdaPrimitive{+}}\AgdaSpace{}%
\AgdaBound{x}\<%
\end{code}
}
\and
\codeblock{
\begin{code}%
\>[2]\AgdaFunction{`foo}%
\>[138I]\AgdaSymbol{=}\AgdaSpace{}%
\AgdaInductiveConstructor{Definition.function}\<%
\\
\>[.][@{}l@{}]\<[138I]%
\>[7]\AgdaOperator{\AgdaFunction{\$}}%
\>[140I]\AgdaInductiveConstructor{Clause.clause}\<%
\\
\>[140I][@{}l@{\AgdaIndent{0}}]%
\>[11]\AgdaInductiveConstructor{[]}\<%
\\
\>[11]\AgdaSymbol{(}\AgdaInductiveConstructor{vArg}\AgdaSpace{}%
\AgdaSymbol{(}\AgdaInductiveConstructor{Pattern.con}\AgdaSpace{}%
\AgdaSymbol{(}\AgdaKeyword{quote}\AgdaSpace{}%
\AgdaInductiveConstructor{ℕ.zero}\AgdaSymbol{)}\AgdaSpace{}%
\AgdaInductiveConstructor{[]}\AgdaSymbol{)}\AgdaSpace{}%
\AgdaOperator{\AgdaInductiveConstructor{∷}}\AgdaSpace{}%
\AgdaInductiveConstructor{[]}\AgdaSymbol{)}\<%
\\
\>[11]\AgdaSymbol{(}\AgdaInductiveConstructor{Term.con}\AgdaSpace{}%
\AgdaSymbol{(}\AgdaKeyword{quote}\AgdaSpace{}%
\AgdaInductiveConstructor{ℕ.zero}\AgdaSymbol{)}\AgdaSpace{}%
\AgdaInductiveConstructor{[]}\AgdaSymbol{)}\<%
\\
\>[7]\AgdaOperator{\AgdaInductiveConstructor{∷}}%
\>[150I]\AgdaInductiveConstructor{Clause.clause}\<%
\\
\>[150I][@{}l@{\AgdaIndent{0}}]%
\>[11]\AgdaSymbol{((}\AgdaString{"x"}\AgdaSpace{}%
\AgdaOperator{\AgdaInductiveConstructor{,}}\AgdaSpace{}%
\AgdaInductiveConstructor{vArg}\AgdaSpace{}%
\AgdaSymbol{(}\AgdaInductiveConstructor{Term.def}\AgdaSpace{}%
\AgdaSymbol{(}\AgdaKeyword{quote}\AgdaSpace{}%
\AgdaDatatype{ℕ}\AgdaSymbol{)}\AgdaSpace{}%
\AgdaInductiveConstructor{[]}\AgdaSymbol{))}\AgdaSpace{}%
\AgdaOperator{\AgdaInductiveConstructor{∷}}\AgdaSpace{}%
\AgdaInductiveConstructor{[]}\AgdaSymbol{)}\<%
\\
\>[11]\AgdaSymbol{(}\AgdaInductiveConstructor{vArg}\AgdaSpace{}%
\AgdaSymbol{(}\AgdaInductiveConstructor{Pattern.con}\AgdaSpace{}%
\AgdaSymbol{(}\AgdaKeyword{quote}\AgdaSpace{}%
\AgdaInductiveConstructor{ℕ.suc}\AgdaSymbol{)}\AgdaSpace{}%
\AgdaSymbol{(}\AgdaInductiveConstructor{vArg}\AgdaSpace{}%
\AgdaSymbol{(}\AgdaInductiveConstructor{Pattern.var}\AgdaSpace{}%
\AgdaNumber{0}\AgdaSymbol{)}\AgdaSpace{}%
\AgdaOperator{\AgdaInductiveConstructor{∷}}\AgdaSpace{}%
\AgdaInductiveConstructor{[]}\AgdaSymbol{))}\AgdaSpace{}%
\AgdaOperator{\AgdaInductiveConstructor{∷}}\AgdaSpace{}%
\AgdaInductiveConstructor{[]}\AgdaSymbol{)}\<%
\\
\>[11]\AgdaSymbol{(}\AgdaInductiveConstructor{def}\AgdaSpace{}%
\AgdaSymbol{(}\AgdaKeyword{quote}\AgdaSpace{}%
\AgdaOperator{\AgdaPrimitive{\AgdaUnderscore{}+\AgdaUnderscore{}}}\AgdaSymbol{)}\AgdaSpace{}%
\AgdaSymbol{(}\AgdaInductiveConstructor{vArg}\AgdaSpace{}%
\AgdaSymbol{(}\AgdaInductiveConstructor{Term.var}\AgdaSpace{}%
\AgdaNumber{0}\AgdaSpace{}%
\AgdaInductiveConstructor{[]}\AgdaSymbol{)}\AgdaSpace{}%
\AgdaOperator{\AgdaInductiveConstructor{∷}}\AgdaSpace{}%
\AgdaInductiveConstructor{vArg}\AgdaSpace{}%
\AgdaSymbol{(}\AgdaInductiveConstructor{Term.var}\AgdaSpace{}%
\AgdaNumber{0}\AgdaSpace{}%
\AgdaInductiveConstructor{[]}\AgdaSymbol{)}\AgdaSpace{}%
\AgdaOperator{\AgdaInductiveConstructor{∷}}\AgdaSpace{}%
\AgdaInductiveConstructor{[]}\AgdaSymbol{))}\<%
\\
\>[7]\AgdaOperator{\AgdaInductiveConstructor{∷}}\AgdaSpace{}%
\AgdaInductiveConstructor{[]}\<%
\end{code}
}
\end{mathpar}
The reflected function is defined by the list of clauses \AC{Clause.clause}.  Each
clause has three arguments: i) the telescope, which is a list of free variables
and their types; ii) the list of patterns; and iii) the body of the
clause.  The first clause does not have free variables, so the telescope
is empty. The second clause has one variable called \AB{x}.  The
pattern list in the first clause has one argument;  \AC{vArg} denotes that
it is visible argument (\AC{hArg} is used for hidden arguments).
The actual pattern matches against the \AC{zero} constructor, as expressed
by \AC{Pattern.con}, which has two arguments: the reflected constructor name and the
list of reflected arguments.  The number of reflected arguments must be the
same as the number of the actual arguments, which is none in the case of \AC{zero}.
The \AK{quote} primitive returns
a representation of the name for the given Agda definition or constructor,
which is of type \AD{Name}.
Variables (both in patterns and in terms) are given as de Bruijn indices
into the telescope of the clause.  That is, in the second clause the
de Bruijn index \AN{0} refers to the variable \AS{x}.  Note that we write
\AN{0} instead of \AC{zero}, as numbers are expanded
into their corresponding \AC{zero}/\AC{suc} representations.

Effectively using Agda's reflection API can be challenging because the
syntax it uses matches the \emph{internal}
representation of Agda terms, which differs significantly from the
surface syntax.
Many constructs such as
implicit arguments, instance arguments, \AK{let} and \AK{with} definitions
exist only in the surface language.  Translation from the surface
language is performed by the \emph{elaborator}.
%
%
Following the approach of \emph{elaborator reflection}
introduced by Idris~\cite{idris-refl}, Agda exposes many parts
of the elaborator to the reflection API, including reduction
and normalisation of expressions.
These operations are made available through the \AD{TC} monad, which
takes care of managing the current context of the elaborator.

The key metaprogramming primitives are \AK{quote} and
\AK{unquote}, that operate as follows:
\begin{mathpar}
\codeblock{
\begin{code}%
\>[2]\AgdaFunction{ex}\AgdaSpace{}%
\AgdaSymbol{:}\AgdaSpace{}%
\AgdaSymbol{(}\AgdaBound{a}\AgdaSpace{}%
\AgdaSymbol{:}\AgdaSpace{}%
\AgdaDatatype{Bool}\AgdaSymbol{)}\AgdaSpace{}%
\AgdaSymbol{→}\AgdaSpace{}%
\AgdaOperator{\AgdaFunction{if}}\AgdaSpace{}%
\AgdaBound{a}\AgdaSpace{}%
\AgdaOperator{\AgdaFunction{then}}\AgdaSpace{}%
\AgdaDatatype{ℕ}\AgdaSpace{}%
\AgdaOperator{\AgdaFunction{else}}\AgdaSpace{}%
\AgdaDatatype{Bool}\<%
\\
\>[2]\AgdaFunction{ex}\AgdaSpace{}%
\AgdaInductiveConstructor{true}\AgdaSpace{}%
\AgdaSymbol{=}\AgdaSpace{}%
\AgdaKeyword{unquote}\AgdaSpace{}%
\AgdaFunction{helper}\<%
\\
\>[2][@{}l@{\AgdaIndent{0}}]%
\>[4]\AgdaKeyword{where}%
\>[198I]\AgdaFunction{helper}\AgdaSpace{}%
\AgdaSymbol{:}\AgdaSpace{}%
\AgdaDatatype{Term}\AgdaSpace{}%
\AgdaSymbol{→}\AgdaSpace{}%
\AgdaPostulate{TC}\AgdaSpace{}%
\AgdaRecord{⊤}\<%
\\
\>[.][@{}l@{}]\<[198I]%
\>[10]\AgdaFunction{helper}\AgdaSpace{}%
\AgdaBound{h}\AgdaSpace{}%
\AgdaSymbol{=}\AgdaSpace{}%
\AgdaPostulate{unify}\AgdaSpace{}%
\AgdaBound{h}\AgdaSpace{}%
\AgdaSymbol{(}\AgdaInductiveConstructor{lit}\AgdaSpace{}%
\AgdaSymbol{(}\AgdaInductiveConstructor{nat}\AgdaSpace{}%
\AgdaNumber{42}\AgdaSymbol{))}\<%
\\
\>[2]\AgdaFunction{ex}\AgdaSpace{}%
\AgdaInductiveConstructor{false}\AgdaSpace{}%
\AgdaSymbol{=}\AgdaSpace{}%
\AgdaInductiveConstructor{false}\<%
\end{code}
}
\and
\codeblock{
\begin{code}%
\>[2]\AgdaKeyword{macro}%
\>[214I]\AgdaFunction{getDef}\AgdaSpace{}%
\AgdaSymbol{:}\AgdaSpace{}%
\AgdaPostulate{Name}\AgdaSpace{}%
\AgdaSymbol{→}\AgdaSpace{}%
\AgdaSymbol{(}\AgdaDatatype{Term}\AgdaSpace{}%
\AgdaSymbol{→}\AgdaSpace{}%
\AgdaPostulate{TC}\AgdaSpace{}%
\AgdaRecord{⊤}\AgdaSymbol{)}\<%
\\
\>[.][@{}l@{}]\<[214I]%
\>[8]\AgdaFunction{getDef}\AgdaSpace{}%
\AgdaBound{n}\AgdaSpace{}%
\AgdaBound{h}\AgdaSpace{}%
\AgdaSymbol{=}\AgdaSpace{}%
\AgdaKeyword{do}\<%
\\
\>[8][@{}l@{\AgdaIndent{0}}]%
\>[9]\AgdaBound{d}\AgdaSpace{}%
\AgdaOperator{\AgdaFunction{←}}\AgdaSpace{}%
\AgdaPostulate{getDefinition}\AgdaSpace{}%
\AgdaBound{n}\<%
\\
\>[9]\AgdaBound{t}\AgdaSpace{}%
\AgdaOperator{\AgdaFunction{←}}\AgdaSpace{}%
\AgdaPostulate{quoteTC}\AgdaSpace{}%
\AgdaBound{d}\<%
\\
\>[9]\AgdaPostulate{unify}\AgdaSpace{}%
\AgdaBound{h}\AgdaSpace{}%
\AgdaBound{t}\<%
\\
\>[2]\AgdaFunction{``foo}\AgdaSpace{}%
\AgdaSymbol{=}\AgdaSpace{}%
\AgdaMacro{getDef}\AgdaSpace{}%
\AgdaFunction{foo}\<%
\end{code}
}
\end{mathpar}
In \AF{ex}, the \AK{unquote} occurs in the \AC{true} clause of the
function.
%
The argument to \AK{unquote} is expected to be a
function of type \AD{Term} → \AD{TC} \AD{⊤}, where \AD{⊤} is the unit
type. During elaboration,  Agda creates a
metavariable \AB{h} of type \AD{ℕ}, quotes it, and passes it to the
function \AF{helper}. In the body of \AF{helper}, we call the \AF{TC}
operation \AF{unify} \AB{h} (\AC{lit} (\AC{nat} \AN{42})) to
unify the two expressions, instantiating
\AB{h} to the value 42.  Finally, Agda replaces the
expression \AK{unquote} \AF{helper} expression with the instantiated
value of \AB{h}.  Overall, the effect of \AK{unquote} \AF{helper} is
identical to just writing \AN{42}.  However, the expression inside the
\AF{helper} can be arbitrarily complex and can depend on the syntactic
structure of the term \AB{h} as well as information obtained through
operations in the \AF{TC} monad.

Instead of quoting/unquoting explicitly, we can use the \AK{macro}
keyword to wrap any function with return type \AD{Term} →
\AD{TC} \AD{⊤}.  This takes care of quoting the arguments and unquoting
the result.  On the right, we define a macro \AF{getDef} that
obtains the definition of a name.  The macro calls three functions
from the reflection API.  Firstly, \AF{getDefinition} obtains the definition
of the object with the name \AB{n}.  Secondly, \AF{quoteTC} quotes the
previously obtained definition (resulting in a doubly-quoted expression).
Finally, we \AF{unify} the quoted hole and the doubly quoted definition,
so that after unquoting we get the reflected definition (and not the
original one).  We apply the macro in the last line, and as can be seen,
no \AK{quote}/\AK{unquote} is needed.
More details on reflection in Agda can be found in the user
manual~\cite{agda}.

%
%

\begin{code}[hide]%
\>[0]\AgdaSymbol{\{-\#}\AgdaSpace{}%
\AgdaKeyword{OPTIONS}\AgdaSpace{}%
\AgdaPragma{--rewriting}\AgdaSpace{}%
\AgdaSymbol{\#-\}}\<%
\\
\\[\AgdaEmptyExtraSkip]%
\>[0]\AgdaComment{--open\ import\ Data.Nat\ as\ ℕ\ hiding\ (\AgdaUnderscore{}≟\AgdaUnderscore{})}\<%
\\
\>[0]\AgdaKeyword{open}\AgdaSpace{}%
\AgdaKeyword{import}\AgdaSpace{}%
\AgdaModule{Data.String}\AgdaSpace{}%
\AgdaKeyword{using}\AgdaSpace{}%
\AgdaSymbol{(}\AgdaPostulate{String}\AgdaSymbol{)}\AgdaSpace{}%
\AgdaComment{--;\ length)}\<%
\\
\>[0]\AgdaKeyword{open}\AgdaSpace{}%
\AgdaKeyword{import}\AgdaSpace{}%
\AgdaModule{Data.List}\AgdaSpace{}%
\AgdaSymbol{as}\AgdaSpace{}%
\AgdaModule{𝕃}\AgdaSpace{}%
\AgdaKeyword{using}\AgdaSpace{}%
\AgdaSymbol{(}\AgdaDatatype{List}\AgdaSymbol{;}\AgdaSpace{}%
\AgdaInductiveConstructor{[]}\AgdaSymbol{;}\AgdaSpace{}%
\AgdaOperator{\AgdaInductiveConstructor{\AgdaUnderscore{}∷\AgdaUnderscore{}}}\AgdaSymbol{;}\AgdaSpace{}%
\AgdaOperator{\AgdaFunction{[\AgdaUnderscore{}]}}\AgdaSymbol{)}\<%
\\
\>[0]\AgdaKeyword{open}\AgdaSpace{}%
\AgdaKeyword{import}\AgdaSpace{}%
\AgdaModule{Data.Unit}\AgdaSpace{}%
\AgdaKeyword{hiding}\AgdaSpace{}%
\AgdaSymbol{(}\AgdaOperator{\AgdaFunction{\AgdaUnderscore{}≟\AgdaUnderscore{}}}\AgdaSymbol{)}\<%
\\
\>[0]\AgdaComment{--open\ import\ Data.Bool\ as\ 𝔹\ hiding\ (\AgdaUnderscore{}<\AgdaUnderscore{};\ \AgdaUnderscore{}≟\AgdaUnderscore{};\ T)}\<%
\\
\>[0]\AgdaKeyword{open}\AgdaSpace{}%
\AgdaKeyword{import}\AgdaSpace{}%
\AgdaModule{Function}\<%
\\
\>[0]\AgdaKeyword{module}\AgdaSpace{}%
\AgdaModule{\AgdaUnderscore{}}\AgdaSpace{}%
\AgdaKeyword{where}\<%
\\
\>[0]\AgdaKeyword{postulate}\<%
\\
\>[0][@{}l@{\AgdaIndent{0}}]%
\>[2]\AgdaPostulate{⋯}\AgdaSpace{}%
\AgdaSymbol{:}\AgdaSpace{}%
\AgdaSymbol{∀}\AgdaSpace{}%
\AgdaSymbol{\{}\AgdaBound{a}\AgdaSymbol{\}\{}\AgdaBound{A}\AgdaSpace{}%
\AgdaSymbol{:}\AgdaSpace{}%
\AgdaPrimitive{Set}\AgdaSpace{}%
\AgdaBound{a}\AgdaSymbol{\}}\AgdaSpace{}%
\AgdaSymbol{→}\AgdaSpace{}%
\AgdaBound{A}\<%
\end{code}
\section{\label{sec:basic-extr}Basic Extraction}


In this section we demonstrate basic mechanisms that are necessary when
implementing reflection-based extractors.  To make our examples
concrete, we use a minimalist language called
Kaleidoscope~\cite{kaleidoscope} as our target.  We explain the challenges
and demonstrate Agda code snippets of the actual extractor for Kaleidoscope.

\subsection{Framework Overview}

The extraction process naturally splits into a language-dependent and
a language-independent part. We start by explaining the reusable
language-independent module, which we refer to the as the
``framework''.
The entry point of the framework is a parametrised module \AM{Extract} that
contains a single macro called \AF{kompile}:
\begin{code}[hide]%
\>[0]\AgdaKeyword{module}\AgdaSpace{}%
\AgdaModule{ExtrMod}\AgdaSpace{}%
\AgdaKeyword{where}\<%
\\
\>[0][@{}l@{\AgdaIndent{0}}]%
\>[2]\AgdaKeyword{open}\AgdaSpace{}%
\AgdaKeyword{import}\AgdaSpace{}%
\AgdaModule{Reflection}\<%
\\
\>[2]\AgdaKeyword{open}\AgdaSpace{}%
\AgdaKeyword{import}\AgdaSpace{}%
\AgdaModule{Data.Nat}\AgdaSpace{}%
\AgdaSymbol{as}\AgdaSpace{}%
\AgdaModule{ℕ}\AgdaSpace{}%
\AgdaKeyword{hiding}\AgdaSpace{}%
\AgdaSymbol{(}\AgdaOperator{\AgdaFunction{\AgdaUnderscore{}≟\AgdaUnderscore{}}}\AgdaSymbol{)}\AgdaSpace{}%
\AgdaKeyword{public}\<%
\\
\>[2]\AgdaFunction{SKS}\AgdaSpace{}%
\AgdaSymbol{:}\AgdaSpace{}%
\AgdaPrimitive{Set}\AgdaSpace{}%
\AgdaSymbol{→}\AgdaSpace{}%
\AgdaPrimitive{Set}\AgdaSpace{}%
\AgdaSymbol{;}\AgdaSpace{}%
\AgdaFunction{SKS}\AgdaSpace{}%
\AgdaSymbol{=}\AgdaSpace{}%
\AgdaPostulate{⋯}\<%
\\
\>[2]\AgdaFunction{Prog}\AgdaSpace{}%
\AgdaSymbol{:}\AgdaSpace{}%
\AgdaPrimitive{Set}\AgdaSpace{}%
\AgdaSymbol{;}\AgdaSpace{}%
\AgdaFunction{Prog}\AgdaSpace{}%
\AgdaSymbol{=}\AgdaSpace{}%
\AgdaPostulate{⋯}\<%
\\
\>[2]\AgdaKeyword{module}\AgdaSpace{}%
\AgdaModule{Kaleid}\AgdaSpace{}%
\AgdaKeyword{where}\<%
\\
\>[2][@{}l@{\AgdaIndent{0}}]%
\>[4]\AgdaFunction{kompile-fun}\AgdaSpace{}%
\AgdaSymbol{:}\AgdaSpace{}%
\AgdaFunction{Type}\AgdaSpace{}%
\AgdaSymbol{→}\AgdaSpace{}%
\AgdaDatatype{Term}\AgdaSpace{}%
\AgdaSymbol{→}\AgdaSpace{}%
\AgdaPostulate{Name}\AgdaSpace{}%
\AgdaSymbol{→}\AgdaSpace{}%
\AgdaFunction{SKS}\AgdaSpace{}%
\AgdaFunction{Prog}\<%
\\
\>[4]\AgdaFunction{kompile-fun}\AgdaSpace{}%
\AgdaSymbol{=}\AgdaSpace{}%
\AgdaPostulate{⋯}\<%
\end{code}
\begin{code}%
\>[2]\AgdaKeyword{module}\AgdaSpace{}%
\AgdaModule{Extract}\AgdaSpace{}%
\AgdaSymbol{(}\AgdaBound{kompile-fun}\AgdaSpace{}%
\AgdaSymbol{:}\AgdaSpace{}%
\AgdaFunction{Type}\AgdaSpace{}%
\AgdaSymbol{→}\AgdaSpace{}%
\AgdaDatatype{Term}\AgdaSpace{}%
\AgdaSymbol{→}\AgdaSpace{}%
\AgdaPostulate{Name}\AgdaSpace{}%
\AgdaSymbol{→}\AgdaSpace{}%
\AgdaFunction{SKS}\AgdaSpace{}%
\AgdaFunction{Prog}\AgdaSymbol{)}\AgdaSpace{}%
\AgdaKeyword{where}\<%
\\
\>[2][@{}l@{\AgdaIndent{0}}]%
\>[4]\AgdaKeyword{macro}\<%
\\
\>[4][@{}l@{\AgdaIndent{0}}]%
\>[6]\AgdaFunction{kompile}\AgdaSpace{}%
\AgdaSymbol{:}\AgdaSpace{}%
\AgdaPostulate{Name}\AgdaSpace{}%
\AgdaSymbol{→}\AgdaSpace{}%
\AgdaFunction{Names}\AgdaSpace{}%
\AgdaSymbol{→}\AgdaSpace{}%
\AgdaFunction{Names}\AgdaSpace{}%
\AgdaSymbol{→}\AgdaSpace{}%
\AgdaSymbol{(}\AgdaDatatype{Term}\AgdaSpace{}%
\AgdaSymbol{→}\AgdaSpace{}%
\AgdaPostulate{TC}\AgdaSpace{}%
\AgdaRecord{⊤}\AgdaSymbol{)}\<%
\\
\>[6]\AgdaFunction{kompile}\AgdaSpace{}%
\AgdaBound{n}\AgdaSpace{}%
\AgdaBound{base}\AgdaSpace{}%
\AgdaBound{skip}\AgdaSpace{}%
\AgdaBound{hole}%
\>[98I]\AgdaSymbol{=}\AgdaSpace{}%
\AgdaComment{--}\<%
\end{code}
\begin{code}[hide]%
\>[98I][@{}l@{\AgdaIndent{1}}]%
\>[32]\AgdaPostulate{quoteTC}\AgdaSpace{}%
\AgdaNumber{42}\AgdaSpace{}%
\AgdaOperator{\AgdaFunction{>>=}}\AgdaSpace{}%
\AgdaPostulate{unify}\AgdaSpace{}%
\AgdaBound{hole}\<%
\end{code}
The \AB{kompile-fun} parameter is a language-specific function
specifying how to compile a single Agda function, given by
its type, body and name.
It operates within the \AF{SKS} state monad, which it can use to
register other functions that should also be extracted, and returns
either a representation of the extracted function in the target
language in case extraction succeeded, or an error message, as
specified by the \AD{Prog} type.
\begin{mathpar}
\codeblock{
\begin{code}%
\>[0]\AgdaKeyword{data}\AgdaSpace{}%
\AgdaDatatype{Err}\AgdaSpace{}%
\AgdaSymbol{\{}\AgdaBound{ℓ}\AgdaSymbol{\}}\AgdaSpace{}%
\AgdaSymbol{(}\AgdaBound{A}\AgdaSpace{}%
\AgdaSymbol{:}\AgdaSpace{}%
\AgdaPrimitive{Set}\AgdaSpace{}%
\AgdaBound{ℓ}\AgdaSymbol{)}\AgdaSpace{}%
\AgdaSymbol{:}\AgdaSpace{}%
\AgdaPrimitive{Set}\AgdaSpace{}%
\AgdaBound{ℓ}\AgdaSpace{}%
\AgdaKeyword{where}\<%
\\
\>[0][@{}l@{\AgdaIndent{0}}]%
\>[2]\AgdaInductiveConstructor{error}\AgdaSpace{}%
\AgdaSymbol{:}\AgdaSpace{}%
\AgdaPostulate{String}\AgdaSpace{}%
\AgdaSymbol{→}\AgdaSpace{}%
\AgdaDatatype{Err}\AgdaSpace{}%
\AgdaBound{A}\<%
\\
\>[2]\AgdaInductiveConstructor{ok}%
\>[8]\AgdaSymbol{:}\AgdaSpace{}%
\AgdaBound{A}\AgdaSpace{}%
\AgdaSymbol{→}\AgdaSpace{}%
\AgdaDatatype{Err}\AgdaSpace{}%
\AgdaBound{A}\<%
\end{code}
}
\and
\codeblock{
\begin{code}[hide]%
\>[0]\AgdaKeyword{module}\AgdaSpace{}%
\AgdaModule{ExtrStructMod}\AgdaSpace{}%
\AgdaKeyword{where}\<%
\\
\>[0][@{}l@{\AgdaIndent{0}}]%
\>[2]\AgdaKeyword{open}\AgdaSpace{}%
\AgdaKeyword{import}\AgdaSpace{}%
\AgdaModule{Data.Nat}\AgdaSpace{}%
\AgdaSymbol{as}\AgdaSpace{}%
\AgdaModule{ℕ}\AgdaSpace{}%
\AgdaKeyword{hiding}\AgdaSpace{}%
\AgdaSymbol{(}\AgdaOperator{\AgdaFunction{\AgdaUnderscore{}≟\AgdaUnderscore{}}}\AgdaSymbol{)}\<%
\\
\>[2]\AgdaKeyword{open}\AgdaSpace{}%
\AgdaModule{ExtrMod}\AgdaSpace{}%
\AgdaKeyword{using}\AgdaSpace{}%
\AgdaSymbol{(}\AgdaKeyword{module}\AgdaSpace{}%
\AgdaModule{Kaleid}\AgdaSymbol{;}\AgdaSpace{}%
\AgdaKeyword{module}\AgdaSpace{}%
\AgdaModule{Extract}\AgdaSymbol{)}\<%
\end{code}
\begin{code}%
\>[2]\AgdaFunction{SKS}\AgdaSpace{}%
\AgdaSymbol{:}\AgdaSpace{}%
\AgdaPrimitive{Set}\AgdaSpace{}%
\AgdaSymbol{→}\AgdaSpace{}%
\AgdaPrimitive{Set}%
\>[21]\AgdaComment{--\ State\ Monad\ (KS)}\<%
\\
\>[2]\AgdaFunction{TC}%
\>[6]\AgdaSymbol{:}\AgdaSpace{}%
\AgdaPrimitive{Set}\AgdaSpace{}%
\AgdaSymbol{→}\AgdaSpace{}%
\AgdaPrimitive{Set}%
\>[21]\AgdaComment{--\ TypeChecking\ Monad}\<%
\\
\>[2]\AgdaFunction{Prog}\AgdaSpace{}%
\AgdaSymbol{=}\AgdaSpace{}%
\AgdaDatatype{Err}\AgdaSpace{}%
\AgdaPostulate{String}%
\>[21]\AgdaComment{--\ String\ or\ Error}\<%
\end{code}
}
\end{mathpar}
\begin{code}[hide]%
\>[2]\AgdaFunction{SKS}\AgdaSpace{}%
\AgdaSymbol{=}\AgdaSpace{}%
\AgdaPostulate{⋯}\<%
\\
\>[2]\AgdaFunction{TC}\AgdaSpace{}%
\AgdaSymbol{=}\AgdaSpace{}%
\AgdaPostulate{⋯}\<%
\end{code}

Assuming \AF{kompile-fun} is defined
in the module \AM{Kaleid} here is how to instantiate the framework:
\begin{code}%
\>[2]\AgdaKeyword{open}\AgdaSpace{}%
\AgdaModule{Extract}\AgdaSpace{}%
\AgdaSymbol{(}\AgdaFunction{Kaleid.kompile-fun}\AgdaSymbol{)}\<%
\\
\>[2]\AgdaFunction{foo}\AgdaSpace{}%
\AgdaSymbol{:}\AgdaSpace{}%
\AgdaDatatype{ℕ}\AgdaSpace{}%
\AgdaSymbol{→}\AgdaSpace{}%
\AgdaDatatype{ℕ}\AgdaSpace{}%
\AgdaSymbol{;}\AgdaSpace{}%
\AgdaFunction{foo}\AgdaSpace{}%
\AgdaSymbol{=}\AgdaSpace{}%
\AgdaPostulate{⋯}\AgdaSpace{}%
\AgdaSymbol{;}\AgdaSpace{}%
\AgdaFunction{base-functions}\AgdaSpace{}%
\AgdaSymbol{=}\AgdaSpace{}%
\AgdaPostulate{⋯}\AgdaSpace{}%
\AgdaSymbol{;}\AgdaSpace{}%
\AgdaFunction{skip-functions}\AgdaSpace{}%
\AgdaSymbol{=}\AgdaSpace{}%
\AgdaPostulate{⋯}\<%
\\
\>[2]\AgdaFunction{extracted}\AgdaSpace{}%
\AgdaSymbol{=}\AgdaSpace{}%
\AgdaMacro{kompile}\AgdaSpace{}%
\AgdaFunction{foo}\AgdaSpace{}%
\AgdaFunction{base-functions}\AgdaSpace{}%
\AgdaFunction{skip-functions}\<%
\end{code}
The first argument to the \AF{kompile} macro is the name of the main
function that we want to extract, in this case \AF{foo}. The second
and the third parameters of \AF{kompile} are lists of names that
control function inlining in the extracted terms and traversal into
the definitions found in the call graph (which are added by the
\AF{kompile-term} function explained in Section~\ref{sec:translating-terms}).
The \AF{kompile}
macro obtains the normalised type and body of the main function, runs
\AF{kompile-fun} for the actual extraction and recursively extracts
any functions that have been registered for extraction during the
processing of \AF{foo}.

To avoid repeated extraction of the same function, \AF{kompile}
keeps track of already compiled functions.%
\footnote{We had to postulate termination of this traversal as
  since the reflection API of Agda currently does not
  provide the guarantee that there is only a finite number of function
  symbols.}
After all required functions have been compiled, the bodies
of all extracted functions are concatenated and
returned as the result of extraction.

\subsection{Kaleidoscope}
We borrow the Kaleidoscope example language from the tutorial on
building frontends to LLVM~\cite{kaleidoscope}. Kaleidoscope is a minimalist first-order
language with a single data type, the type of natural numbers\footnote{The original
version of Kaleidoscope used floats, but natural numbers are easier in the context of Agda
as we can prove more properties of them.},
with basic arithmetic operations and comparisons. Following C convention, boolean values
are encoded as numbers where 0 is false, and any other value is
true.  Function calls and conditionals operate as usual, and `let'
makes it possible
to bind immutable values to variables.  We extend Kaleidoscope with a one-argument \AF{assert}
operator that terminates the program if its argument evaluates to zero.
Functions are defined by giving a name, a list of arguments and the
expression for its body.  External functions are defined by giving a name and a list of its
arguments. We encode Kaleidoskope's AST as follows:

\begin{code}[hide]%
\>[0]\AgdaKeyword{module}\AgdaSpace{}%
\AgdaModule{Kaleid}\AgdaSpace{}%
\AgdaKeyword{where}\<%
\\
\>[0][@{}l@{\AgdaIndent{0}}]%
\>[2]\AgdaKeyword{open}\AgdaSpace{}%
\AgdaKeyword{import}\AgdaSpace{}%
\AgdaModule{Data.Nat}\AgdaSpace{}%
\AgdaSymbol{as}\AgdaSpace{}%
\AgdaModule{ℕ}\AgdaSpace{}%
\AgdaKeyword{hiding}\AgdaSpace{}%
\AgdaSymbol{(}\AgdaOperator{\AgdaFunction{\AgdaUnderscore{}≟\AgdaUnderscore{}}}\AgdaSymbol{)}\<%
\end{code}
\begin{mathpar}
\codeblock{%
\begin{code}%
\>[2]\AgdaFunction{Id}\AgdaSpace{}%
\AgdaSymbol{=}\AgdaSpace{}%
\AgdaPostulate{String}\<%
\\
\\[\AgdaEmptyExtraSkip]%
\>[2]\AgdaKeyword{data}\AgdaSpace{}%
\AgdaDatatype{Op}\AgdaSpace{}%
\AgdaSymbol{:}\AgdaSpace{}%
\AgdaPrimitive{Set}\AgdaSpace{}%
\AgdaKeyword{where}\<%
\\
\>[2][@{}l@{\AgdaIndent{0}}]%
\>[4]\AgdaInductiveConstructor{Plus}\AgdaSpace{}%
\AgdaInductiveConstructor{Minus}\AgdaSpace{}%
\AgdaInductiveConstructor{Times}\AgdaSpace{}%
\AgdaInductiveConstructor{Divide}\AgdaSpace{}%
\AgdaSymbol{:}\AgdaSpace{}%
\AgdaDatatype{Op}\<%
\\
\>[4]\AgdaInductiveConstructor{Eq}\AgdaSpace{}%
\AgdaInductiveConstructor{Neq}\AgdaSpace{}%
\AgdaInductiveConstructor{And}\AgdaSpace{}%
\AgdaInductiveConstructor{Gt}\AgdaSpace{}%
\AgdaInductiveConstructor{Lt}\AgdaSpace{}%
\AgdaSymbol{:}\AgdaSpace{}%
\AgdaDatatype{Op}\<%
\\
\\[\AgdaEmptyExtraSkip]%
\>[2]\AgdaKeyword{data}\AgdaSpace{}%
\AgdaDatatype{Expr}\AgdaSpace{}%
\AgdaSymbol{:}\AgdaSpace{}%
\AgdaPrimitive{Set}\AgdaSpace{}%
\AgdaKeyword{where}\<%
\\
\>[2][@{}l@{\AgdaIndent{0}}]%
\>[4]\AgdaInductiveConstructor{Nat}%
\>[13]\AgdaSymbol{:}\AgdaSpace{}%
\AgdaDatatype{ℕ}\AgdaSpace{}%
\AgdaSymbol{→}\AgdaSpace{}%
\AgdaDatatype{Expr}\<%
\\
\>[4]\AgdaInductiveConstructor{BinOp}%
\>[13]\AgdaSymbol{:}\AgdaSpace{}%
\AgdaDatatype{Op}\AgdaSpace{}%
\AgdaSymbol{→}\AgdaSpace{}%
\AgdaDatatype{Expr}\AgdaSpace{}%
\AgdaSymbol{→}\AgdaSpace{}%
\AgdaDatatype{Expr}\AgdaSpace{}%
\AgdaSymbol{→}\AgdaSpace{}%
\AgdaDatatype{Expr}\<%
\\
\>[4]\AgdaInductiveConstructor{Var}%
\>[13]\AgdaSymbol{:}\AgdaSpace{}%
\AgdaPostulate{String}\AgdaSpace{}%
\AgdaSymbol{→}\AgdaSpace{}%
\AgdaDatatype{Expr}\<%
\\
\>[4]\AgdaInductiveConstructor{Call}%
\>[13]\AgdaSymbol{:}\AgdaSpace{}%
\AgdaFunction{Id}\AgdaSpace{}%
\AgdaSymbol{→}\AgdaSpace{}%
\AgdaDatatype{List}\AgdaSpace{}%
\AgdaDatatype{Expr}\AgdaSpace{}%
\AgdaSymbol{→}\AgdaSpace{}%
\AgdaDatatype{Expr}\<%
\\
\>[4]\AgdaInductiveConstructor{Function}\AgdaSpace{}%
\AgdaSymbol{:}\AgdaSpace{}%
\AgdaFunction{Id}\AgdaSpace{}%
\AgdaSymbol{→}\AgdaSpace{}%
\AgdaDatatype{List}\AgdaSpace{}%
\AgdaFunction{Id}\AgdaSpace{}%
\AgdaSymbol{→}\AgdaSpace{}%
\AgdaDatatype{Expr}\AgdaSpace{}%
\AgdaSymbol{→}\AgdaSpace{}%
\AgdaDatatype{Expr}\<%
\\
\>[4]\AgdaInductiveConstructor{Extern}%
\>[13]\AgdaSymbol{:}\AgdaSpace{}%
\AgdaFunction{Id}\AgdaSpace{}%
\AgdaSymbol{→}\AgdaSpace{}%
\AgdaDatatype{List}\AgdaSpace{}%
\AgdaFunction{Id}\AgdaSpace{}%
\AgdaSymbol{→}\AgdaSpace{}%
\AgdaDatatype{Expr}\<%
\\
\>[4]\AgdaInductiveConstructor{Let}%
\>[13]\AgdaSymbol{:}\AgdaSpace{}%
\AgdaFunction{Id}\AgdaSpace{}%
\AgdaSymbol{→}\AgdaSpace{}%
\AgdaDatatype{Expr}\AgdaSpace{}%
\AgdaSymbol{→}\AgdaSpace{}%
\AgdaDatatype{Expr}\AgdaSpace{}%
\AgdaSymbol{→}\AgdaSpace{}%
\AgdaDatatype{Expr}\<%
\\
\>[4]\AgdaInductiveConstructor{Assert}%
\>[13]\AgdaSymbol{:}\AgdaSpace{}%
\AgdaDatatype{Expr}\AgdaSpace{}%
\AgdaSymbol{→}\AgdaSpace{}%
\AgdaDatatype{Expr}\<%
\\
\>[4]\AgdaInductiveConstructor{If}%
\>[13]\AgdaSymbol{:}\AgdaSpace{}%
\AgdaDatatype{Expr}\AgdaSpace{}%
\AgdaSymbol{→}\AgdaSpace{}%
\AgdaDatatype{Expr}\AgdaSpace{}%
\AgdaSymbol{→}\AgdaSpace{}%
\AgdaDatatype{Expr}\AgdaSpace{}%
\AgdaSymbol{→}\AgdaSpace{}%
\AgdaDatatype{Expr}\<%
\end{code}
}
\and
\codeblock{\begin{code}%
\>[2]\AgdaComment{--\ Recursive\ Fibonacci\ function:}\<%
\\
\>[2]\AgdaFunction{fib}\AgdaSpace{}%
\AgdaSymbol{=}%
\>[255I]\AgdaInductiveConstructor{Function}\AgdaSpace{}%
\AgdaString{"fib"}\AgdaSpace{}%
\AgdaSymbol{(}\AgdaString{"n"}\AgdaSpace{}%
\AgdaOperator{\AgdaInductiveConstructor{∷}}\AgdaSpace{}%
\AgdaInductiveConstructor{[]}\AgdaSymbol{)}\AgdaSpace{}%
\AgdaOperator{\AgdaFunction{\$}}\<%
\\
\>[.][@{}l@{}]\<[255I]%
\>[8]\AgdaInductiveConstructor{If}%
\>[261I]\AgdaSymbol{(}\AgdaInductiveConstructor{BinOp}\AgdaSpace{}%
\AgdaInductiveConstructor{Lt}\AgdaSpace{}%
\AgdaSymbol{(}\AgdaInductiveConstructor{Var}\AgdaSpace{}%
\AgdaString{"n"}\AgdaSymbol{)}\AgdaSpace{}%
\AgdaSymbol{(}\AgdaInductiveConstructor{Nat}\AgdaSpace{}%
\AgdaNumber{2}\AgdaSymbol{))}\<%
\\
\>[.][@{}l@{}]\<[261I]%
\>[11]\AgdaSymbol{(}\AgdaInductiveConstructor{Nat}\AgdaSpace{}%
\AgdaNumber{1}\AgdaSymbol{)}\<%
\\
\>[11]\AgdaSymbol{(}\AgdaInductiveConstructor{BinOp}\AgdaSpace{}%
\AgdaInductiveConstructor{Plus}\<%
\\
\>[11][@{}l@{\AgdaIndent{0}}]%
\>[12]\AgdaSymbol{(}\AgdaInductiveConstructor{Call}%
\>[269I]\AgdaString{"fib"}\<%
\\
\>[.][@{}l@{}]\<[269I]%
\>[18]\AgdaOperator{\AgdaFunction{[}}\AgdaSpace{}%
\AgdaInductiveConstructor{BinOp}\AgdaSpace{}%
\AgdaInductiveConstructor{Minus}\AgdaSpace{}%
\AgdaSymbol{(}\AgdaInductiveConstructor{Var}\AgdaSpace{}%
\AgdaString{"n"}\AgdaSymbol{)}\AgdaSpace{}%
\AgdaSymbol{(}\AgdaInductiveConstructor{Nat}\AgdaSpace{}%
\AgdaNumber{2}\AgdaSymbol{)}\AgdaOperator{\AgdaFunction{]}}\AgdaSymbol{)}\<%
\\
\>[12]\AgdaSymbol{(}\AgdaInductiveConstructor{Call}%
\>[276I]\AgdaString{"fib"}\<%
\\
\>[.][@{}l@{}]\<[276I]%
\>[18]\AgdaOperator{\AgdaFunction{[}}\AgdaSpace{}%
\AgdaInductiveConstructor{BinOp}\AgdaSpace{}%
\AgdaInductiveConstructor{Minus}\AgdaSpace{}%
\AgdaSymbol{(}\AgdaInductiveConstructor{Var}\AgdaSpace{}%
\AgdaString{"n"}\AgdaSymbol{)}\AgdaSpace{}%
\AgdaSymbol{(}\AgdaInductiveConstructor{Nat}\AgdaSpace{}%
\AgdaNumber{1}\AgdaSymbol{)}\AgdaOperator{\AgdaFunction{]}}\AgdaSymbol{))}\<%
\end{code}
}
\end{mathpar}

\subsection{A shallow embedding of Kaleidoscope in Agda}
To extract a Kaleidoscope program from an Agda program,
we first need to identify what subset of Agda can be sensibly translated to Kaleidoscope.
Let us start with the types. First, we need the natural
number type \AD{ℕ} as it is the main data type of Kaleidoscope. To
describe invariants we also support the type  \AD{Fin n} of natural number
strictly less than \AD{n}, as well as the identity type
\AD{\_≡\_} and the inequality type \AD{\_<\_} on natural numbers.
The \AD{Fin} type is mapped to numbers in the target language, while
all proofs of \AD{\_≡\_} and \AD{\_<\_} are mapped to the constant \AN{1}.
We also allow the decidability predicates
\AD{Dec (a ≡ b)} and \AD{Dec (a < b)}, which carry a boolean
value and a proof that the relation holds or does not hold,
depending on the value of the boolean.
We map \AC{true} to \AN{1} and \AC{false} to \AN{0}, ignoring the proof. First order
functions of the above types such as basic arithmetic \AD{\_+\_}, \AD{\_-\_}, \etc{}
are mapped to corresponding functions in the target language.

While it is tempting to say that any Agda term of the above types could
be translated into Kaleidoscope, this is not the case.  For example, consider
a function:
\begin{code}[hide]%
\>[0]\AgdaKeyword{module}\AgdaSpace{}%
\AgdaModule{Problem}\AgdaSpace{}%
\AgdaKeyword{where}\<%
\\
\>[0][@{}l@{\AgdaIndent{0}}]%
\>[2]\AgdaKeyword{open}\AgdaSpace{}%
\AgdaKeyword{import}\AgdaSpace{}%
\AgdaModule{Data.Nat.Show}\AgdaSpace{}%
\AgdaKeyword{renaming}\AgdaSpace{}%
\AgdaSymbol{(}\AgdaFunction{show}\AgdaSpace{}%
\AgdaSymbol{to}\AgdaSpace{}%
\AgdaFunction{showNat}\AgdaSymbol{)}\<%
\\
\>[2]\AgdaKeyword{open}\AgdaSpace{}%
\AgdaKeyword{import}\AgdaSpace{}%
\AgdaModule{Data.String}\AgdaSpace{}%
\AgdaKeyword{using}\AgdaSpace{}%
\AgdaSymbol{(}\AgdaFunction{length}\AgdaSymbol{)}\<%
\\
\>[2]\AgdaKeyword{open}\AgdaSpace{}%
\AgdaKeyword{import}\AgdaSpace{}%
\AgdaModule{Data.Nat}\AgdaSpace{}%
\AgdaSymbol{as}\AgdaSpace{}%
\AgdaModule{ℕ}\AgdaSpace{}%
\AgdaKeyword{hiding}\AgdaSpace{}%
\AgdaSymbol{(}\AgdaOperator{\AgdaFunction{\AgdaUnderscore{}≟\AgdaUnderscore{}}}\AgdaSymbol{)}\<%
\end{code}
\begin{code}%
\>[2]\AgdaFunction{ex}\AgdaSpace{}%
\AgdaSymbol{:}\AgdaSpace{}%
\AgdaDatatype{ℕ}\AgdaSpace{}%
\AgdaSymbol{→}\AgdaSpace{}%
\AgdaDatatype{ℕ}\<%
\\
\>[2]\AgdaFunction{ex}\AgdaSpace{}%
\AgdaBound{x}\AgdaSpace{}%
\AgdaSymbol{=}\AgdaSpace{}%
\AgdaFunction{length}\AgdaSpace{}%
\AgdaSymbol{(}\AgdaFunction{showNat}\AgdaSpace{}%
\AgdaBound{x}\AgdaSymbol{)}\<%
\end{code}
where \AF{showNat} returns a string representation of
the given number.  Neither \AF{length} nor \AF{showNat}
are representable in Kaleidoscope, as there is no notion of strings
in the language.
To pin down precisely what fragment of Agda we can extract,
we would have to restrict what
types are allowed in embedded functions and what terms can appear
in function bodies, taking us away from the world of
shallow embeddings and into the world of deep embeddings.
While it is certainly possible to define strongly typed deep
embeddings in a dependently typed host language, all current solutions
are very heavyweight when one has to deal with an embedded language
that uses dependent types, as we do here (recall that we allow
\AF{\_<\_}, \AF{\_≡\_}, \etc{}). In particular, one needs to encode
not only types and terms of the embedded language, but also contexts
and explicit substitutions, turning even the simplest programs into
large and non-trivial terms.

It is still an open question whether
there exists a satisfying middle ground between shallow and
deep embedding.
%
Our solution in this paper is to avoid the encoding problem entirely
and rely instead on metaprogramming to extract a subset of Agda into
our target language.  An Agda term is defined to belong to the
embedding if the extractor does not fail on it.


\subsection{Normalisation} \label{sec:normalisation}
Working with a shallow embedding gives us an important benefit: we may use any host
language constructs that are not present in the embedding, as long as they can
be eliminated prior to extraction.  For example, the target language may not
support polymorphic or higher-order functions, yet we could write
programs such as:
\begin{mathpar}
\codeblock{\begin{code}[hide]%
\>[0]\AgdaKeyword{module}\AgdaSpace{}%
\AgdaModule{NormMod}\AgdaSpace{}%
\AgdaKeyword{where}\<%
\\
\>[0][@{}l@{\AgdaIndent{0}}]%
\>[2]\AgdaKeyword{open}\AgdaSpace{}%
\AgdaKeyword{import}\AgdaSpace{}%
\AgdaModule{Data.String}\AgdaSpace{}%
\AgdaKeyword{using}\AgdaSpace{}%
\AgdaSymbol{(}\AgdaFunction{length}\AgdaSymbol{)}\<%
\\
\>[2]\AgdaKeyword{open}\AgdaSpace{}%
\AgdaKeyword{import}\AgdaSpace{}%
\AgdaModule{Data.Product}\<%
\\
\>[2]\AgdaKeyword{open}\AgdaSpace{}%
\AgdaKeyword{import}\AgdaSpace{}%
\AgdaModule{Data.Nat}\AgdaSpace{}%
\AgdaSymbol{as}\AgdaSpace{}%
\AgdaModule{ℕ}\AgdaSpace{}%
\AgdaKeyword{hiding}\AgdaSpace{}%
\AgdaSymbol{(}\AgdaOperator{\AgdaFunction{\AgdaUnderscore{}≟\AgdaUnderscore{}}}\AgdaSymbol{)}\<%
\end{code}
\begin{code}%
\>[2]\AgdaFunction{ex}\AgdaSpace{}%
\AgdaSymbol{:}\AgdaSpace{}%
\AgdaSymbol{(}\AgdaBound{n}\AgdaSpace{}%
\AgdaSymbol{:}\AgdaSpace{}%
\AgdaDatatype{ℕ}\AgdaSymbol{)}\AgdaSpace{}%
\AgdaSymbol{→}\AgdaSpace{}%
\AgdaBound{n}\AgdaSpace{}%
\AgdaOperator{\AgdaFunction{<}}\AgdaSpace{}%
\AgdaFunction{length}\AgdaSpace{}%
\AgdaString{"999"}\AgdaSpace{}%
\AgdaSymbol{→}\AgdaSpace{}%
\AgdaDatatype{ℕ}\<%
\\
\>[2]\AgdaFunction{ex}\AgdaSpace{}%
\AgdaSymbol{=}\AgdaSpace{}%
\AgdaPostulate{⋯}\<%
\end{code}}
\and
\codeblock{\begin{code}%
\>[2]\AgdaFunction{fib}\AgdaSpace{}%
\AgdaSymbol{:}\AgdaSpace{}%
\AgdaSymbol{(}\AgdaBound{k}\AgdaSpace{}%
\AgdaBound{m}\AgdaSpace{}%
\AgdaBound{n}\AgdaSpace{}%
\AgdaSymbol{:}\AgdaSpace{}%
\AgdaDatatype{ℕ}\AgdaSymbol{)}\AgdaSpace{}%
\AgdaSymbol{→}\AgdaSpace{}%
\AgdaDatatype{ℕ}\<%
\\
\>[2]\AgdaFunction{fib}\AgdaSpace{}%
\AgdaNumber{0}%
\>[15]\AgdaBound{m}\AgdaSpace{}%
\AgdaBound{n}%
\>[20]\AgdaSymbol{=}\AgdaSpace{}%
\AgdaBound{m}\<%
\\
\>[2]\AgdaFunction{fib}\AgdaSpace{}%
\AgdaSymbol{(}\AgdaInductiveConstructor{suc}\AgdaSpace{}%
\AgdaBound{k}\AgdaSymbol{)}%
\>[15]\AgdaBound{m}\AgdaSpace{}%
\AgdaBound{n}%
\>[20]\AgdaSymbol{=}%
\>[351I]\AgdaKeyword{let}%
\>[27]\AgdaBound{m'}\AgdaSpace{}%
\AgdaOperator{\AgdaInductiveConstructor{,}}\AgdaSpace{}%
\AgdaBound{n'}\AgdaSpace{}%
\AgdaSymbol{=}\AgdaSpace{}%
\AgdaBound{n}\AgdaSpace{}%
\AgdaOperator{\AgdaInductiveConstructor{,}}\AgdaSpace{}%
\AgdaBound{m}\AgdaSpace{}%
\AgdaOperator{\AgdaPrimitive{+}}\AgdaSpace{}%
\AgdaBound{n}\<%
\\
\>[.][@{}l@{}]\<[351I]%
\>[22]\AgdaKeyword{in}%
\>[27]\AgdaFunction{fib}\AgdaSpace{}%
\AgdaBound{k}\AgdaSpace{}%
\AgdaBound{m'}\AgdaSpace{}%
\AgdaBound{n'}\<%
\end{code}}
\end{mathpar}
In the type of \AF{ex}, \AF{length} is a function from \AD{String} to \AD{ℕ}, but
it is applied to a constant string.  In the second clause of \AF{fib} we
create a tuple (\AB{n} \AC{,} \AB{m} \AF{+} \AB{n})
and immediately destruct it via pattern matching. Note that Kaleidoscope
supports neither strings nor tuples, so neither \AF{length} nor \AF{\_,\_}
can be part of the final
extracted Kaleidoscope code.
However, if we simplify the terms, the result no longer contains any
about strings or tuples, and hence can be extracted safely.

Such a simplification can be conveniently achieved by
normalising the terms, \ie{}~by applying reduction rules to (sub)terms until
they turn into values or neutral terms.
%
Agda's reflection API offers a function \AF{normalise} for this purpose.
However, this only normalises the term itself and not the body of
functions used in this term.
This is a technical limitation that has to do with the internal
representation of pattern-matching functions.
To work around this limitation, we also recursively traverse the
definition of each function used in the term and normalise all terms
in their bodies.

During the implementation of this traversal, we were faced with
the challenge of reconstructing the right typing context for each clause.
Agda constructs this context internally during elaboration of the clauses,
but the reflection API did not provide access to it. Rather than going
through the painful and error-prone process of reconstructing this
context, we instead extended the reflection API to provide it for us
(see \url{https://github.com/agda/agda/pull/4722} for the full story).

\subsection{Controlling Reduction} \label{sec:controlling-reduction}

Fully normalising a term sometimes leads to undesirable results.
Consider the following program:
\begin{code}[hide]%
\>[0]\AgdaKeyword{module}\AgdaSpace{}%
\AgdaModule{RedMod}\AgdaSpace{}%
\AgdaKeyword{where}\<%
\\
\>[0][@{}l@{\AgdaIndent{0}}]%
\>[2]\AgdaKeyword{open}\AgdaSpace{}%
\AgdaKeyword{import}\AgdaSpace{}%
\AgdaModule{Relation.Nullary}\<%
\\
\>[2]\AgdaKeyword{open}\AgdaSpace{}%
\AgdaKeyword{import}\AgdaSpace{}%
\AgdaModule{Data.Nat}\AgdaSpace{}%
\AgdaSymbol{as}\AgdaSpace{}%
\AgdaModule{ℕ}\<%
\end{code}
\begin{code}%
\>[2]\AgdaFunction{ex₅}\AgdaSpace{}%
\AgdaSymbol{:}\AgdaSpace{}%
\AgdaDatatype{ℕ}\AgdaSpace{}%
\AgdaSymbol{→}\AgdaSpace{}%
\AgdaDatatype{ℕ}\<%
\\
\>[2]\AgdaFunction{ex₅}\AgdaSpace{}%
\AgdaBound{x}\AgdaSpace{}%
\AgdaKeyword{with}\AgdaSpace{}%
\AgdaBound{x}\AgdaSpace{}%
\AgdaOperator{\AgdaFunction{≟}}\AgdaSpace{}%
\AgdaNumber{42}\<%
\\
\>[2]\AgdaSymbol{...}\AgdaSpace{}%
\AgdaSymbol{|}\AgdaSpace{}%
\AgdaInductiveConstructor{yes}\AgdaSpace{}%
\AgdaSymbol{\AgdaUnderscore{}}\AgdaSpace{}%
\AgdaSymbol{=}\AgdaSpace{}%
\AgdaNumber{10}\<%
\\
\>[2]\AgdaSymbol{...}\AgdaSpace{}%
\AgdaSymbol{|}\AgdaSpace{}%
\AgdaInductiveConstructor{no}%
\>[12]\AgdaSymbol{\AgdaUnderscore{}}\AgdaSpace{}%
\AgdaSymbol{=}\AgdaSpace{}%
\AgdaNumber{20}\<%
\end{code}
The definition of \AF{\_≟\_} in the standard library is quite complex:
\begin{code}[hide]%
\>[0]\AgdaKeyword{module}\AgdaSpace{}%
\AgdaModule{RedMod2}\AgdaSpace{}%
\AgdaKeyword{where}\<%
\\
\>[0][@{}l@{\AgdaIndent{0}}]%
\>[2]\AgdaKeyword{open}\AgdaSpace{}%
\AgdaKeyword{import}\AgdaSpace{}%
\AgdaModule{Relation.Binary.PropositionalEquality}\<%
\\
\>[2]\AgdaKeyword{open}\AgdaSpace{}%
\AgdaKeyword{import}\AgdaSpace{}%
\AgdaModule{Relation.Binary}\<%
\\
\>[2]\AgdaKeyword{open}\AgdaSpace{}%
\AgdaKeyword{import}\AgdaSpace{}%
\AgdaModule{Relation.Nullary}\<%
\\
\>[2]\AgdaKeyword{open}\AgdaSpace{}%
\AgdaKeyword{import}\AgdaSpace{}%
\AgdaModule{Data.Bool}\AgdaSpace{}%
\AgdaKeyword{using}\AgdaSpace{}%
\AgdaSymbol{(}\AgdaFunction{T}\AgdaSymbol{)}\<%
\\
\>[2]\AgdaKeyword{open}\AgdaSpace{}%
\AgdaKeyword{import}\AgdaSpace{}%
\AgdaModule{Data.Nat}\AgdaSpace{}%
\AgdaSymbol{as}\AgdaSpace{}%
\AgdaModule{ℕ}\AgdaSpace{}%
\AgdaKeyword{hiding}\AgdaSpace{}%
\AgdaSymbol{(}\AgdaOperator{\AgdaFunction{\AgdaUnderscore{}≟\AgdaUnderscore{}}}\AgdaSymbol{)}\<%
\\
\>[2]\AgdaKeyword{postulate}\<%
\\
\>[2][@{}l@{\AgdaIndent{0}}]%
\>[4]\AgdaPostulate{≡ᵇ⇒≡}\AgdaSpace{}%
\AgdaSymbol{:}\AgdaSpace{}%
\AgdaSymbol{∀}\AgdaSpace{}%
\AgdaBound{m}\AgdaSpace{}%
\AgdaBound{n}\AgdaSpace{}%
\AgdaSymbol{→}\AgdaSpace{}%
\AgdaFunction{T}\AgdaSpace{}%
\AgdaSymbol{(}\AgdaBound{m}\AgdaSpace{}%
\AgdaOperator{\AgdaPrimitive{≡ᵇ}}\AgdaSpace{}%
\AgdaBound{n}\AgdaSymbol{)}\AgdaSpace{}%
\AgdaSymbol{→}\AgdaSpace{}%
\AgdaBound{m}\AgdaSpace{}%
\AgdaOperator{\AgdaDatatype{≡}}\AgdaSpace{}%
\AgdaBound{n}\<%
\\
\>[4]\AgdaPostulate{≡⇒≡ᵇ}\AgdaSpace{}%
\AgdaSymbol{:}\AgdaSpace{}%
\AgdaSymbol{∀}\AgdaSpace{}%
\AgdaBound{m}\AgdaSpace{}%
\AgdaBound{n}\AgdaSpace{}%
\AgdaSymbol{→}\AgdaSpace{}%
\AgdaBound{m}\AgdaSpace{}%
\AgdaOperator{\AgdaDatatype{≡}}\AgdaSpace{}%
\AgdaBound{n}\AgdaSpace{}%
\AgdaSymbol{→}\AgdaSpace{}%
\AgdaFunction{T}\AgdaSpace{}%
\AgdaSymbol{(}\AgdaBound{m}\AgdaSpace{}%
\AgdaOperator{\AgdaPrimitive{≡ᵇ}}\AgdaSpace{}%
\AgdaBound{n}\AgdaSymbol{)}\<%
\\
\>[4]\AgdaPostulate{T?}\AgdaSpace{}%
\AgdaSymbol{:}\AgdaSpace{}%
\AgdaSymbol{∀}\AgdaSpace{}%
\AgdaBound{x}\AgdaSpace{}%
\AgdaSymbol{→}\AgdaSpace{}%
\AgdaRecord{Dec}\AgdaSpace{}%
\AgdaSymbol{(}\AgdaFunction{T}\AgdaSpace{}%
\AgdaBound{x}\AgdaSymbol{)}\<%
\\
\>[2]\AgdaOperator{\AgdaFunction{\AgdaUnderscore{}≟\AgdaUnderscore{}}}\AgdaSpace{}%
\AgdaSymbol{:}\AgdaSpace{}%
\AgdaFunction{Decidable}\AgdaSpace{}%
\AgdaSymbol{\{}\AgdaArgument{A}\AgdaSpace{}%
\AgdaSymbol{=}\AgdaSpace{}%
\AgdaDatatype{ℕ}\AgdaSymbol{\}}\AgdaSpace{}%
\AgdaOperator{\AgdaDatatype{\AgdaUnderscore{}≡\AgdaUnderscore{}}}\<%
\\
\>[2]\AgdaFunction{map′}\AgdaSpace{}%
\AgdaSymbol{:}\AgdaSpace{}%
\AgdaSymbol{∀}\AgdaSpace{}%
\AgdaSymbol{\{}\AgdaBound{P}\AgdaSpace{}%
\AgdaBound{Q}\AgdaSpace{}%
\AgdaSymbol{:}\AgdaSpace{}%
\AgdaPrimitive{Set}\AgdaSymbol{\}}\AgdaSpace{}%
\AgdaSymbol{→}\AgdaSpace{}%
\AgdaSymbol{(}\AgdaBound{P}\AgdaSpace{}%
\AgdaSymbol{→}\AgdaSpace{}%
\AgdaBound{Q}\AgdaSymbol{)}\AgdaSpace{}%
\AgdaSymbol{→}\AgdaSpace{}%
\AgdaSymbol{(}\AgdaBound{Q}\AgdaSpace{}%
\AgdaSymbol{→}\AgdaSpace{}%
\AgdaBound{P}\AgdaSymbol{)}\AgdaSpace{}%
\AgdaSymbol{→}\AgdaSpace{}%
\AgdaRecord{Dec}\AgdaSpace{}%
\AgdaBound{P}\AgdaSpace{}%
\AgdaSymbol{→}\AgdaSpace{}%
\AgdaRecord{Dec}\AgdaSpace{}%
\AgdaBound{Q}\AgdaSpace{}%
\AgdaSymbol{;}\AgdaSpace{}%
\AgdaFunction{map′}\AgdaSpace{}%
\AgdaSymbol{=}\AgdaSpace{}%
\AgdaPostulate{⋯}\<%
\end{code}
\begin{code}%
\>[2]\AgdaBound{m}\AgdaSpace{}%
\AgdaOperator{\AgdaFunction{≟}}\AgdaSpace{}%
\AgdaBound{n}\AgdaSpace{}%
\AgdaSymbol{=}\AgdaSpace{}%
\AgdaFunction{map′}\AgdaSpace{}%
\AgdaSymbol{(}\AgdaPostulate{≡ᵇ⇒≡}\AgdaSpace{}%
\AgdaBound{m}\AgdaSpace{}%
\AgdaBound{n}\AgdaSymbol{)}\AgdaSpace{}%
\AgdaSymbol{(}\AgdaPostulate{≡⇒≡ᵇ}\AgdaSpace{}%
\AgdaBound{m}\AgdaSpace{}%
\AgdaBound{n}\AgdaSymbol{)}\AgdaSpace{}%
\AgdaSymbol{(}\AgdaPostulate{T?}\AgdaSpace{}%
\AgdaSymbol{(}\AgdaBound{m}\AgdaSpace{}%
\AgdaOperator{\AgdaPrimitive{≡ᵇ}}\AgdaSpace{}%
\AgdaBound{n}\AgdaSymbol{))}\<%
\end{code}
The four functions used in the body (\eg{} \AF{map′}, \AF{≡ᵇ⇒≡}, \etc{}) are not
representable in Kaleidoscope, but comparison of natural numbers is.
More generally, there is a common pattern where the host language
represents a concept in a radically different way than in the target
language.
In such cases, we can customise the extractor by hard-coding the
mapping of the Agda function to the target language function.  For
example, in this case we map \AF{\_≟\_} to \AC{Eq}.

To do this, we have to make sure that normalisation does not expand
certain definitions.  This is what the second argument (base) to our interface
function \AF{kompile} is used for --- to specify the list of functions that
must not reduce during normalisation.
This functionality was not previously available in Agda, so we added two new primitives
to the reflection API --- \AF{dontReduceDefs} and \AF{onlyReduceDefs} --- with pull
request \url{https://github.com/agda/agda/pull/4978}.
These functions give us an environment where any call to
\AF{reduce} or \AD{normalise} will avoid reducing any function that is
in the list (for \AF{dontReduceDefs}) or not in the list (for
\AF{onlyReduceDefs}).
This new feature is used by the \AF{kompile} macro to avoid reducing
functions for which we have a fixed translation.


\subsection{\label{sec:maptypes}Mapping Agda Types to Kaleidoscope Assertions}

The next step after normalisation is to verify and translate the type
signature of the embedded function into the target language.  Kaleidoscope
does not support type annotations, but we
still need to verify that the function is first-order, and that the
argument and return types are from the right universe.
This is implemented in the \AF{kompile-ty} function, of which we present a
small fragment here:
\begin{code}[hide]%
\>[0]\AgdaKeyword{module}\AgdaSpace{}%
\AgdaModule{KompTy}\AgdaSpace{}%
\AgdaKeyword{where}\<%
\\
\>[0][@{}l@{\AgdaIndent{0}}]%
\>[2]\AgdaKeyword{open}\AgdaSpace{}%
\AgdaKeyword{import}\AgdaSpace{}%
\AgdaModule{Reflection}\AgdaSpace{}%
\AgdaKeyword{hiding}\AgdaSpace{}%
\AgdaSymbol{(}\AgdaPostulate{TC}\AgdaSymbol{;}\AgdaSpace{}%
\AgdaPostulate{return}\AgdaSymbol{;}\AgdaSpace{}%
\AgdaOperator{\AgdaFunction{\AgdaUnderscore{}>>=\AgdaUnderscore{}}}\AgdaSymbol{;}\AgdaSpace{}%
\AgdaOperator{\AgdaFunction{\AgdaUnderscore{}>>\AgdaUnderscore{}}}\AgdaSymbol{)}\<%
\\
\>[2]\AgdaKeyword{open}\AgdaSpace{}%
\AgdaKeyword{import}\AgdaSpace{}%
\AgdaModule{Reflection.Term}\<%
\\
\>[2]\AgdaKeyword{open}\AgdaSpace{}%
\AgdaKeyword{import}\AgdaSpace{}%
\AgdaModule{Data.Fin}\<%
\\
\>[2]\AgdaKeyword{open}\AgdaSpace{}%
\AgdaKeyword{import}\AgdaSpace{}%
\AgdaModule{Data.Bool}\AgdaSpace{}%
\AgdaKeyword{using}\AgdaSpace{}%
\AgdaSymbol{(}\AgdaDatatype{Bool}\AgdaSymbol{;}\AgdaSpace{}%
\AgdaInductiveConstructor{true}\AgdaSymbol{)}\<%
\\
\>[2]\AgdaKeyword{open}\AgdaSpace{}%
\AgdaKeyword{import}\AgdaSpace{}%
\AgdaModule{Relation.Binary.PropositionalEquality}\<%
\\
\>[2]\AgdaKeyword{open}\AgdaSpace{}%
\AgdaKeyword{import}\AgdaSpace{}%
\AgdaModule{Data.Product}\<%
\\
\>[2]\AgdaKeyword{open}\AgdaSpace{}%
\AgdaKeyword{import}\AgdaSpace{}%
\AgdaModule{Data.Nat}\AgdaSpace{}%
\AgdaSymbol{as}\AgdaSpace{}%
\AgdaModule{ℕ}\AgdaSpace{}%
\AgdaKeyword{hiding}\AgdaSpace{}%
\AgdaSymbol{(}\AgdaOperator{\AgdaFunction{\AgdaUnderscore{}≟\AgdaUnderscore{}}}\AgdaSymbol{)}\<%
\\
\>[2]\AgdaKeyword{open}\AgdaSpace{}%
\AgdaModule{Kaleid}\<%
\\
\\[\AgdaEmptyExtraSkip]%
\>[2]\AgdaFunction{SPS}\AgdaSpace{}%
\AgdaSymbol{:}\AgdaSpace{}%
\AgdaPrimitive{Set}\AgdaSpace{}%
\AgdaSymbol{→}\AgdaSpace{}%
\AgdaPrimitive{Set}\AgdaSpace{}%
\AgdaSymbol{;}\AgdaSpace{}%
\AgdaFunction{SPS}\AgdaSpace{}%
\AgdaSymbol{=}\AgdaSpace{}%
\AgdaPostulate{⋯}\<%
\\
\>[2]\AgdaFunction{sps-kompile-term}\AgdaSpace{}%
\AgdaSymbol{:}\AgdaSpace{}%
\AgdaDatatype{Term}\AgdaSpace{}%
\AgdaSymbol{→}\AgdaSpace{}%
\AgdaFunction{SPS}\AgdaSpace{}%
\AgdaOperator{\AgdaFunction{\$}}\AgdaSpace{}%
\AgdaDatatype{Err}\AgdaSpace{}%
\AgdaDatatype{Expr}\AgdaSpace{}%
\AgdaSymbol{;}\AgdaSpace{}%
\AgdaFunction{sps-kompile-term}\AgdaSpace{}%
\AgdaSymbol{=}\AgdaSpace{}%
\AgdaPostulate{⋯}\<%
\\
\\[\AgdaEmptyExtraSkip]%
\>[2]\AgdaKeyword{record}\AgdaSpace{}%
\AgdaRecord{PS}\AgdaSpace{}%
\AgdaSymbol{:}\AgdaSpace{}%
\AgdaPrimitive{Set}\AgdaSpace{}%
\AgdaKeyword{where}\<%
\\
\>[2][@{}l@{\AgdaIndent{0}}]%
\>[4]\AgdaKeyword{field}\AgdaSpace{}%
\AgdaField{cur}\AgdaSpace{}%
\AgdaSymbol{:}\AgdaSpace{}%
\AgdaPostulate{String}\<%
\\
\\[\AgdaEmptyExtraSkip]%
\>[2]\AgdaKeyword{record}\AgdaSpace{}%
\AgdaRecord{Assrt}\AgdaSpace{}%
\AgdaSymbol{:}\AgdaSpace{}%
\AgdaPrimitive{Set}\<%
\\
\>[2]\AgdaKeyword{infixl}\AgdaSpace{}%
\AgdaNumber{4}\AgdaSpace{}%
\AgdaOperator{\AgdaFunction{\AgdaUnderscore{}<\$>\AgdaUnderscore{}}}\<%
\\
\>[2]\AgdaOperator{\AgdaFunction{\AgdaUnderscore{}<\$>\AgdaUnderscore{}}}\AgdaSpace{}%
\AgdaSymbol{:}\AgdaSpace{}%
\AgdaSymbol{∀\{}\AgdaBound{A}\AgdaSpace{}%
\AgdaBound{B}\AgdaSpace{}%
\AgdaSymbol{:}\AgdaSpace{}%
\AgdaPrimitive{Set}\AgdaSymbol{\}}\AgdaSpace{}%
\AgdaSymbol{→}\AgdaSpace{}%
\AgdaSymbol{(}\AgdaBound{A}\AgdaSpace{}%
\AgdaSymbol{→}\AgdaSpace{}%
\AgdaBound{B}\AgdaSymbol{)}\AgdaSpace{}%
\AgdaSymbol{→}\AgdaSpace{}%
\AgdaFunction{SPS}\AgdaSpace{}%
\AgdaBound{A}\AgdaSpace{}%
\AgdaSymbol{→}\AgdaSpace{}%
\AgdaFunction{SPS}\AgdaSpace{}%
\AgdaBound{B}\AgdaSpace{}%
\AgdaSymbol{;}\AgdaSpace{}%
\AgdaOperator{\AgdaFunction{\AgdaUnderscore{}<\$>\AgdaUnderscore{}}}\AgdaSpace{}%
\AgdaSymbol{=}\AgdaSpace{}%
\AgdaPostulate{⋯}\<%
\\
\>[2]\AgdaFunction{get}\AgdaSpace{}%
\AgdaSymbol{:}\AgdaSpace{}%
\AgdaFunction{SPS}\AgdaSpace{}%
\AgdaRecord{PS}\AgdaSpace{}%
\AgdaSymbol{;}\AgdaSpace{}%
\AgdaFunction{get}\AgdaSpace{}%
\AgdaSymbol{=}\AgdaSpace{}%
\AgdaPostulate{⋯}\<%
\\
\>[2]\AgdaFunction{modify}\AgdaSpace{}%
\AgdaSymbol{:}\AgdaSpace{}%
\AgdaSymbol{∀}\AgdaSpace{}%
\AgdaSymbol{\{}\AgdaBound{i}\AgdaSpace{}%
\AgdaSymbol{:}\AgdaSpace{}%
\AgdaRecord{⊤}\AgdaSymbol{\}}\AgdaSpace{}%
\AgdaSymbol{→}\AgdaSpace{}%
\AgdaSymbol{(}\AgdaRecord{PS}\AgdaSpace{}%
\AgdaSymbol{→}\AgdaSpace{}%
\AgdaRecord{PS}\AgdaSymbol{)}\AgdaSpace{}%
\AgdaSymbol{→}\AgdaSpace{}%
\AgdaFunction{SPS}\AgdaSpace{}%
\AgdaRecord{⊤}\AgdaSpace{}%
\AgdaSymbol{;}\AgdaSpace{}%
\AgdaFunction{modify}\AgdaSpace{}%
\AgdaSymbol{=}\AgdaSpace{}%
\AgdaPostulate{⋯}\<%
\\
\>[2]\AgdaFunction{ke}\AgdaSpace{}%
\AgdaSymbol{:}\AgdaSpace{}%
\AgdaSymbol{∀}\AgdaSpace{}%
\AgdaSymbol{\{}\AgdaBound{X}\AgdaSymbol{\}}\AgdaSpace{}%
\AgdaSymbol{→}\AgdaSpace{}%
\AgdaPostulate{String}\AgdaSpace{}%
\AgdaSymbol{→}\AgdaSpace{}%
\AgdaFunction{SPS}\AgdaSpace{}%
\AgdaSymbol{(}\AgdaDatatype{Err}\AgdaSpace{}%
\AgdaBound{X}\AgdaSymbol{)}\AgdaSpace{}%
\AgdaSymbol{;}\AgdaSpace{}%
\AgdaFunction{ke}\AgdaSpace{}%
\AgdaSymbol{=}\AgdaSpace{}%
\AgdaPostulate{⋯}\<%
\\
\>[2]\AgdaFunction{return}\AgdaSpace{}%
\AgdaSymbol{:}\AgdaSpace{}%
\AgdaSymbol{∀}\AgdaSpace{}%
\AgdaSymbol{\{}\AgdaBound{X}\AgdaSymbol{\}}\AgdaSpace{}%
\AgdaSymbol{→}\AgdaSpace{}%
\AgdaBound{X}\AgdaSpace{}%
\AgdaSymbol{→}\AgdaSpace{}%
\AgdaFunction{SPS}\AgdaSpace{}%
\AgdaBound{X}\AgdaSpace{}%
\AgdaSymbol{;}\AgdaSpace{}%
\AgdaFunction{return}\AgdaSpace{}%
\AgdaSymbol{=}\AgdaSpace{}%
\AgdaPostulate{⋯}\<%
\\
\>[2]\AgdaOperator{\AgdaFunction{\AgdaUnderscore{}>>=\AgdaUnderscore{}}}\AgdaSpace{}%
\AgdaSymbol{:}\AgdaSpace{}%
\AgdaSymbol{∀}\AgdaSpace{}%
\AgdaSymbol{\{}\AgdaBound{X}\AgdaSpace{}%
\AgdaBound{Y}\AgdaSymbol{\}}\AgdaSpace{}%
\AgdaSymbol{→}\AgdaSpace{}%
\AgdaFunction{SPS}\AgdaSpace{}%
\AgdaBound{X}\AgdaSpace{}%
\AgdaSymbol{→}\AgdaSpace{}%
\AgdaSymbol{(}\AgdaBound{X}\AgdaSpace{}%
\AgdaSymbol{→}\AgdaSpace{}%
\AgdaFunction{SPS}\AgdaSpace{}%
\AgdaBound{Y}\AgdaSymbol{)}\AgdaSpace{}%
\AgdaSymbol{→}\AgdaSpace{}%
\AgdaFunction{SPS}\AgdaSpace{}%
\AgdaBound{Y}\AgdaSpace{}%
\AgdaSymbol{;}\AgdaSpace{}%
\AgdaOperator{\AgdaFunction{\AgdaUnderscore{}>>=\AgdaUnderscore{}}}\AgdaSpace{}%
\AgdaSymbol{=}\AgdaSpace{}%
\AgdaPostulate{⋯}\<%
\\
\>[2]\AgdaOperator{\AgdaFunction{\AgdaUnderscore{}>>\AgdaUnderscore{}}}\AgdaSpace{}%
\AgdaSymbol{:}\AgdaSpace{}%
\AgdaSymbol{∀}\AgdaSpace{}%
\AgdaSymbol{\{}\AgdaBound{X}\AgdaSpace{}%
\AgdaBound{Y}\AgdaSymbol{\}}\AgdaSpace{}%
\AgdaSymbol{→}\AgdaSpace{}%
\AgdaFunction{SPS}\AgdaSpace{}%
\AgdaBound{X}\AgdaSpace{}%
\AgdaSymbol{→}\AgdaSpace{}%
\AgdaFunction{SPS}\AgdaSpace{}%
\AgdaBound{Y}\AgdaSpace{}%
\AgdaSymbol{→}\AgdaSpace{}%
\AgdaFunction{SPS}\AgdaSpace{}%
\AgdaBound{Y}\AgdaSpace{}%
\AgdaSymbol{;}\AgdaSpace{}%
\AgdaBound{x}\AgdaSpace{}%
\AgdaOperator{\AgdaFunction{>>}}\AgdaSpace{}%
\AgdaBound{y}\AgdaSpace{}%
\AgdaSymbol{=}\AgdaSpace{}%
\AgdaBound{x}\AgdaSpace{}%
\AgdaOperator{\AgdaFunction{>>=}}\AgdaSpace{}%
\AgdaFunction{const}\AgdaSpace{}%
\AgdaBound{y}\<%
\end{code}
\begin{code}%
\>[2]\AgdaKeyword{record}\AgdaSpace{}%
\AgdaRecord{Assrt}\AgdaSpace{}%
\AgdaKeyword{where}\<%
\\
\>[2][@{}l@{\AgdaIndent{0}}]%
\>[4]\AgdaKeyword{constructor}\AgdaSpace{}%
\AgdaInductiveConstructor{mk}\<%
\\
\>[4]\AgdaKeyword{field}\AgdaSpace{}%
\AgdaField{v}\AgdaSpace{}%
\AgdaSymbol{:}\AgdaSpace{}%
\AgdaFunction{Id}\AgdaSpace{}%
\AgdaSymbol{;}\AgdaSpace{}%
\AgdaField{a}\AgdaSpace{}%
\AgdaSymbol{:}\AgdaSpace{}%
\AgdaDatatype{Expr}\<%
\\
\\[\AgdaEmptyExtraSkip]%
\>[2]\AgdaOperator{\AgdaFunction{\AgdaUnderscore{}p+=a\AgdaUnderscore{}}}\AgdaSpace{}%
\AgdaSymbol{:}\AgdaSpace{}%
\AgdaRecord{PS}\AgdaSpace{}%
\AgdaSymbol{→}\AgdaSpace{}%
\AgdaRecord{Assrt}\AgdaSpace{}%
\AgdaSymbol{→}\AgdaSpace{}%
\AgdaRecord{PS}\AgdaSpace{}%
\AgdaSymbol{;}\AgdaSpace{}%
\AgdaOperator{\AgdaFunction{\AgdaUnderscore{}p+=a\AgdaUnderscore{}}}\AgdaSpace{}%
\AgdaSymbol{=}\AgdaSpace{}%
\AgdaPostulate{⋯}\<%
\\
\\[\AgdaEmptyExtraSkip]%
\>[2]\AgdaFunction{kompile-ty}\AgdaSpace{}%
\AgdaSymbol{:}\AgdaSpace{}%
\AgdaFunction{Type}\AgdaSpace{}%
\AgdaSymbol{→}\AgdaSpace{}%
\AgdaSymbol{(}\AgdaBound{pi-ok}\AgdaSpace{}%
\AgdaSymbol{:}\AgdaSpace{}%
\AgdaDatatype{Bool}\AgdaSymbol{)}\AgdaSpace{}%
\AgdaSymbol{→}\AgdaSpace{}%
\AgdaFunction{SPS}\AgdaSpace{}%
\AgdaSymbol{(}\AgdaDatatype{Err}\AgdaSpace{}%
\AgdaRecord{⊤}\AgdaSymbol{)}\<%
\\
\>[2]\AgdaFunction{kompile-ty}\AgdaSpace{}%
\AgdaSymbol{(}\AgdaInductiveConstructor{def}\AgdaSpace{}%
\AgdaSymbol{(}\AgdaKeyword{quote}\AgdaSpace{}%
\AgdaDatatype{ℕ}\AgdaSymbol{)}\AgdaSpace{}%
\AgdaBound{args}\AgdaSymbol{)}\AgdaSpace{}%
\AgdaSymbol{\AgdaUnderscore{}}\AgdaSpace{}%
\AgdaSymbol{=}\AgdaSpace{}%
\AgdaFunction{return}\AgdaSpace{}%
\AgdaOperator{\AgdaFunction{\$}}\AgdaSpace{}%
\AgdaInductiveConstructor{ok}\AgdaSpace{}%
\AgdaInductiveConstructor{tt}\<%
\\
\>[2]\AgdaFunction{kompile-ty}\AgdaSpace{}%
\AgdaSymbol{(}\AgdaInductiveConstructor{def}\AgdaSpace{}%
\AgdaSymbol{(}\AgdaKeyword{quote}\AgdaSpace{}%
\AgdaDatatype{Fin}\AgdaSymbol{)}\AgdaSpace{}%
\AgdaSymbol{(}\AgdaInductiveConstructor{arg}\AgdaSpace{}%
\AgdaSymbol{\AgdaUnderscore{}}\AgdaSpace{}%
\AgdaBound{x}\AgdaSpace{}%
\AgdaOperator{\AgdaInductiveConstructor{∷}}\AgdaSpace{}%
\AgdaInductiveConstructor{[]}\AgdaSymbol{))}\AgdaSpace{}%
\AgdaSymbol{\AgdaUnderscore{}}\AgdaSpace{}%
\AgdaSymbol{=}\AgdaSpace{}%
\AgdaKeyword{do}\<%
\\
\>[2][@{}l@{\AgdaIndent{0}}]%
\>[4]\AgdaInductiveConstructor{ok}\AgdaSpace{}%
\AgdaBound{p}%
\>[10]\AgdaOperator{\AgdaFunction{←}}\AgdaSpace{}%
\AgdaFunction{sps-kompile-term}\AgdaSpace{}%
\AgdaBound{x}\AgdaSpace{}%
\AgdaKeyword{where}\AgdaSpace{}%
\AgdaInductiveConstructor{error}\AgdaSpace{}%
\AgdaBound{x}\AgdaSpace{}%
\AgdaSymbol{→}\AgdaSpace{}%
\AgdaFunction{ke}\AgdaSpace{}%
\AgdaBound{x}\<%
\\
\>[4]\AgdaBound{v}%
\>[10]\AgdaOperator{\AgdaFunction{←}}\AgdaSpace{}%
\AgdaField{PS.cur}\AgdaSpace{}%
\AgdaOperator{\AgdaFunction{<\$>}}\AgdaSpace{}%
\AgdaFunction{get}\<%
\\
\>[4]\AgdaFunction{modify}\AgdaSpace{}%
\AgdaOperator{\AgdaFunction{\$}}\AgdaSpace{}%
\AgdaOperator{\AgdaFunction{\AgdaUnderscore{}p+=a}}\AgdaSpace{}%
\AgdaSymbol{(}\AgdaInductiveConstructor{mk}\AgdaSpace{}%
\AgdaBound{v}\AgdaSpace{}%
\AgdaSymbol{(}\AgdaInductiveConstructor{BinOp}\AgdaSpace{}%
\AgdaInductiveConstructor{Lt}\AgdaSpace{}%
\AgdaSymbol{(}\AgdaInductiveConstructor{Var}\AgdaSpace{}%
\AgdaBound{v}\AgdaSymbol{)}\AgdaSpace{}%
\AgdaBound{p}\AgdaSymbol{))}\<%
\\
\>[4]\AgdaFunction{return}\AgdaSpace{}%
\AgdaOperator{\AgdaFunction{\$}}\AgdaSpace{}%
\AgdaInductiveConstructor{ok}\AgdaSpace{}%
\AgdaInductiveConstructor{tt}\<%
\\
\>[2]\AgdaCatchallClause{\AgdaFunction{kompile-ty}}\AgdaSpace{}%
\AgdaCatchallClause{\AgdaSymbol{\AgdaUnderscore{}}}\AgdaSpace{}%
\AgdaCatchallClause{\AgdaSymbol{\AgdaUnderscore{}}}\AgdaSpace{}%
\AgdaSymbol{=}\AgdaSpace{}%
\AgdaPostulate{⋯}\<%
\end{code}
It operates within the state monad \AD{SPS} where the state is given
by the type \AD{PS} (pi-type state).  As we traverse the type signature
of a function, for non-dependent types such as \AD{ℕ} we only verify
whether the type is supported.

Dependent types such as \AD{Fin} can be seen as encoding some
predicate on their arguments as well as the element of the dependent
type itself.  We preserve this information by mapping
them to assertions in the target language, essentially trading static
checks for dynamic ones.
Preserving this information is is important for two reasons.
First, the toolchain can
use this information for more aggressive optimisations.  For example,
the Kaleidoscope expression \texttt{if a > b: x else: y} can be simplified
to \texttt{x} if we know from the types that \texttt{a} is greater than \texttt{b}.
Second, extraction can be applied to partial programs.  When extracted
functions are being called externally, the calls are not typechecked, so
the preconditions on the arguments that would normally be enforced by
the type system may not hold.  For example, given
\begin{code}[inline]%
\>[2]\AgdaFunction{f}\AgdaSpace{}%
\AgdaSymbol{:}\AgdaSpace{}%
\AgdaSymbol{(}\AgdaBound{x}\AgdaSpace{}%
\AgdaSymbol{:}\AgdaSpace{}%
\AgdaDatatype{ℕ}\AgdaSymbol{)}\AgdaSpace{}%
\AgdaSymbol{→}\AgdaSpace{}%
\AgdaBound{x}\AgdaSpace{}%
\AgdaOperator{\AgdaFunction{>}}\AgdaSpace{}%
\AgdaNumber{0}\AgdaSpace{}%
\AgdaSymbol{→}\AgdaSpace{}%
\AgdaDatatype{ℕ}\<%
\end{code}
\begin{code}[hide]%
\>[2]\AgdaFunction{f}\AgdaSpace{}%
\AgdaSymbol{=}\AgdaSpace{}%
\AgdaPostulate{⋯}\<%
\end{code}, its first argument must not be zero.  However, as \AB{x} \AF{>}
\AN{0} cannot be represented, external calls may pass zero to
the extracted version of \AF{f}.

In the case for \AF{Fin} in the definition of \AF{kompile-ty}, we
first extract the argument \AB{x} (obtaining a Kaleidoscope
expression).  Then we get the name of the function argument by
referring to the \AR{PS.cur} field of the state.  Finally, we generate
an assertion that checks whether the encoded witness is less than the
argument to \AD{Fin} (the upper bound), and add it to the state.

In general, to translate a dependent type $\AF{P} : \AF{I} →
\AF{Set}$, we start by picking basic types \AD{TI} and \AD{TP} that
encode \AD{I} and \AD{P}, together with encoding functions $\AF{enc-i}
: \AD{I} → \AD{TI}$ and $\AF{enc-p} : ∀ \{i\} (p : \AD{P}\ i) →
\AD{TP}$.  We then introduce a function $\AF{assrt-p} : (\AB{ti} :
\AD{TI})(\AB{tp} : \AD{TP}) → \AD{Bool}$ that decides whether the
encoded arguments \AB{ti} and \AB{tp} come from a valid input, that is
whether \AB{tp} is the image under \AF{enc-p} of some $\AB{p} :
\AD{P}\ i$ where \AB{ti} is the image of \AB{i}. Finally, we prove
soundness and completeness of the encoding:
\begin{code}[hide]%
\>[2]\AgdaFunction{I}\AgdaSpace{}%
\AgdaSymbol{:}\AgdaSpace{}%
\AgdaPrimitive{Set}\AgdaSpace{}%
\AgdaSymbol{;}\AgdaSpace{}%
\AgdaFunction{TP}\AgdaSpace{}%
\AgdaSymbol{:}\AgdaSpace{}%
\AgdaPrimitive{Set}\AgdaSpace{}%
\AgdaSymbol{;}\AgdaSpace{}%
\AgdaFunction{TI}\AgdaSpace{}%
\AgdaSymbol{:}\AgdaSpace{}%
\AgdaPrimitive{Set}\<%
\\
\>[2]\AgdaFunction{P}\AgdaSpace{}%
\AgdaSymbol{:}\AgdaSpace{}%
\AgdaFunction{I}\AgdaSpace{}%
\AgdaSymbol{→}\AgdaSpace{}%
\AgdaPrimitive{Set}\AgdaSpace{}%
\AgdaSymbol{;}\AgdaSpace{}%
\AgdaFunction{T}\AgdaSpace{}%
\AgdaSymbol{:}\AgdaSpace{}%
\AgdaPrimitive{Set}\AgdaSpace{}%
\AgdaSymbol{→}\AgdaSpace{}%
\AgdaPrimitive{Set}\AgdaSpace{}%
\AgdaComment{--\ T\ I\ ≡\ TI\ ;\ T\ (P\ i)\ ≡\ TP}\<%
\\
\>[2]\AgdaFunction{enc-i}\AgdaSpace{}%
\AgdaSymbol{:}\AgdaSpace{}%
\AgdaFunction{I}\AgdaSpace{}%
\AgdaSymbol{→}\AgdaSpace{}%
\AgdaFunction{TI}\AgdaSpace{}%
\AgdaSymbol{;}\AgdaSpace{}%
\AgdaFunction{enc-p}\AgdaSpace{}%
\AgdaSymbol{:}\AgdaSpace{}%
\AgdaSymbol{∀}\AgdaSpace{}%
\AgdaSymbol{\{}\AgdaBound{i}\AgdaSymbol{\}}\AgdaSpace{}%
\AgdaSymbol{(}\AgdaBound{p}\AgdaSpace{}%
\AgdaSymbol{:}\AgdaSpace{}%
\AgdaFunction{P}\AgdaSpace{}%
\AgdaBound{i}\AgdaSymbol{)}\AgdaSpace{}%
\AgdaSymbol{→}\AgdaSpace{}%
\AgdaFunction{TP}\<%
\\
\>[2]\AgdaFunction{assrt-p}\AgdaSpace{}%
\AgdaSymbol{:}\AgdaSpace{}%
\AgdaSymbol{(}\AgdaBound{ti}\AgdaSpace{}%
\AgdaSymbol{:}\AgdaSpace{}%
\AgdaFunction{TI}\AgdaSymbol{)}\AgdaSpace{}%
\AgdaSymbol{(}\AgdaBound{tp}\AgdaSpace{}%
\AgdaSymbol{:}\AgdaSpace{}%
\AgdaFunction{TP}\AgdaSymbol{)}\AgdaSpace{}%
\AgdaSymbol{→}\AgdaSpace{}%
\AgdaDatatype{Bool}\<%
\end{code}
\begin{code}%
\>[2]\AgdaFunction{sound-p}\AgdaSpace{}%
\AgdaSymbol{:}\AgdaSpace{}%
\AgdaSymbol{∀}\AgdaSpace{}%
\AgdaSymbol{\{}\AgdaBound{i}\AgdaSymbol{\}\{}\AgdaBound{p}\AgdaSpace{}%
\AgdaSymbol{:}\AgdaSpace{}%
\AgdaFunction{P}\AgdaSpace{}%
\AgdaBound{i}\AgdaSymbol{\}}\AgdaSpace{}%
\AgdaSymbol{→}\AgdaSpace{}%
\AgdaFunction{assrt-p}\AgdaSpace{}%
\AgdaSymbol{(}\AgdaFunction{enc-i}\AgdaSpace{}%
\AgdaBound{i}\AgdaSymbol{)}\AgdaSpace{}%
\AgdaSymbol{(}\AgdaFunction{enc-p}\AgdaSpace{}%
\AgdaBound{p}\AgdaSymbol{)}\AgdaSpace{}%
\AgdaOperator{\AgdaDatatype{≡}}\AgdaSpace{}%
\AgdaInductiveConstructor{true}\<%
\\
\>[2]\AgdaFunction{complete-p}\AgdaSpace{}%
\AgdaSymbol{:}\AgdaSpace{}%
\AgdaSymbol{∀}\AgdaSpace{}%
\AgdaBound{ti}\AgdaSpace{}%
\AgdaBound{tp}\AgdaSpace{}%
\AgdaSymbol{→}\AgdaSpace{}%
\AgdaFunction{assrt-p}\AgdaSpace{}%
\AgdaBound{ti}\AgdaSpace{}%
\AgdaBound{tp}\AgdaSpace{}%
\AgdaOperator{\AgdaDatatype{≡}}\AgdaSpace{}%
\AgdaInductiveConstructor{true}\AgdaSpace{}%
\AgdaSymbol{→}\AgdaSpace{}%
\AgdaFunction{∃₂}\AgdaSpace{}%
\AgdaSymbol{λ}\AgdaSpace{}%
\AgdaBound{i}\AgdaSpace{}%
\AgdaSymbol{(}\AgdaBound{p}\AgdaSpace{}%
\AgdaSymbol{:}\AgdaSpace{}%
\AgdaFunction{P}\AgdaSpace{}%
\AgdaBound{i}\AgdaSymbol{)}\AgdaSpace{}%
\AgdaSymbol{→}\AgdaSpace{}%
\AgdaBound{ti}\AgdaSpace{}%
\AgdaOperator{\AgdaDatatype{≡}}\AgdaSpace{}%
\AgdaFunction{enc-i}\AgdaSpace{}%
\AgdaBound{i}\AgdaSpace{}%
\AgdaOperator{\AgdaFunction{×}}\AgdaSpace{}%
\AgdaBound{tp}\AgdaSpace{}%
\AgdaOperator{\AgdaDatatype{≡}}\AgdaSpace{}%
\AgdaFunction{enc-p}\AgdaSpace{}%
\AgdaBound{p}\<%
\end{code}
\begin{code}[hide]%
\>[2]\AgdaFunction{P}\AgdaSpace{}%
\AgdaSymbol{=}\AgdaSpace{}%
\AgdaPostulate{⋯}\AgdaSpace{}%
\AgdaSymbol{;}\AgdaSpace{}%
\AgdaFunction{I}\AgdaSpace{}%
\AgdaSymbol{=}\AgdaSpace{}%
\AgdaPostulate{⋯}\AgdaSpace{}%
\AgdaSymbol{;}\AgdaSpace{}%
\AgdaFunction{TI}\AgdaSpace{}%
\AgdaSymbol{=}\AgdaSpace{}%
\AgdaPostulate{⋯}\AgdaSpace{}%
\AgdaSymbol{;}\AgdaSpace{}%
\AgdaFunction{TP}\AgdaSpace{}%
\AgdaSymbol{=}\AgdaSpace{}%
\AgdaPostulate{⋯}\AgdaSpace{}%
\AgdaSymbol{;}\AgdaSpace{}%
\AgdaFunction{T}\AgdaSpace{}%
\AgdaSymbol{=}\AgdaSpace{}%
\AgdaPostulate{⋯}\<%
\\
\>[2]\AgdaFunction{enc-i}\AgdaSpace{}%
\AgdaSymbol{=}\AgdaSpace{}%
\AgdaPostulate{⋯}\AgdaSpace{}%
\AgdaSymbol{;}\AgdaSpace{}%
\AgdaFunction{enc-p}\AgdaSpace{}%
\AgdaSymbol{=}\AgdaSpace{}%
\AgdaPostulate{⋯}\AgdaSpace{}%
\AgdaSymbol{;}\AgdaSpace{}%
\AgdaFunction{assrt-p}\AgdaSpace{}%
\AgdaSymbol{=}\AgdaSpace{}%
\AgdaPostulate{⋯}\AgdaSpace{}%
\AgdaSymbol{;}\AgdaSpace{}%
\AgdaFunction{sound-p}\AgdaSpace{}%
\AgdaSymbol{=}\AgdaSpace{}%
\AgdaPostulate{⋯}\AgdaSpace{}%
\AgdaSymbol{;}\AgdaSpace{}%
\AgdaFunction{complete-p}\AgdaSpace{}%
\AgdaSymbol{=}\AgdaSpace{}%
\AgdaPostulate{⋯}\<%
\end{code}
Dependent types with multiple arguments can be
treated in exactly the same way.

%


Now, for decidable relations, we can entirely avoid encoding the proof
object, as long as the computational behaviour of the function does not
depend on the structure of the proof. This is in particular always the
case for proof-irrelevant types such as \AD{\_≡\_} and \AD{\_<\_}.
We encode the elements of these types with the unit value (natural number \AN{1}).
We then generate an assertion that uses the decision procedure. This
decision procedure returns an element of the \AD{Dec} type
which we interpret as a boolean value: \AN{1} for \AC{yes} and \AN{0} for \AC{no}.

If a function returns a value of a dependent type, we also generate an assertion
using the same rules. We organise the body of the extracted function
such that there is a fixed variable binding the return value,
and attach the assertion to that variable. Here is an example:
\begin{code}[hide]%
\>[0]\AgdaKeyword{module}\AgdaSpace{}%
\AgdaModule{ExFin}\AgdaSpace{}%
\AgdaKeyword{where}\<%
\\
\>[0][@{}l@{\AgdaIndent{0}}]%
\>[2]\AgdaKeyword{open}\AgdaSpace{}%
\AgdaKeyword{import}\AgdaSpace{}%
\AgdaModule{Data.Fin}\AgdaSpace{}%
\AgdaKeyword{using}\AgdaSpace{}%
\AgdaSymbol{(}\AgdaDatatype{Fin}\AgdaSymbol{;}\AgdaSpace{}%
\AgdaFunction{fromℕ<}\AgdaSymbol{)}\<%
\\
\>[2]\AgdaKeyword{open}\AgdaSpace{}%
\AgdaKeyword{import}\AgdaSpace{}%
\AgdaModule{Data.Nat.Properties}\<%
\\
\>[2]\AgdaKeyword{open}\AgdaSpace{}%
\AgdaKeyword{import}\AgdaSpace{}%
\AgdaModule{Data.Nat}\AgdaSpace{}%
\AgdaSymbol{as}\AgdaSpace{}%
\AgdaModule{ℕ}\AgdaSpace{}%
\AgdaKeyword{hiding}\AgdaSpace{}%
\AgdaSymbol{(}\AgdaOperator{\AgdaFunction{\AgdaUnderscore{}≟\AgdaUnderscore{}}}\AgdaSymbol{)}\<%
\end{code}
\begin{code}%
\>[2]\AgdaFunction{ex₇}\AgdaSpace{}%
\AgdaSymbol{:}\AgdaSpace{}%
\AgdaSymbol{(}\AgdaBound{n}\AgdaSpace{}%
\AgdaSymbol{:}\AgdaSpace{}%
\AgdaDatatype{ℕ}\AgdaSymbol{)}\AgdaSpace{}%
\AgdaSymbol{→}\AgdaSpace{}%
\AgdaDatatype{Fin}\AgdaSpace{}%
\AgdaSymbol{(}\AgdaNumber{1}\AgdaSpace{}%
\AgdaOperator{\AgdaPrimitive{+}}\AgdaSpace{}%
\AgdaBound{n}\AgdaSpace{}%
\AgdaOperator{\AgdaPrimitive{*}}\AgdaSpace{}%
\AgdaBound{n}\AgdaSymbol{)}%
\>[38]\AgdaComment{--\ def\ ex7\ (x\AgdaUnderscore{}1):}\<%
\\
\>[2]\AgdaFunction{ex₇}\AgdaSpace{}%
\AgdaBound{n}\AgdaSpace{}%
\AgdaSymbol{=}\AgdaSpace{}%
\AgdaFunction{fromℕ<}\AgdaSpace{}%
\AgdaOperator{\AgdaFunction{\$}}\AgdaSpace{}%
\AgdaFunction{n<1+n}\AgdaSpace{}%
\AgdaSymbol{(}\AgdaBound{n}\AgdaSpace{}%
\AgdaOperator{\AgdaPrimitive{*}}\AgdaSpace{}%
\AgdaBound{n}\AgdaSymbol{)}%
\>[38]\AgdaComment{--\ \ \ let\ n\ =\ x\AgdaUnderscore{}1\ ;\ \AgdaUnderscore{}\AgdaUnderscore{}ret\ =\ n\ *\ n}\<%
\\
\>[38]\AgdaComment{--\ \ \ let\ \AgdaUnderscore{}\AgdaUnderscore{}ret\AgdaUnderscore{}assrt\ =\ assert\ (\AgdaUnderscore{}\AgdaUnderscore{}ret\ <\ 1\ +\ x\AgdaUnderscore{}1\ *\ x\AgdaUnderscore{}1)}\<%
\\
\>[38]\AgdaComment{--\ \ \ \AgdaUnderscore{}\AgdaUnderscore{}ret}\<%
\end{code}
First, note the
assertion for the return value, which has been generated from the return
type of \AF{ex₇}. Next, recall the types of \AF{fromℕ<} and \AF{n<1+n}:
\begin{code}[hide]%
\>[0]\AgdaKeyword{module}\AgdaSpace{}%
\AgdaModule{Signatures}\AgdaSpace{}%
\AgdaKeyword{where}\<%
\\
\>[0][@{}l@{\AgdaIndent{0}}]%
\>[2]\AgdaKeyword{open}\AgdaSpace{}%
\AgdaKeyword{import}\AgdaSpace{}%
\AgdaModule{Data.Fin}\AgdaSpace{}%
\AgdaKeyword{using}\AgdaSpace{}%
\AgdaSymbol{(}\AgdaDatatype{Fin}\AgdaSymbol{)}\<%
\\
\>[2]\AgdaKeyword{open}\AgdaSpace{}%
\AgdaKeyword{import}\AgdaSpace{}%
\AgdaModule{Data.Nat}\AgdaSpace{}%
\AgdaSymbol{as}\AgdaSpace{}%
\AgdaModule{ℕ}\AgdaSpace{}%
\AgdaKeyword{hiding}\AgdaSpace{}%
\AgdaSymbol{(}\AgdaOperator{\AgdaFunction{\AgdaUnderscore{}≟\AgdaUnderscore{}}}\AgdaSymbol{)}\<%
\end{code}
\begin{code}%
\>[2]\AgdaFunction{fromℕ<}%
\>[10]\AgdaSymbol{:}\AgdaSpace{}%
\AgdaSymbol{∀}\AgdaSpace{}%
\AgdaSymbol{\{}\AgdaBound{m}\AgdaSpace{}%
\AgdaBound{n}\AgdaSymbol{\}}\AgdaSpace{}%
\AgdaSymbol{→}\AgdaSpace{}%
\AgdaBound{m}\AgdaSpace{}%
\AgdaOperator{\AgdaFunction{<}}\AgdaSpace{}%
\AgdaBound{n}\AgdaSpace{}%
\AgdaSymbol{→}\AgdaSpace{}%
\AgdaDatatype{Fin}\AgdaSpace{}%
\AgdaBound{n}\<%
\\
\>[2]\AgdaFunction{n<1+n}%
\>[10]\AgdaSymbol{:}\AgdaSpace{}%
\AgdaSymbol{∀}\AgdaSpace{}%
\AgdaBound{n}\AgdaSpace{}%
\AgdaSymbol{→}\AgdaSpace{}%
\AgdaBound{n}\AgdaSpace{}%
\AgdaOperator{\AgdaFunction{<}}\AgdaSpace{}%
\AgdaNumber{1}\AgdaSpace{}%
\AgdaOperator{\AgdaPrimitive{+}}\AgdaSpace{}%
\AgdaBound{n}\<%
\end{code}
\begin{code}[hide]%
\>[2]\AgdaFunction{fromℕ<}\AgdaSpace{}%
\AgdaSymbol{=}\AgdaSpace{}%
\AgdaPostulate{⋯}\<%
\\
\>[2]\AgdaFunction{n<1+n}\AgdaSpace{}%
\AgdaSymbol{=}\AgdaSpace{}%
\AgdaPostulate{⋯}\<%
\end{code}
The \AF{fromℕ<} function turns a proof of \AB{m} \AF{<} \AB{n} into an element of type
\AF{Fin} \AB{n}.  As we are encoding \AF{Fin}s as natural numbers,
the extracted version of \AF{fromℕ<} just returns the first argument of
\AF{\_<\_}, which is the implicit argument \AB{m}.
Note that by doing so we are not losing any information,
as the proof here is merely asserting that \AB{m} fits the specification.
In general, it is always possible to extract the type-level arguments
of a dependent type such as \AF{\_<\_}, as long as we avoid using
features that distinguish between type-level and term-level arguments
such as parametricity or run-time irrelevance.

It might seem that the assertion on the result is unnecessary, since it is guaranteed to be satisfied
by construction. However, by inserting this assertion we
pass on information (that might be undecidable to recompute) further down the toolchain.
This may be used for example by the compiler of the target language to perform more optimizations.
All these assertions can
be turned off if a programmer or a compiler decides so, but this is
not a concern of the extractor.

\subsection{Pattern Matching}
Many target languages do not support function
definitions in a pattern-matching style, whereas in Agda it is the primary
way of defining functions.
Hence extractors often need to transform a definition by pattern matching
into one using conditionals.
In this section we show how to do this, and
demonstrate how implementing the extractor in Agda can lead to
extra safety guarantees.

%

The problem of compiling a definition by pattern matching splits
naturally into two subproblems: computing a condition from
each given clause, and joining all such conditions in a single conditional.
Let us start with the latter.
We implement the algorithm in the \AF{kompile-cls} function
of the following type:
\begin{code}[hide]%
\>[0]\AgdaKeyword{module}\AgdaSpace{}%
\AgdaModule{Cls}\AgdaSpace{}%
\AgdaKeyword{where}\<%
\\
\>[0][@{}l@{\AgdaIndent{0}}]%
\>[2]\AgdaKeyword{open}\AgdaSpace{}%
\AgdaKeyword{import}\AgdaSpace{}%
\AgdaModule{Reflection}\<%
\\
\>[2]\AgdaKeyword{open}\AgdaSpace{}%
\AgdaKeyword{import}\AgdaSpace{}%
\AgdaModule{Data.Nat}\AgdaSpace{}%
\AgdaSymbol{as}\AgdaSpace{}%
\AgdaModule{ℕ}\AgdaSpace{}%
\AgdaKeyword{hiding}\AgdaSpace{}%
\AgdaSymbol{(}\AgdaOperator{\AgdaFunction{\AgdaUnderscore{}≟\AgdaUnderscore{}}}\AgdaSymbol{)}\<%
\\
\>[2]\AgdaFunction{Strings}\AgdaSpace{}%
\AgdaSymbol{=}\AgdaSpace{}%
\AgdaDatatype{List}\AgdaSpace{}%
\AgdaPostulate{String}\<%
\\
\>[2]\AgdaKeyword{open}\AgdaSpace{}%
\AgdaModule{Kaleid}\<%
\\
\>[2]\AgdaKeyword{open}\AgdaSpace{}%
\AgdaModule{ExtrStructMod}\AgdaSpace{}%
\AgdaKeyword{using}\AgdaSpace{}%
\AgdaSymbol{(}\AgdaFunction{SKS}\AgdaSymbol{)}\<%
\\
\>[0]\<%
\end{code}
\begin{code}%
\>[0][@{}l@{\AgdaIndent{1}}]%
\>[2]\AgdaFunction{kompile-cls}\AgdaSpace{}%
\AgdaSymbol{:}\AgdaSpace{}%
\AgdaSymbol{(}\AgdaBound{cls}\AgdaSpace{}%
\AgdaSymbol{:}\AgdaSpace{}%
\AgdaFunction{Clauses}\AgdaSymbol{)}\AgdaSpace{}%
\AgdaSymbol{→}\AgdaSpace{}%
\AgdaSymbol{(}\AgdaBound{vars}\AgdaSpace{}%
\AgdaSymbol{:}\AgdaSpace{}%
\AgdaFunction{Strings}\AgdaSymbol{)}\AgdaSpace{}%
\AgdaSymbol{→}\AgdaSpace{}%
\AgdaSymbol{(}\AgdaBound{ret}\AgdaSpace{}%
\AgdaSymbol{:}\AgdaSpace{}%
\AgdaPostulate{String}\AgdaSymbol{)}\AgdaSpace{}%
\AgdaSymbol{→}\AgdaSpace{}%
\AgdaFunction{SKS}\AgdaSpace{}%
\AgdaSymbol{(}\AgdaDatatype{Err}\AgdaSpace{}%
\AgdaDatatype{Expr}\AgdaSymbol{)}\<%
\end{code}
\begin{code}[hide]%
\>[2]\AgdaFunction{kompile-cls}\AgdaSpace{}%
\AgdaSymbol{=}\AgdaSpace{}%
\AgdaPostulate{⋯}\<%
\end{code}
The first argument is the list of clauses, the second argument is the
list of variable names, and the last argument is the name of the variable
we assign the return value to.  As we are extracting the body of the clauses,
we need to propagate the state of extraction, so the function operates in the \AD{SKS}
monad.  The function traverses clauses in the order they appear
in the definition and combines them in a nested if-then-else chain as
in the following example:
\begin{code}%
\>[2]\AgdaFunction{ack}\AgdaSpace{}%
\AgdaSymbol{:}\AgdaSpace{}%
\AgdaDatatype{ℕ}\AgdaSpace{}%
\AgdaSymbol{→}\AgdaSpace{}%
\AgdaDatatype{ℕ}\AgdaSpace{}%
\AgdaSymbol{→}\AgdaSpace{}%
\AgdaDatatype{ℕ}%
\>[52]\AgdaComment{--\ def\ ack\ (x1,\ x2):}\<%
\\
\>[2]\AgdaFunction{ack}\AgdaSpace{}%
\AgdaNumber{0}%
\>[14]\AgdaBound{n}%
\>[23]\AgdaSymbol{=}\AgdaSpace{}%
\AgdaNumber{1}\AgdaSpace{}%
\AgdaOperator{\AgdaPrimitive{+}}\AgdaSpace{}%
\AgdaBound{n}%
\>[52]\AgdaComment{--\ \ \ if\ x1\ ==\ 0:\ 1\ +\ x2}\<%
\\
\>[2]\AgdaFunction{ack}\AgdaSpace{}%
\AgdaSymbol{(}\AgdaInductiveConstructor{suc}\AgdaSpace{}%
\AgdaBound{m}\AgdaSymbol{)}\AgdaSpace{}%
\AgdaNumber{0}%
\>[23]\AgdaSymbol{=}\AgdaSpace{}%
\AgdaFunction{ack}\AgdaSpace{}%
\AgdaBound{m}\AgdaSpace{}%
\AgdaNumber{1}%
\>[52]\AgdaComment{--\ \ \ else\ if\ x1\ >\ 0\ \&\&\ x2\ ==\ 0:\ ack\ (x1-1)\ 1}\<%
\\
\>[2]\AgdaFunction{ack}\AgdaSpace{}%
\AgdaSymbol{(}\AgdaInductiveConstructor{suc}\AgdaSpace{}%
\AgdaBound{m}\AgdaSymbol{)}\AgdaSpace{}%
\AgdaSymbol{(}\AgdaInductiveConstructor{suc}\AgdaSpace{}%
\AgdaBound{n}\AgdaSymbol{)}%
\>[23]\AgdaSymbol{=}\AgdaSpace{}%
\AgdaFunction{ack}\AgdaSpace{}%
\AgdaBound{m}\AgdaSpace{}%
\AgdaSymbol{(}\AgdaFunction{ack}\AgdaSpace{}%
\AgdaSymbol{(}\AgdaInductiveConstructor{suc}\AgdaSpace{}%
\AgdaBound{m}\AgdaSymbol{)}\AgdaSpace{}%
\AgdaBound{n}\AgdaSymbol{)}%
\>[52]\AgdaComment{--\ \ \ else:\ ack\ (x1-1)\ (ack\ x1\ (x2-1))}\<%
\end{code}
Note that in the second conditional, we have an explicit check that
\AB{x1} \AF{>} \AN{0}, which is redundant.  However, if the first
and the second clauses were swapped, such a comparison with zero must be
present.  Our current implementation takes a minimalist approach and generates
predicates for each clause separately, without taking the previous clauses
into account.
Many target languages will optimize away such redundant checks.

To compile definitions with absurd clauses, we need a way to abort
computation. For example:
\begin{code}%
\>[2]\AgdaFunction{ex₁₁}\AgdaSpace{}%
\AgdaSymbol{:}\AgdaSpace{}%
\AgdaSymbol{∀}\AgdaSpace{}%
\AgdaBound{n}\AgdaSpace{}%
\AgdaSymbol{→}\AgdaSpace{}%
\AgdaBound{n}\AgdaSpace{}%
\AgdaOperator{\AgdaFunction{<}}\AgdaSpace{}%
\AgdaNumber{0}\AgdaSpace{}%
\AgdaSymbol{→}\AgdaSpace{}%
\AgdaDatatype{ℕ}%
\>[30]\AgdaComment{--\ def\ ex11\ (x1,\ x2):}\<%
\\
\>[2]\AgdaFunction{ex₁₁}\AgdaSpace{}%
\AgdaBound{n}\AgdaSpace{}%
\AgdaSymbol{()}%
\>[30]\AgdaComment{--\ \ \ assert\ (0)}\<%
\end{code}
%
%
Rather than returning an arbitrary value, we use \AF{assert} (\AN{0})
to abort computation.
When a function has both regular and absurd clauses, we are faced with
a design decision. We can either preserve the absurd clauses and use
\AF{assert} (\AN{0}) as their body, or eliminate them entirely.
While eliminating them is sound, leaving them in can provide valuable
information for the compiler of the target language. Consider the following example:
\begin{code}%
\>[2]\AgdaFunction{ex₁₀}\AgdaSpace{}%
\AgdaSymbol{:}\AgdaSpace{}%
\AgdaKeyword{let}\AgdaSpace{}%
\AgdaBound{P}\AgdaSpace{}%
\AgdaBound{x}\AgdaSpace{}%
\AgdaSymbol{=}\AgdaSpace{}%
\AgdaBound{x}\AgdaSpace{}%
\AgdaOperator{\AgdaPrimitive{*}}\AgdaSpace{}%
\AgdaBound{x}\AgdaSpace{}%
\AgdaOperator{\AgdaPrimitive{+}}\AgdaSpace{}%
\AgdaNumber{2}\AgdaSpace{}%
\AgdaOperator{\AgdaPrimitive{∸}}\AgdaSpace{}%
\AgdaNumber{3}\AgdaSpace{}%
\AgdaOperator{\AgdaPrimitive{*}}\AgdaSpace{}%
\AgdaBound{x}\AgdaSpace{}%
\AgdaKeyword{in}\AgdaSpace{}%
\AgdaSymbol{∀}\AgdaSpace{}%
\AgdaBound{x}\AgdaSpace{}%
\AgdaSymbol{→}\AgdaSpace{}%
\AgdaNumber{0}\AgdaSpace{}%
\AgdaOperator{\AgdaFunction{<}}\AgdaSpace{}%
\AgdaBound{P}\AgdaSpace{}%
\AgdaBound{x}\AgdaSpace{}%
\AgdaSymbol{→}\AgdaSpace{}%
\AgdaDatatype{ℕ}\<%
\\
\>[2]\AgdaFunction{ex₁₀}\AgdaSpace{}%
\AgdaNumber{1}\AgdaSpace{}%
\AgdaSymbol{()}\AgdaSpace{}%
\AgdaSymbol{;}\AgdaSpace{}%
\AgdaFunction{ex₁₀}\AgdaSpace{}%
\AgdaNumber{2}\AgdaSpace{}%
\AgdaSymbol{()}\AgdaSpace{}%
\AgdaSymbol{;}\AgdaSpace{}%
\AgdaCatchallClause{\AgdaFunction{ex₁₀}}\AgdaSpace{}%
\AgdaCatchallClause{\AgdaBound{x}}\AgdaSpace{}%
\AgdaCatchallClause{\AgdaBound{pf}}\AgdaSpace{}%
\AgdaSymbol{=}\AgdaSpace{}%
\AgdaPostulate{⋯}\<%
\end{code}
We can generate an assertion that \AF{P} \AB{x} is greater than 0, and
after eliminating first two cases, the body of the function would
reduce to a single statement over variables \AB{x} and \AB{pf}.
However, deducing that \AB{x} does not equal \AN{1} or \AN{2} is not
straightforward.  Instead, we preserve this information as an
assertion, so the compiler of the target language can make use of
it.

The actual translation from patterns to predicates is implemented in
\AF{kompile-clpats}. Here we show only two clauses for compiling a
function that matches on the \AF{Fin} constructors \AC{zero} and
\AC{suc}, respectively.

\begin{code}[hide]%
\>[0]\AgdaKeyword{module}\AgdaSpace{}%
\AgdaModule{ClPats}\AgdaSpace{}%
\AgdaKeyword{where}\<%
\\
\>[0][@{}l@{\AgdaIndent{0}}]%
\>[2]\AgdaKeyword{open}\AgdaSpace{}%
\AgdaKeyword{import}\AgdaSpace{}%
\AgdaModule{Reflection}\AgdaSpace{}%
\AgdaKeyword{hiding}\AgdaSpace{}%
\AgdaSymbol{(}\AgdaOperator{\AgdaFunction{\AgdaUnderscore{}>>=\AgdaUnderscore{}}}\AgdaSymbol{;}\AgdaSpace{}%
\AgdaPostulate{return}\AgdaSymbol{;}\AgdaSpace{}%
\AgdaOperator{\AgdaFunction{\AgdaUnderscore{}>>\AgdaUnderscore{}}}\AgdaSymbol{)}\<%
\\
\>[2]\AgdaKeyword{open}\AgdaSpace{}%
\AgdaKeyword{import}\AgdaSpace{}%
\AgdaModule{Reflection.Term}\AgdaSpace{}%
\AgdaKeyword{using}\AgdaSpace{}%
\AgdaSymbol{(}\AgdaFunction{Telescope}\AgdaSymbol{;}\AgdaSpace{}%
\AgdaInductiveConstructor{con}\AgdaSymbol{)}\<%
\\
\>[2]\AgdaKeyword{open}\AgdaSpace{}%
\AgdaKeyword{import}\AgdaSpace{}%
\AgdaModule{Data.Fin}\AgdaSpace{}%
\AgdaSymbol{as}\AgdaSpace{}%
\AgdaModule{F}\AgdaSpace{}%
\AgdaKeyword{using}\AgdaSpace{}%
\AgdaSymbol{(}\AgdaDatatype{Fin}\AgdaSymbol{;}\AgdaSpace{}%
\AgdaInductiveConstructor{zero}\AgdaSymbol{;}\AgdaSpace{}%
\AgdaInductiveConstructor{suc}\AgdaSymbol{)}\<%
\\
\>[2]\AgdaKeyword{open}\AgdaSpace{}%
\AgdaKeyword{import}\AgdaSpace{}%
\AgdaModule{Data.Product}\<%
\\
\>[2]\AgdaKeyword{open}\AgdaSpace{}%
\AgdaKeyword{import}\AgdaSpace{}%
\AgdaModule{Data.List}\AgdaSpace{}%
\AgdaKeyword{using}\AgdaSpace{}%
\AgdaSymbol{(}\AgdaOperator{\AgdaFunction{\AgdaUnderscore{}++\AgdaUnderscore{}}}\AgdaSymbol{)}\<%
\\
\>[2]\AgdaKeyword{open}\AgdaSpace{}%
\AgdaKeyword{import}\AgdaSpace{}%
\AgdaModule{Data.Nat}\AgdaSpace{}%
\AgdaSymbol{as}\AgdaSpace{}%
\AgdaModule{ℕ}\AgdaSpace{}%
\AgdaKeyword{hiding}\AgdaSpace{}%
\AgdaSymbol{(}\AgdaOperator{\AgdaFunction{\AgdaUnderscore{}≟\AgdaUnderscore{}}}\AgdaSymbol{)}\<%
\\
\>[2]\AgdaKeyword{open}\AgdaSpace{}%
\AgdaModule{Kaleid}\<%
\\
\>[2]\AgdaKeyword{record}\AgdaSpace{}%
\AgdaRecord{PatSt}\AgdaSpace{}%
\AgdaSymbol{:}\AgdaSpace{}%
\AgdaPrimitive{Set}\AgdaSpace{}%
\AgdaKeyword{where}\<%
\\
\>[2][@{}l@{\AgdaIndent{0}}]%
\>[4]\AgdaKeyword{field}\AgdaSpace{}%
\AgdaField{conds}\AgdaSpace{}%
\AgdaSymbol{:}\AgdaSpace{}%
\AgdaDatatype{List}\AgdaSpace{}%
\AgdaDatatype{Expr}\<%
\\
\\[\AgdaEmptyExtraSkip]%
\>[2]\AgdaKeyword{infixl}\AgdaSpace{}%
\AgdaNumber{10}\AgdaSpace{}%
\AgdaOperator{\AgdaFunction{\AgdaUnderscore{}+=c\AgdaUnderscore{}}}\<%
\\
\>[2]\AgdaOperator{\AgdaFunction{\AgdaUnderscore{}+=c\AgdaUnderscore{}}}\AgdaSpace{}%
\AgdaSymbol{:}\AgdaSpace{}%
\AgdaRecord{PatSt}\AgdaSpace{}%
\AgdaSymbol{→}\AgdaSpace{}%
\AgdaDatatype{Expr}\AgdaSpace{}%
\AgdaSymbol{→}\AgdaSpace{}%
\AgdaRecord{PatSt}\<%
\\
\>[2]\AgdaBound{p}\AgdaSpace{}%
\AgdaOperator{\AgdaFunction{+=c}}\AgdaSpace{}%
\AgdaBound{c}\AgdaSpace{}%
\AgdaSymbol{=}\AgdaSpace{}%
\AgdaKeyword{record}\AgdaSpace{}%
\AgdaBound{p}\AgdaSpace{}%
\AgdaSymbol{\{}\AgdaSpace{}%
\AgdaField{conds}\AgdaSpace{}%
\AgdaSymbol{=}\AgdaSpace{}%
\AgdaField{PatSt.conds}\AgdaSpace{}%
\AgdaBound{p}\AgdaSpace{}%
\AgdaOperator{\AgdaFunction{++}}\AgdaSpace{}%
\AgdaOperator{\AgdaFunction{[}}\AgdaSpace{}%
\AgdaBound{c}\AgdaSpace{}%
\AgdaOperator{\AgdaFunction{]}}\AgdaSpace{}%
\AgdaSymbol{\}}\<%
\\
\>[2]\AgdaFunction{pst-fresh}\AgdaSpace{}%
\AgdaSymbol{:}\AgdaSpace{}%
\AgdaRecord{PatSt}\AgdaSpace{}%
\AgdaSymbol{→}\AgdaSpace{}%
\AgdaPostulate{String}\AgdaSpace{}%
\AgdaSymbol{→}\AgdaSpace{}%
\AgdaDatatype{Err}\AgdaSpace{}%
\AgdaOperator{\AgdaFunction{\$}}\AgdaSpace{}%
\AgdaPostulate{String}\AgdaSpace{}%
\AgdaOperator{\AgdaFunction{×}}\AgdaSpace{}%
\AgdaRecord{PatSt}\<%
\\
\>[2]\AgdaFunction{pst-fresh}\AgdaSpace{}%
\AgdaSymbol{=}\AgdaSpace{}%
\AgdaPostulate{⋯}\<%
\\
\>[2]\AgdaOperator{\AgdaFunction{\AgdaUnderscore{}>>=\AgdaUnderscore{}}}\AgdaSpace{}%
\AgdaSymbol{:}\AgdaSpace{}%
\AgdaSymbol{∀}\AgdaSpace{}%
\AgdaSymbol{\{}\AgdaBound{X}\AgdaSpace{}%
\AgdaBound{Y}\AgdaSpace{}%
\AgdaSymbol{:}\AgdaSpace{}%
\AgdaPrimitive{Set}\AgdaSymbol{\}}\AgdaSpace{}%
\AgdaSymbol{→}\AgdaSpace{}%
\AgdaDatatype{Err}\AgdaSpace{}%
\AgdaBound{X}\AgdaSpace{}%
\AgdaSymbol{→}\AgdaSpace{}%
\AgdaSymbol{(}\AgdaBound{X}\AgdaSpace{}%
\AgdaSymbol{→}\AgdaSpace{}%
\AgdaDatatype{Err}\AgdaSpace{}%
\AgdaBound{Y}\AgdaSymbol{)}\AgdaSpace{}%
\AgdaSymbol{→}\AgdaSpace{}%
\AgdaDatatype{Err}\AgdaSpace{}%
\AgdaBound{Y}\<%
\\
\>[2]\AgdaOperator{\AgdaFunction{\AgdaUnderscore{}>>=\AgdaUnderscore{}}}\AgdaSpace{}%
\AgdaSymbol{=}\AgdaSpace{}%
\AgdaPostulate{⋯}\<%
\\
\>[2]\AgdaFunction{return}\AgdaSpace{}%
\AgdaSymbol{:}\AgdaSpace{}%
\AgdaSymbol{∀}\AgdaSpace{}%
\AgdaSymbol{\{}\AgdaBound{X}\AgdaSpace{}%
\AgdaSymbol{:}\AgdaSpace{}%
\AgdaPrimitive{Set}\AgdaSymbol{\}}\AgdaSpace{}%
\AgdaSymbol{→}\AgdaSpace{}%
\AgdaBound{X}\AgdaSpace{}%
\AgdaSymbol{→}\AgdaSpace{}%
\AgdaDatatype{Err}\AgdaSpace{}%
\AgdaBound{X}\<%
\\
\>[2]\AgdaFunction{return}\AgdaSpace{}%
\AgdaSymbol{=}\AgdaSpace{}%
\AgdaPostulate{⋯}\<%
\end{code}
\begin{code}%
\>[2]\AgdaSymbol{\{-\#}\AgdaSpace{}%
\AgdaKeyword{TERMINATING}\AgdaSpace{}%
\AgdaSymbol{\#-\}}\<%
\\
\>[2]\AgdaFunction{kompile-clpats}\AgdaSpace{}%
\AgdaSymbol{:}\AgdaSpace{}%
\AgdaFunction{Telescope}\AgdaSpace{}%
\AgdaSymbol{→}\AgdaSpace{}%
\AgdaSymbol{(}\AgdaBound{pats}\AgdaSpace{}%
\AgdaSymbol{:}\AgdaSpace{}%
\AgdaDatatype{List}\AgdaSpace{}%
\AgdaSymbol{(}\AgdaDatatype{Arg}\AgdaSpace{}%
\AgdaDatatype{Pattern}\AgdaSymbol{))}\AgdaSpace{}%
\AgdaSymbol{→}\AgdaSpace{}%
\AgdaSymbol{(}\AgdaBound{exprs}\AgdaSpace{}%
\AgdaSymbol{:}\AgdaSpace{}%
\AgdaDatatype{List}\AgdaSpace{}%
\AgdaDatatype{Expr}\AgdaSymbol{)}\AgdaSpace{}%
\AgdaSymbol{→}\AgdaSpace{}%
\AgdaRecord{PatSt}\AgdaSpace{}%
\AgdaSymbol{→}\AgdaSpace{}%
\AgdaDatatype{Err}\AgdaSpace{}%
\AgdaRecord{PatSt}\<%
\\
\>[2]\AgdaFunction{kompile-clpats}\AgdaSpace{}%
\AgdaBound{tel}\AgdaSpace{}%
\AgdaSymbol{(}\AgdaInductiveConstructor{arg}\AgdaSpace{}%
\AgdaBound{i}\AgdaSpace{}%
\AgdaSymbol{(}\AgdaInductiveConstructor{con}\AgdaSpace{}%
\AgdaSymbol{(}\AgdaKeyword{quote}\AgdaSpace{}%
\AgdaInductiveConstructor{Fin.zero}\AgdaSymbol{)}\AgdaSpace{}%
\AgdaBound{ps}\AgdaSymbol{)}\AgdaSpace{}%
\AgdaOperator{\AgdaInductiveConstructor{∷}}\AgdaSpace{}%
\AgdaBound{l}\AgdaSymbol{)}\AgdaSpace{}%
\AgdaSymbol{(}\AgdaBound{e}\AgdaSpace{}%
\AgdaOperator{\AgdaInductiveConstructor{∷}}\AgdaSpace{}%
\AgdaBound{ctx}\AgdaSymbol{)}\AgdaSpace{}%
\AgdaBound{pst}\AgdaSpace{}%
\AgdaSymbol{=}\<%
\\
\>[2][@{}l@{\AgdaIndent{0}}]%
\>[4]\AgdaFunction{kompile-clpats}\AgdaSpace{}%
\AgdaBound{tel}\AgdaSpace{}%
\AgdaBound{l}\AgdaSpace{}%
\AgdaBound{ctx}\AgdaSpace{}%
\AgdaOperator{\AgdaFunction{\$}}\AgdaSpace{}%
\AgdaBound{pst}\AgdaSpace{}%
\AgdaOperator{\AgdaFunction{+=c}}\AgdaSpace{}%
\AgdaSymbol{(}\AgdaInductiveConstructor{BinOp}\AgdaSpace{}%
\AgdaInductiveConstructor{Eq}\AgdaSpace{}%
\AgdaBound{e}\AgdaSpace{}%
\AgdaSymbol{(}\AgdaInductiveConstructor{Nat}\AgdaSpace{}%
\AgdaNumber{0}\AgdaSymbol{))}\<%
\\
\>[2]\AgdaFunction{kompile-clpats}\AgdaSpace{}%
\AgdaBound{tel}\AgdaSpace{}%
\AgdaSymbol{(}\AgdaInductiveConstructor{arg}\AgdaSpace{}%
\AgdaBound{i}\AgdaSpace{}%
\AgdaSymbol{(}\AgdaInductiveConstructor{con}\AgdaSpace{}%
\AgdaSymbol{(}\AgdaKeyword{quote}\AgdaSpace{}%
\AgdaInductiveConstructor{Fin.suc}\AgdaSymbol{)}\AgdaSpace{}%
\AgdaBound{ps}\AgdaSymbol{@(\AgdaUnderscore{}}\AgdaSpace{}%
\AgdaOperator{\AgdaInductiveConstructor{∷}}\AgdaSpace{}%
\AgdaSymbol{\AgdaUnderscore{}}\AgdaSpace{}%
\AgdaOperator{\AgdaInductiveConstructor{∷}}\AgdaSpace{}%
\AgdaInductiveConstructor{[]}\AgdaSymbol{))}\AgdaSpace{}%
\AgdaOperator{\AgdaInductiveConstructor{∷}}\AgdaSpace{}%
\AgdaBound{l}\AgdaSymbol{)}\AgdaSpace{}%
\AgdaSymbol{(}\AgdaBound{e}\AgdaSpace{}%
\AgdaOperator{\AgdaInductiveConstructor{∷}}\AgdaSpace{}%
\AgdaBound{ctx}\AgdaSymbol{)}\AgdaSpace{}%
\AgdaBound{pst}\AgdaSpace{}%
\AgdaSymbol{=}\AgdaSpace{}%
\AgdaKeyword{do}\<%
\\
\>[2][@{}l@{\AgdaIndent{0}}]%
\>[4]\AgdaSymbol{(}\AgdaBound{ub}\AgdaSpace{}%
\AgdaOperator{\AgdaInductiveConstructor{,}}\AgdaSpace{}%
\AgdaBound{pst}\AgdaSymbol{)}\AgdaSpace{}%
\AgdaOperator{\AgdaFunction{←}}\AgdaSpace{}%
\AgdaFunction{pst-fresh}\AgdaSpace{}%
\AgdaBound{pst}\AgdaSpace{}%
\AgdaString{"ub\AgdaUnderscore{}"}\<%
\\
\>[4]\AgdaFunction{kompile-clpats}%
\>[1318I]\AgdaBound{tel}\AgdaSpace{}%
\AgdaSymbol{(}\AgdaBound{ps}\AgdaSpace{}%
\AgdaOperator{\AgdaFunction{++}}\AgdaSpace{}%
\AgdaBound{l}\AgdaSymbol{)}\AgdaSpace{}%
\AgdaSymbol{(}\AgdaInductiveConstructor{Var}\AgdaSpace{}%
\AgdaBound{ub}\AgdaSpace{}%
\AgdaOperator{\AgdaInductiveConstructor{∷}}\AgdaSpace{}%
\AgdaSymbol{(}\AgdaInductiveConstructor{BinOp}\AgdaSpace{}%
\AgdaInductiveConstructor{Minus}\AgdaSpace{}%
\AgdaBound{e}\AgdaSpace{}%
\AgdaSymbol{(}\AgdaInductiveConstructor{Nat}\AgdaSpace{}%
\AgdaNumber{1}\AgdaSymbol{))}\AgdaSpace{}%
\AgdaOperator{\AgdaInductiveConstructor{∷}}\AgdaSpace{}%
\AgdaBound{ctx}\AgdaSymbol{)}\<%
\\
\>[.][@{}l@{}]\<[1318I]%
\>[19]\AgdaOperator{\AgdaFunction{\$}}\AgdaSpace{}%
\AgdaBound{pst}\AgdaSpace{}%
\AgdaOperator{\AgdaFunction{+=c}}\AgdaSpace{}%
\AgdaSymbol{(}\AgdaInductiveConstructor{BinOp}\AgdaSpace{}%
\AgdaInductiveConstructor{Gt}\AgdaSpace{}%
\AgdaBound{e}\AgdaSpace{}%
\AgdaSymbol{(}\AgdaInductiveConstructor{Nat}\AgdaSpace{}%
\AgdaNumber{0}\AgdaSymbol{))}\<%
\\
\>[2]\AgdaCatchallClause{\AgdaFunction{kompile-clpats}}\AgdaSpace{}%
\AgdaCatchallClause{\AgdaBound{tel}}\AgdaSpace{}%
\AgdaCatchallClause{\AgdaBound{ps}}\AgdaSpace{}%
\AgdaCatchallClause{\AgdaBound{ctx}}\AgdaSpace{}%
\AgdaCatchallClause{\AgdaBound{pst}}\AgdaSpace{}%
\AgdaSymbol{=}\AgdaSpace{}%
\AgdaPostulate{⋯}\<%
\end{code}
The function take four arguments:
the telescope mapping variables used in the pattern list to their names
and types;
the list of patterns; the list of expressions that are being matched
by the patterns; and the state record \AD{PatSt}. When compiling a clause,
the list of expressions is initialised with the function arguments. The state record
contains the counter for fresh variables, the list of conditions
accumulated so far, local assignments, and so on.

For each constructor pattern we produce a condition that holds only
when the encoded value in the target language represents the value
that was built using the given constructor.  For example, as we
represent \AF{Fin} with natural numbers, the conditions for matching
\AB{e} against a pattern \AC{zero} \{\AB{ub}\} constructor is \AB{e}
\AF{==} \AN{0} (note that the precondition generated from the type of
\AB{e} already enforces that \AB{e} \AF{<} \AB{ub}, so we do not need
to check it again).
Correspondingly, the pattern \AC{suc} \{\AB{ub}\} \AB{x} yields the condition
\AB{e} \AF{>} \AN{0}.
The \AF{pst-fresh} function generates a variable with the unique
(within the clause) name, and \AF{\_+=c\_} adds the new condition to
the state.

Note that we marked this function as terminating.  We had to do so as
the recursive call to \AF{kompile-clpats} with argument \AB{ps}
\AF{++} \AB{l} is not structurally decreasing.
%
Nevertheless, the function is terminating because the \AF{Pattern}
type is well-founded, and all the objects of that type are finite.
The full version of our code works around this limitation of the
termination checker by adding an extra argument expressing the
overall size of the pattern, but we omit it here for simplicity.

As a side note, Agda internally represents definitions by pattern
matching as \emph{case trees} where most redundant checks have been
eliminated. However, unfortunately this representation is currently
not available via the reflection API. However, if we wanted to
generate more efficient code it would probably be simpler to extend
the reflection API rather than reimplement the translation to a case
tree ourselves.



\subsection{Translating terms} \label{sec:translating-terms}
The actual translation of Agda terms into Kaleidoscope terms is a
mechanical process.  However, as the translation may fail, the use of
monads and \AK{do}-notation for managing errors help us to keep the
code clean:
\begin{code}[hide]%
\>[0]\AgdaKeyword{module}\AgdaSpace{}%
\AgdaModule{KompTerm}\AgdaSpace{}%
\AgdaKeyword{where}\<%
\\
\>[0][@{}l@{\AgdaIndent{0}}]%
\>[2]\AgdaKeyword{open}\AgdaSpace{}%
\AgdaKeyword{import}\AgdaSpace{}%
\AgdaModule{Reflection}\AgdaSpace{}%
\AgdaKeyword{hiding}\AgdaSpace{}%
\AgdaSymbol{(}\AgdaOperator{\AgdaFunction{\AgdaUnderscore{}>>=\AgdaUnderscore{}}}\AgdaSymbol{;}\AgdaSpace{}%
\AgdaPostulate{return}\AgdaSymbol{;}\AgdaSpace{}%
\AgdaOperator{\AgdaFunction{\AgdaUnderscore{}>>\AgdaUnderscore{}}}\AgdaSymbol{)}\<%
\\
\>[2]\AgdaKeyword{open}\AgdaSpace{}%
\AgdaKeyword{import}\AgdaSpace{}%
\AgdaModule{Reflection.Term}\<%
\\
\>[2]\AgdaKeyword{open}\AgdaSpace{}%
\AgdaKeyword{import}\AgdaSpace{}%
\AgdaModule{Relation.Binary.PropositionalEquality}\AgdaSpace{}%
\AgdaKeyword{using}\AgdaSpace{}%
\AgdaSymbol{(}\AgdaOperator{\AgdaDatatype{\AgdaUnderscore{}≡\AgdaUnderscore{}}}\AgdaSpace{}%
\AgdaSymbol{;}\AgdaSpace{}%
\AgdaInductiveConstructor{refl}\AgdaSymbol{;}\AgdaSpace{}%
\AgdaFunction{cong}\AgdaSymbol{)}\<%
\\
\>[2]\AgdaKeyword{open}\AgdaSpace{}%
\AgdaKeyword{import}\AgdaSpace{}%
\AgdaModule{Relation.Nullary}\<%
\\
\>[2]\AgdaKeyword{open}\AgdaSpace{}%
\AgdaModule{Kaleid}\<%
\\
\>[2]\AgdaKeyword{open}\AgdaSpace{}%
\AgdaModule{ExtrStructMod}\AgdaSpace{}%
\AgdaKeyword{using}\AgdaSpace{}%
\AgdaSymbol{(}\AgdaFunction{SKS}\AgdaSymbol{)}\<%
\\
\>[2]\AgdaKeyword{open}\AgdaSpace{}%
\AgdaKeyword{import}\AgdaSpace{}%
\AgdaModule{Data.Fin}\AgdaSpace{}%
\AgdaSymbol{as}\AgdaSpace{}%
\AgdaModule{F}\AgdaSpace{}%
\AgdaKeyword{using}\AgdaSpace{}%
\AgdaSymbol{()}\<%
\\
\>[2]\AgdaKeyword{open}\AgdaSpace{}%
\AgdaKeyword{import}\AgdaSpace{}%
\AgdaModule{Data.List}\AgdaSpace{}%
\AgdaKeyword{using}\AgdaSpace{}%
\AgdaSymbol{(}\AgdaFunction{length}\AgdaSymbol{;}\AgdaSpace{}%
\AgdaOperator{\AgdaFunction{\AgdaUnderscore{}++\AgdaUnderscore{}}}\AgdaSymbol{)}\<%
\\
\>[2]\AgdaKeyword{open}\AgdaSpace{}%
\AgdaKeyword{import}\AgdaSpace{}%
\AgdaModule{Data.Nat}\AgdaSpace{}%
\AgdaSymbol{as}\AgdaSpace{}%
\AgdaModule{ℕ}\AgdaSpace{}%
\AgdaKeyword{hiding}\AgdaSpace{}%
\AgdaSymbol{(}\AgdaOperator{\AgdaFunction{\AgdaUnderscore{}≟\AgdaUnderscore{}}}\AgdaSymbol{)}\<%
\\
\>[2]\AgdaKeyword{record}\AgdaSpace{}%
\AgdaRecord{KS}\AgdaSpace{}%
\AgdaSymbol{:}\AgdaSpace{}%
\AgdaPrimitive{Set}\AgdaSpace{}%
\AgdaKeyword{where}\<%
\\
\>[2][@{}l@{\AgdaIndent{0}}]%
\>[4]\AgdaKeyword{field}\AgdaSpace{}%
\AgdaField{funs}\AgdaSpace{}%
\AgdaSymbol{:}\AgdaSpace{}%
\AgdaDatatype{List}\AgdaSpace{}%
\AgdaPostulate{Name}\<%
\\
\\[\AgdaEmptyExtraSkip]%
\>[2]\AgdaComment{--SKSE\ =\ SKS\ ∘\ Err}\<%
\\
\>[2]\AgdaOperator{\AgdaFunction{\AgdaUnderscore{}>>=\AgdaUnderscore{}}}\AgdaSpace{}%
\AgdaSymbol{:}\AgdaSpace{}%
\AgdaSymbol{∀}\AgdaSpace{}%
\AgdaSymbol{\{}\AgdaBound{X}\AgdaSpace{}%
\AgdaBound{Y}\AgdaSpace{}%
\AgdaSymbol{:}\AgdaSpace{}%
\AgdaPrimitive{Set}\AgdaSymbol{\}}\AgdaSpace{}%
\AgdaSymbol{→}\AgdaSpace{}%
\AgdaFunction{SKS}\AgdaSpace{}%
\AgdaBound{X}\AgdaSpace{}%
\AgdaSymbol{→}\AgdaSpace{}%
\AgdaSymbol{(}\AgdaBound{X}\AgdaSpace{}%
\AgdaSymbol{→}\AgdaSpace{}%
\AgdaFunction{SKS}\AgdaSpace{}%
\AgdaBound{Y}\AgdaSymbol{)}\AgdaSpace{}%
\AgdaSymbol{→}\AgdaSpace{}%
\AgdaFunction{SKS}\AgdaSpace{}%
\AgdaBound{Y}\<%
\\
\>[2]\AgdaOperator{\AgdaFunction{\AgdaUnderscore{}>>=\AgdaUnderscore{}}}\AgdaSpace{}%
\AgdaSymbol{=}\AgdaSpace{}%
\AgdaPostulate{⋯}\<%
\\
\>[2]\AgdaFunction{return}\AgdaSpace{}%
\AgdaSymbol{:}\AgdaSpace{}%
\AgdaSymbol{∀}\AgdaSpace{}%
\AgdaSymbol{\{}\AgdaBound{X}\AgdaSpace{}%
\AgdaSymbol{:}\AgdaSpace{}%
\AgdaPrimitive{Set}\AgdaSymbol{\}}\AgdaSpace{}%
\AgdaSymbol{→}\AgdaSpace{}%
\AgdaBound{X}\AgdaSpace{}%
\AgdaSymbol{→}\AgdaSpace{}%
\AgdaFunction{SKS}\AgdaSpace{}%
\AgdaBound{X}\<%
\\
\>[2]\AgdaFunction{return}\AgdaSpace{}%
\AgdaSymbol{=}\AgdaSpace{}%
\AgdaPostulate{⋯}\<%
\\
\>[2]\AgdaOperator{\AgdaFunction{\AgdaUnderscore{}>>\AgdaUnderscore{}}}\AgdaSpace{}%
\AgdaSymbol{:}\AgdaSpace{}%
\AgdaSymbol{∀}\AgdaSpace{}%
\AgdaSymbol{\{}\AgdaBound{A}\AgdaSpace{}%
\AgdaBound{B}\AgdaSpace{}%
\AgdaSymbol{:}\AgdaSpace{}%
\AgdaPrimitive{Set}\AgdaSymbol{\}}\AgdaSpace{}%
\AgdaSymbol{→}\AgdaSpace{}%
\AgdaFunction{SKS}\AgdaSpace{}%
\AgdaBound{A}\AgdaSpace{}%
\AgdaSymbol{→}\AgdaSpace{}%
\AgdaFunction{SKS}\AgdaSpace{}%
\AgdaBound{B}\AgdaSpace{}%
\AgdaSymbol{→}\AgdaSpace{}%
\AgdaFunction{SKS}\AgdaSpace{}%
\AgdaBound{B}\AgdaSpace{}%
\AgdaSymbol{;}\AgdaSpace{}%
\AgdaOperator{\AgdaFunction{\AgdaUnderscore{}>>\AgdaUnderscore{}}}\AgdaSpace{}%
\AgdaSymbol{=}\AgdaSpace{}%
\AgdaPostulate{⋯}\<%
\\
\>[2]\AgdaKeyword{infixl}\AgdaSpace{}%
\AgdaNumber{4}\AgdaSpace{}%
\AgdaOperator{\AgdaFunction{\AgdaUnderscore{}<\$>\AgdaUnderscore{}}}\AgdaSpace{}%
\AgdaOperator{\AgdaFunction{\AgdaUnderscore{}⊛\AgdaUnderscore{}}}\<%
\\
\>[2]\AgdaOperator{\AgdaFunction{\AgdaUnderscore{}<\$>\AgdaUnderscore{}}}\AgdaSpace{}%
\AgdaSymbol{:}\AgdaSpace{}%
\AgdaSymbol{∀\{}\AgdaBound{A}\AgdaSpace{}%
\AgdaBound{B}\AgdaSpace{}%
\AgdaSymbol{:}\AgdaSpace{}%
\AgdaPrimitive{Set}\AgdaSymbol{\}}\AgdaSpace{}%
\AgdaSymbol{→}\AgdaSpace{}%
\AgdaSymbol{(}\AgdaBound{A}\AgdaSpace{}%
\AgdaSymbol{→}\AgdaSpace{}%
\AgdaBound{B}\AgdaSymbol{)}\AgdaSpace{}%
\AgdaSymbol{→}\AgdaSpace{}%
\AgdaDatatype{Err}\AgdaSpace{}%
\AgdaBound{A}\AgdaSpace{}%
\AgdaSymbol{→}\AgdaSpace{}%
\AgdaDatatype{Err}\AgdaSpace{}%
\AgdaBound{B}\AgdaSpace{}%
\AgdaSymbol{;}\AgdaSpace{}%
\AgdaOperator{\AgdaFunction{\AgdaUnderscore{}<\$>\AgdaUnderscore{}}}\AgdaSpace{}%
\AgdaSymbol{=}\AgdaSpace{}%
\AgdaPostulate{⋯}\<%
\\
\>[2]\AgdaOperator{\AgdaFunction{\AgdaUnderscore{}⊛\AgdaUnderscore{}}}\AgdaSpace{}%
\AgdaSymbol{:}\AgdaSpace{}%
\AgdaSymbol{∀}\AgdaSpace{}%
\AgdaSymbol{\{}\AgdaBound{A}\AgdaSpace{}%
\AgdaBound{B}\AgdaSpace{}%
\AgdaSymbol{:}\AgdaSpace{}%
\AgdaPrimitive{Set}\AgdaSymbol{\}}\AgdaSpace{}%
\AgdaSymbol{→}\AgdaSpace{}%
\AgdaDatatype{Err}\AgdaSpace{}%
\AgdaSymbol{(}\AgdaBound{A}\AgdaSpace{}%
\AgdaSymbol{→}\AgdaSpace{}%
\AgdaBound{B}\AgdaSymbol{)}\AgdaSpace{}%
\AgdaSymbol{→}\AgdaSpace{}%
\AgdaDatatype{Err}\AgdaSpace{}%
\AgdaBound{A}\AgdaSpace{}%
\AgdaSymbol{→}\AgdaSpace{}%
\AgdaDatatype{Err}\AgdaSpace{}%
\AgdaBound{B}\AgdaSpace{}%
\AgdaSymbol{;}\AgdaSpace{}%
\AgdaOperator{\AgdaFunction{\AgdaUnderscore{}⊛\AgdaUnderscore{}}}\AgdaSpace{}%
\AgdaSymbol{=}\AgdaSpace{}%
\AgdaPostulate{⋯}\<%
\\
\>[2]\AgdaFunction{modify}\AgdaSpace{}%
\AgdaSymbol{:}\AgdaSpace{}%
\AgdaSymbol{∀}\AgdaSpace{}%
\AgdaSymbol{\{}\AgdaBound{i}\AgdaSpace{}%
\AgdaSymbol{:}\AgdaSpace{}%
\AgdaRecord{⊤}\AgdaSymbol{\}}\AgdaSpace{}%
\AgdaSymbol{→}\AgdaSpace{}%
\AgdaSymbol{(}\AgdaRecord{KS}\AgdaSpace{}%
\AgdaSymbol{→}\AgdaSpace{}%
\AgdaRecord{KS}\AgdaSymbol{)}\AgdaSpace{}%
\AgdaSymbol{→}\AgdaSpace{}%
\AgdaFunction{SKS}\AgdaSpace{}%
\AgdaRecord{⊤}\AgdaSpace{}%
\AgdaSymbol{;}\AgdaSpace{}%
\AgdaFunction{modify}\AgdaSpace{}%
\AgdaSymbol{=}\AgdaSpace{}%
\AgdaPostulate{⋯}\<%
\\
\>[2]\AgdaFunction{kt}\AgdaSpace{}%
\AgdaSymbol{:}\AgdaSpace{}%
\AgdaSymbol{∀}\AgdaSpace{}%
\AgdaSymbol{\{}\AgdaBound{X}\AgdaSymbol{\}}\AgdaSpace{}%
\AgdaSymbol{→}\AgdaSpace{}%
\AgdaPostulate{String}\AgdaSpace{}%
\AgdaSymbol{→}\AgdaSpace{}%
\AgdaFunction{SKS}\AgdaSpace{}%
\AgdaSymbol{(}\AgdaDatatype{Err}\AgdaSpace{}%
\AgdaBound{X}\AgdaSymbol{)}\AgdaSpace{}%
\AgdaSymbol{;}\AgdaSpace{}%
\AgdaFunction{kt}\AgdaSpace{}%
\AgdaSymbol{=}\AgdaSpace{}%
\AgdaPostulate{⋯}\<%
\\
\>[2]\AgdaFunction{kompile-arglist}\AgdaSpace{}%
\AgdaSymbol{:}\AgdaSpace{}%
\AgdaDatatype{List}\AgdaSpace{}%
\AgdaOperator{\AgdaFunction{\$}}\AgdaSpace{}%
\AgdaDatatype{Arg}\AgdaSpace{}%
\AgdaDatatype{Term}\AgdaSpace{}%
\AgdaSymbol{→}\AgdaSpace{}%
\AgdaDatatype{List}\AgdaSpace{}%
\AgdaDatatype{ℕ}\AgdaSpace{}%
\AgdaSymbol{→}\AgdaSpace{}%
\AgdaFunction{Telescope}\AgdaSpace{}%
\AgdaSymbol{→}\AgdaSpace{}%
\AgdaFunction{SKS}\AgdaSpace{}%
\AgdaOperator{\AgdaFunction{\$}}\AgdaSpace{}%
\AgdaDatatype{Err}\AgdaSpace{}%
\AgdaSymbol{(}\AgdaDatatype{List}\AgdaSpace{}%
\AgdaDatatype{Expr}\AgdaSymbol{)}\<%
\\
\>[2]\AgdaFunction{kompile-arglist}\AgdaSpace{}%
\AgdaSymbol{=}\AgdaSpace{}%
\AgdaPostulate{⋯}\<%
\\
\>[2]\AgdaFunction{mk-iota-mask}\AgdaSpace{}%
\AgdaSymbol{:}\AgdaSpace{}%
\AgdaDatatype{ℕ}\AgdaSpace{}%
\AgdaSymbol{→}\AgdaSpace{}%
\AgdaDatatype{List}\AgdaSpace{}%
\AgdaDatatype{ℕ}\AgdaSpace{}%
\AgdaSymbol{;}\AgdaSpace{}%
\AgdaFunction{mk-iota-mask}\AgdaSpace{}%
\AgdaSymbol{=}\AgdaSpace{}%
\AgdaPostulate{⋯}\<%
\\
\>[2]\AgdaFunction{normalise-name}\AgdaSpace{}%
\AgdaSymbol{:}\AgdaSpace{}%
\AgdaPostulate{String}\AgdaSpace{}%
\AgdaSymbol{→}\AgdaSpace{}%
\AgdaPostulate{String}\AgdaSpace{}%
\AgdaSymbol{;}\AgdaSpace{}%
\AgdaFunction{normalise-name}\AgdaSpace{}%
\AgdaSymbol{=}\AgdaSpace{}%
\AgdaPostulate{⋯}\<%
\end{code}
\begin{code}%
\>[2]\AgdaFunction{kompile-term}\AgdaSpace{}%
\AgdaSymbol{:}\AgdaSpace{}%
\AgdaDatatype{Term}\AgdaSpace{}%
\AgdaSymbol{→}\AgdaSpace{}%
\AgdaFunction{Telescope}\AgdaSpace{}%
\AgdaSymbol{→}\AgdaSpace{}%
\AgdaFunction{SKS}\AgdaSpace{}%
\AgdaSymbol{(}\AgdaDatatype{Err}\AgdaSpace{}%
\AgdaDatatype{Expr}\AgdaSymbol{)}\<%
\\
\>[2]\AgdaFunction{kompile-term}\AgdaSpace{}%
\AgdaSymbol{(}\AgdaInductiveConstructor{def}\AgdaSpace{}%
\AgdaSymbol{(}\AgdaKeyword{quote}\AgdaSpace{}%
\AgdaOperator{\AgdaPrimitive{\AgdaUnderscore{}+\AgdaUnderscore{}}}\AgdaSymbol{)}\AgdaSpace{}%
\AgdaBound{args}\AgdaSymbol{@(}\AgdaInductiveConstructor{arg}\AgdaSpace{}%
\AgdaSymbol{\AgdaUnderscore{}}\AgdaSpace{}%
\AgdaBound{a}\AgdaSpace{}%
\AgdaOperator{\AgdaInductiveConstructor{∷}}\AgdaSpace{}%
\AgdaInductiveConstructor{arg}\AgdaSpace{}%
\AgdaSymbol{\AgdaUnderscore{}}\AgdaSpace{}%
\AgdaBound{b}\AgdaSpace{}%
\AgdaOperator{\AgdaInductiveConstructor{∷}}\AgdaSpace{}%
\AgdaInductiveConstructor{[]}\AgdaSymbol{))}\AgdaSpace{}%
\AgdaBound{vars}\AgdaSpace{}%
\AgdaSymbol{=}\AgdaSpace{}%
\AgdaKeyword{do}\<%
\\
\>[2][@{}l@{\AgdaIndent{0}}]%
\>[4]\AgdaBound{a}\AgdaSpace{}%
\AgdaOperator{\AgdaFunction{←}}\AgdaSpace{}%
\AgdaFunction{kompile-term}\AgdaSpace{}%
\AgdaBound{a}\AgdaSpace{}%
\AgdaBound{vars}\<%
\\
\>[4]\AgdaBound{b}\AgdaSpace{}%
\AgdaOperator{\AgdaFunction{←}}\AgdaSpace{}%
\AgdaFunction{kompile-term}\AgdaSpace{}%
\AgdaBound{b}\AgdaSpace{}%
\AgdaBound{vars}\<%
\\
\>[4]\AgdaFunction{return}\AgdaSpace{}%
\AgdaOperator{\AgdaFunction{\$}}\AgdaSpace{}%
\AgdaInductiveConstructor{BinOp}\AgdaSpace{}%
\AgdaOperator{\AgdaFunction{<\$>}}\AgdaSpace{}%
\AgdaInductiveConstructor{ok}\AgdaSpace{}%
\AgdaInductiveConstructor{Plus}\AgdaSpace{}%
\AgdaOperator{\AgdaFunction{⊛}}\AgdaSpace{}%
\AgdaBound{a}\AgdaSpace{}%
\AgdaOperator{\AgdaFunction{⊛}}\AgdaSpace{}%
\AgdaBound{b}\<%
\\
\>[2]\AgdaFunction{kompile-term}\AgdaSpace{}%
\AgdaSymbol{(}\AgdaInductiveConstructor{def}\AgdaSpace{}%
\AgdaSymbol{(}\AgdaKeyword{quote}\AgdaSpace{}%
\AgdaFunction{F.fromℕ<}\AgdaSymbol{)}\AgdaSpace{}%
\AgdaBound{args}\AgdaSymbol{)}\AgdaSpace{}%
\AgdaBound{vars}\AgdaSpace{}%
\AgdaSymbol{=}\AgdaSpace{}%
\AgdaKeyword{do}\<%
\\
\>[2][@{}l@{\AgdaIndent{0}}]%
\>[4]\AgdaInductiveConstructor{ok}\AgdaSpace{}%
\AgdaSymbol{(}\AgdaBound{x}\AgdaSpace{}%
\AgdaOperator{\AgdaInductiveConstructor{∷}}\AgdaSpace{}%
\AgdaInductiveConstructor{[]}\AgdaSymbol{)}\AgdaSpace{}%
\AgdaOperator{\AgdaFunction{←}}%
\>[1912I]\AgdaFunction{kompile-arglist}\AgdaSpace{}%
\AgdaBound{args}\AgdaSpace{}%
\AgdaSymbol{(}\AgdaNumber{0}\AgdaSpace{}%
\AgdaOperator{\AgdaInductiveConstructor{∷}}\AgdaSpace{}%
\AgdaInductiveConstructor{[]}\AgdaSymbol{)}\AgdaSpace{}%
\AgdaBound{vars}\<%
\\
\>[.][@{}l@{}]\<[1912I]%
\>[18]\AgdaKeyword{where}\AgdaSpace{}%
\AgdaCatchallClause{\AgdaSymbol{\AgdaUnderscore{}}}\AgdaSpace{}%
\AgdaSymbol{→}\AgdaSpace{}%
\AgdaFunction{kt}\AgdaSpace{}%
\AgdaString{"wrong\ assumptions\ about\ arguments\ of\ fromℕ<"}\<%
\\
\>[4]\AgdaFunction{return}\AgdaSpace{}%
\AgdaOperator{\AgdaFunction{\$}}\AgdaSpace{}%
\AgdaInductiveConstructor{ok}\AgdaSpace{}%
\AgdaBound{x}\<%
\\
\>[2]\AgdaCatchallClause{\AgdaFunction{kompile-term}}\AgdaSpace{}%
\AgdaCatchallClause{\AgdaSymbol{(}}\AgdaCatchallClause{\AgdaInductiveConstructor{def}}\AgdaSpace{}%
\AgdaCatchallClause{\AgdaBound{n}}\AgdaSpace{}%
\AgdaCatchallClause{\AgdaBound{args}}\AgdaCatchallClause{\AgdaSymbol{@(\AgdaUnderscore{}}}\AgdaSpace{}%
\AgdaCatchallClause{\AgdaOperator{\AgdaInductiveConstructor{∷}}}\AgdaSpace{}%
\AgdaCatchallClause{\AgdaSymbol{\AgdaUnderscore{}))}}\AgdaSpace{}%
\AgdaCatchallClause{\AgdaBound{vars}}\AgdaSpace{}%
\AgdaSymbol{=}\AgdaSpace{}%
\AgdaKeyword{do}\<%
\\
\>[2][@{}l@{\AgdaIndent{0}}]%
\>[4]\AgdaFunction{modify}\AgdaSpace{}%
\AgdaSymbol{λ}\AgdaSpace{}%
\AgdaBound{k}\AgdaSpace{}%
\AgdaSymbol{→}\AgdaSpace{}%
\AgdaKeyword{record}\AgdaSpace{}%
\AgdaBound{k}\AgdaSpace{}%
\AgdaSymbol{\{}\AgdaSpace{}%
\AgdaField{funs}\AgdaSpace{}%
\AgdaSymbol{=}\AgdaSpace{}%
\AgdaField{KS.funs}\AgdaSpace{}%
\AgdaBound{k}\AgdaSpace{}%
\AgdaOperator{\AgdaFunction{++}}\AgdaSpace{}%
\AgdaOperator{\AgdaFunction{[}}\AgdaSpace{}%
\AgdaBound{n}\AgdaSpace{}%
\AgdaOperator{\AgdaFunction{]}}\AgdaSpace{}%
\AgdaSymbol{\}}\<%
\\
\>[4]\AgdaBound{args}\AgdaSpace{}%
\AgdaOperator{\AgdaFunction{←}}\AgdaSpace{}%
\AgdaFunction{kompile-arglist}\AgdaSpace{}%
\AgdaBound{args}\AgdaSpace{}%
\AgdaSymbol{(}\AgdaFunction{mk-iota-mask}\AgdaSpace{}%
\AgdaOperator{\AgdaFunction{\$}}\AgdaSpace{}%
\AgdaFunction{length}\AgdaSpace{}%
\AgdaBound{args}\AgdaSymbol{)}\AgdaSpace{}%
\AgdaBound{vars}\<%
\\
\>[4]\AgdaFunction{return}\AgdaSpace{}%
\AgdaOperator{\AgdaFunction{\$}}\AgdaSpace{}%
\AgdaInductiveConstructor{Call}\AgdaSpace{}%
\AgdaOperator{\AgdaFunction{<\$>}}\AgdaSpace{}%
\AgdaInductiveConstructor{ok}\AgdaSpace{}%
\AgdaSymbol{(}\AgdaFunction{normalise-name}\AgdaSpace{}%
\AgdaOperator{\AgdaFunction{\$}}\AgdaSpace{}%
\AgdaPrimitive{showName}\AgdaSpace{}%
\AgdaBound{n}\AgdaSymbol{)}\AgdaSpace{}%
\AgdaOperator{\AgdaFunction{⊛}}\AgdaSpace{}%
\AgdaBound{args}\<%
\\
\>[2]\AgdaCatchallClause{\AgdaFunction{kompile-term}}\AgdaSpace{}%
\AgdaCatchallClause{\AgdaBound{t}}\AgdaSpace{}%
\AgdaCatchallClause{\AgdaBound{vars}}\AgdaSpace{}%
\AgdaSymbol{=}\AgdaSpace{}%
\AgdaPostulate{⋯}\<%
\end{code}
We demonstrate three representative clauses of the term extracting function.
First, we turn \AD{SKS} and \AD{Err} into monads by defining their bind and
return actions.  As each monad is an applicative functor, we get \AF{\_<\$>\_}
and \AF{\_⊛\_} operations for free.  The instance resolution
mechanism\footnote{For the details see the ``Instance Arguments''
  section of Agda's manual~\cite{agda}.}
makes it possible to overload monadic/functorial operations without explicitly
mentioning in which monad we are operating.

A typical case of \AF{kompile-term} extracts the arguments and puts them
together in a corresponding expression of the target language,
as for example in the case for \AF{\_+\_}.

For some constructions it is convenient to handle arguments without
explicitly pattern-matching them, \eg{} some constructor with a long argument
list where we are interested only in a particular subset.  For such reasons we
introduce the \AF{kompile-arglist} function where the first argument
is the list of \AC{Arg}uments; and the second argument is the mask that specifies
indices into the first argument.  The function extracts each argument from the
list as specified by the mask.  In case of \AF{fromℕ<} we use this function to
extract the first argument from the \AB{args} list.

The last clause deals with general function calls that do not require special
treatment.  We first ensure that argument list is non-empty: the \AF{\_ ∷ \_} pattern.
Then we add the name of the function into the \AR{funs} field of the state record,
which is used by \AF{kompile} to extract all the necessary dependencies.
Then we extract the arguments, using
the helper function \AF{mk-iota-mask} that generates indices from \AN{0} to the
length of the argument list.  Finally we create an \AD{Expr}ession for a function
call.  We use the extracted arguments and we normalise the name to get rid of
unicode symbols.

\subsection{Example}

Let us consider the actual output that our extractor generates for a
reasonably complex function:  the binary logarithm.
We take the simplest specification, and we assume that logarithm of zero
is zero.  One difficulty with this function is that it is not structurally
recursive and Agda does not recognise that it terminates. We use a standard
technique of recursing on a well-founded \AF{\_<\_} predicate (inequality
of natural numbers) to prove termination.  Here is the Agda definition and the extracted code
(slightly reformatted) arranged side-by-side:

\begin{code}[hide]%
\>[2]\AgdaKeyword{open}\AgdaSpace{}%
\AgdaKeyword{import}\AgdaSpace{}%
\AgdaModule{Data.Nat.DivMod}\<%
\\
\>[2]\AgdaKeyword{open}\AgdaSpace{}%
\AgdaKeyword{import}\AgdaSpace{}%
\AgdaModule{Data.Nat.Properties}\<%
\end{code}
\begin{code}%
\>[2]\AgdaFunction{x<m⇒sx/2<m}\AgdaSpace{}%
\AgdaSymbol{:}\AgdaSpace{}%
\AgdaSymbol{∀}\AgdaSpace{}%
\AgdaBound{x}\AgdaSpace{}%
\AgdaBound{m}\AgdaSpace{}%
\AgdaSymbol{→}\AgdaSpace{}%
\AgdaBound{x}\AgdaSpace{}%
\AgdaOperator{\AgdaFunction{<}}\AgdaSpace{}%
\AgdaBound{m}\AgdaSpace{}%
\AgdaSymbol{→}\AgdaSpace{}%
\AgdaInductiveConstructor{suc}\AgdaSpace{}%
\AgdaBound{x}\AgdaSpace{}%
\AgdaOperator{\AgdaFunction{/}}\AgdaSpace{}%
\AgdaNumber{2}\AgdaSpace{}%
\AgdaOperator{\AgdaFunction{<}}\AgdaSpace{}%
\AgdaBound{m}\<%
\\
\>[2]\AgdaFunction{x<m⇒sx/2<m}\AgdaSpace{}%
\AgdaBound{x}\AgdaSpace{}%
\AgdaBound{m}\AgdaSpace{}%
\AgdaBound{x<m}\AgdaSpace{}%
\AgdaSymbol{=}\AgdaSpace{}%
\AgdaFunction{≤-trans}\AgdaSpace{}%
\AgdaSymbol{(}\AgdaFunction{m/n<m}\AgdaSpace{}%
\AgdaSymbol{(}\AgdaInductiveConstructor{suc}\AgdaSpace{}%
\AgdaBound{x}\AgdaSymbol{)}\AgdaSpace{}%
\AgdaNumber{2}\AgdaSpace{}%
\AgdaSymbol{(}\AgdaInductiveConstructor{s≤s}\AgdaSpace{}%
\AgdaInductiveConstructor{z≤n}\AgdaSymbol{)}\AgdaSpace{}%
\AgdaFunction{≤-refl}\AgdaSymbol{)}\AgdaSpace{}%
\AgdaBound{x<m}\<%
\\
\>[2]\AgdaComment{--\ Extracted\ with\ command\ :\ kompile\ log₂\ (quote\ ≤-refl\ ∷\ quote\ \AgdaUnderscore{}<\AgdaUnderscore{}\ ∷\ [])\ []}\<%
\\
\>[2]\AgdaFunction{log₂′}\AgdaSpace{}%
\AgdaSymbol{:}\AgdaSpace{}%
\AgdaSymbol{∀}\AgdaSpace{}%
\AgdaSymbol{\{}\AgdaBound{m}\AgdaSymbol{\}}\AgdaSpace{}%
\AgdaSymbol{→}\AgdaSpace{}%
\AgdaSymbol{(}\AgdaBound{n}\AgdaSpace{}%
\AgdaSymbol{:}\AgdaSpace{}%
\AgdaDatatype{ℕ}\AgdaSymbol{)}\AgdaSpace{}%
\AgdaSymbol{→}\AgdaSpace{}%
\AgdaSymbol{(}\AgdaBound{n}\AgdaSpace{}%
\AgdaOperator{\AgdaFunction{<}}\AgdaSpace{}%
\AgdaBound{m}\AgdaSymbol{)}\AgdaSpace{}%
\AgdaSymbol{→}\AgdaSpace{}%
\AgdaDatatype{ℕ}%
\>[41]\AgdaComment{--\ def\ log2'\ (x\AgdaUnderscore{}1,\ x\AgdaUnderscore{}2,\ x\AgdaUnderscore{}3):}\<%
\\
\>[41]\AgdaComment{--\ \ \ let\ x\AgdaUnderscore{}3\AgdaUnderscore{}assrt\ =\ assert\ (x\AgdaUnderscore{}2\ <\ x\AgdaUnderscore{}1)}\<%
\\
\>[2]\AgdaFunction{log₂′}\AgdaSpace{}%
\AgdaSymbol{\{}\AgdaBound{m}\AgdaSymbol{\}}%
\>[16]\AgdaNumber{0}%
\>[26]\AgdaSymbol{\AgdaUnderscore{}}%
\>[30]\AgdaSymbol{=}\AgdaSpace{}%
\AgdaNumber{0}%
\>[41]\AgdaComment{--\ \ \ let\ \AgdaUnderscore{}\AgdaUnderscore{}ret\ =\ if\ (x\AgdaUnderscore{}2\ ==\ 0):}\<%
\\
\>[41]\AgdaComment{--\ \ \ \ \ let\ m\ =\ x\AgdaUnderscore{}1\ ;\ x\ =\ x\AgdaUnderscore{}3}\<%
\\
\>[41]\AgdaComment{--\ \ \ \ \ 0}\<%
\\
\>[2]\AgdaFunction{log₂′}\AgdaSpace{}%
\AgdaSymbol{\{}\AgdaBound{m}\AgdaSymbol{\}}%
\>[16]\AgdaNumber{1}%
\>[26]\AgdaSymbol{\AgdaUnderscore{}}%
\>[30]\AgdaSymbol{=}\AgdaSpace{}%
\AgdaNumber{0}%
\>[41]\AgdaComment{--\ \ \ else\ if\ (x\AgdaUnderscore{}2\ >\ 0)\ \&\&\ (x\AgdaUnderscore{}2\ -\ 1\ ==\ 0):}\<%
\\
\>[41]\AgdaComment{--\ \ \ \ \ let\ m\ =\ x\AgdaUnderscore{}1\ ;\ x\ =\ x\AgdaUnderscore{}3}\<%
\\
\>[41]\AgdaComment{--\ \ \ \ \ 0}\<%
\\
\>[2]\AgdaCatchallClause{\AgdaFunction{log₂′}}\AgdaSpace{}%
\AgdaCatchallClause{\AgdaSymbol{\{}}\AgdaCatchallClause{\AgdaInductiveConstructor{suc}}\AgdaSpace{}%
\AgdaCatchallClause{\AgdaBound{m}}\AgdaCatchallClause{\AgdaSymbol{\}}}\AgdaSpace{}%
\AgdaCatchallClause{\AgdaBound{n}}\AgdaCatchallClause{\AgdaSymbol{@(}}\AgdaCatchallClause{\AgdaInductiveConstructor{suc}}\AgdaSpace{}%
\AgdaCatchallClause{\AgdaBound{x}}\AgdaCatchallClause{\AgdaSymbol{)}}\AgdaSpace{}%
\AgdaCatchallClause{\AgdaBound{n<m}}\AgdaSpace{}%
\AgdaSymbol{=}%
\>[41]\AgdaComment{--\ \ \ else\ if\ (x\AgdaUnderscore{}1\ >\ 0)\ \&\&\ (x\AgdaUnderscore{}2\ >\ 0):}\<%
\\
\>[2][@{}l@{\AgdaIndent{0}}]%
\>[4]\AgdaNumber{1}\AgdaSpace{}%
\AgdaOperator{\AgdaPrimitive{+}}%
\>[2026I]\AgdaFunction{log₂′}\AgdaSpace{}%
\AgdaSymbol{\{}\AgdaArgument{m}\AgdaSpace{}%
\AgdaSymbol{=}\AgdaSpace{}%
\AgdaBound{m}\AgdaSymbol{\}}\AgdaSpace{}%
\AgdaSymbol{(}\AgdaBound{n}\AgdaSpace{}%
\AgdaOperator{\AgdaFunction{/}}\AgdaSpace{}%
\AgdaNumber{2}\AgdaSymbol{)}%
\>[41]\AgdaComment{--\ \ \ \ \ let\ m\ =\ x\AgdaUnderscore{}1\ -\ 1\ ;\ x\ =\ x\AgdaUnderscore{}2\ -\ 1\ ;\ n<m\ =\ x\AgdaUnderscore{}3}\<%
\\
\>[.][@{}l@{}]\<[2026I]%
\>[8]\AgdaSymbol{(}\AgdaFunction{x<m⇒sx/2<m}\AgdaSpace{}%
\AgdaBound{x}\AgdaSpace{}%
\AgdaBound{m}\AgdaSpace{}%
\AgdaOperator{\AgdaFunction{\$}}\AgdaSpace{}%
\AgdaFunction{≤-pred}\AgdaSpace{}%
\AgdaBound{n<m}\AgdaSymbol{)}%
\>[41]\AgdaComment{--\ \ \ \ \ 1\ +\ log2'\ (m,\ 0\ +\ (x\ +\ 1\ -\ 0)\ /\ (1\ +\ 1),\ 1)}\<%
\\
\>[41]\AgdaComment{--\ \ \ else:}\<%
\\
\>[41]\AgdaComment{--\ \ \ \ \ assert\ (0)}\<%
\\
\\[\AgdaEmptyExtraSkip]%
\>[2]\AgdaFunction{log₂}\AgdaSpace{}%
\AgdaSymbol{:}\AgdaSpace{}%
\AgdaDatatype{ℕ}\AgdaSpace{}%
\AgdaSymbol{→}\AgdaSpace{}%
\AgdaDatatype{ℕ}%
\>[41]\AgdaComment{--\ def\ log2\ (x\AgdaUnderscore{}1):}\<%
\\
\>[2]\AgdaFunction{log₂}\AgdaSpace{}%
\AgdaBound{x}\AgdaSpace{}%
\AgdaSymbol{=}\AgdaSpace{}%
\AgdaFunction{log₂′}\AgdaSpace{}%
\AgdaBound{x}\AgdaSpace{}%
\AgdaFunction{≤-refl}%
\>[41]\AgdaComment{--\ \ \ let\ x\ =\ x\AgdaUnderscore{}1\ ;\ \AgdaUnderscore{}\AgdaUnderscore{}ret\ =\ log2'\ (1\ +\ x,\ x,\ 1)}\<%
\\
\>[41]\AgdaComment{--\ \ \ \AgdaUnderscore{}\AgdaUnderscore{}ret}\<%
\end{code}
We define two functions: a wrapper function \AF{log₂}, and a helper \AF{log₂′} where the
actual work happens.  We define two base cases for \AN{0} and \AN{1} and the recursive
case on the argument divided by \AN{2}.  Note that the function
$\AF{\_/\_}~:~(x\ y~:~\AD{ℕ}) \{\AB{≢0}~:~\AF{False} (\AB{y}\ \AD{≟}\ \AN{0})\} → \AD{ℕ}$ takes an implicit
argument that asserts that the divisor is non-zero. Agda is capable to deduce the proof
automatically for constant values such as \AN{2}.  In the extracted code we start with
the assertion that was extracted from the type of \AB{n<m}.  The first case is trivial.
In the second case we see \AB{x\_2} \AF{-} \AN{1} \AF{==} \AN{0},
rather than \AB{x\_2} \AF{==} \AN{1}, which is an artefact of the algorithm
used in \AF{kompile-clpats} and the fact that \AN{1} is represented as
\AC{suc} \AC{zero}.  This is a correct translation, as we ensure that \AB{x\_2}
is greater than zero before subtracting one.  However, this could be further
optimised either by the target language or as a post-extraction step.

In the recursive case, division looks suspiciously complex.
The reason for the complexity of the division
operation is because \AF{\_/\_} in Agda is defined in terms of a helper function
\AF{div-helper} \AB{k} \AB{m} \AB{n} \AB{j} that corresponds to the expression
\AB{k} \AF{+} ((\AB{n}\AF{+}\AB{m}\AF{-}\AB{j}) \AF{/} (\AN{1}\AF{+}\AB{m})).
We could have suppressed the normalisation of \AF{\_/\_}, but this is not
generally desirable, as it prevents evaluation prior to extraction.  For example,
without suppression (\AN{2}\AF{+}\AB{n})\AF{/}\AN{2} normalises to
\AN{1}\AF{+}(\AB{n}\AF{/}\AN{2}), whereas with \AF{\_/\_} suppressed it would be treated
as an opaque object.

Also note how the recursive call uses the value \AN{1} instead of an
actual proof. Per our assumption, \AF{\_<\_} is used as a static assertion
(we cannot pattern-match on its value).  This means that any function that has
a return type \AB{a} \AF{<} \AB{b} can be replaced with the unit value.  This
is valid, because we are extracting function calls that were verified by the
typechecker.  Therefore, we can completely omit the proof part of \AF{\_<\_},
only acknowledging that the type is inhabited.  This particular case relies
on \AF{≤-trans} (from the inlined proof) and \AF{≤-refl} (from the \AF{log₂}
definition) being extracted into unit values.

There is also the final \AK{else} case, which is not specified in the
original code on the left.  The reason for this extra case is that our
pattern matching is not complete.  We are missing the case where
\AB{m} is zero and \AB{n} is greater than \AN{2}.  While Agda's
coverage checker agrees that such case is impossible, it automatically
inserts the missing case into internal representation as an absurd
clause.


\subsection{\label{sec:rewriting}Rewriting}

A common way to define domain-specific compiler optimizations is through
the specification of \emph{rewrite rules} that rewrite terms matching
a given pattern to an equivalent form that is either more efficient
or reveals further optimization opportunities.
By giving a shallow embedding of our target language in Agda, we have
the opportunity to define \emph{verified} rewrite rules, providing
a proof that the left- and right-hand side of the rewrite rule are
equivalent.
To achieve this, we could define our own representation of verified
rewrite rules and integrate them into the extractor.
However, we can avoid the effort of doing so since Agda already has a
built-in concept of rewrite rules.

Rewrite rules were originally introduced to Agda to work around the
limitations of definitional equality in intentional type theory.
For example, it can be used to make $0 + x$ definitionally equal to
$x + 0$.
Since we work with a shallow embedding, these rewrite rules are
equally well suited to optimize the embedded programs we write before
they are extracted.
A typical example of a rewrite rule in Agda rewrites expressions of
the form \AB{x} \AF{+} 0 to \AB{x}:
\begin{code}%
\>[2]\AgdaFunction{plus-0}\AgdaSpace{}%
\AgdaSymbol{:}\AgdaSpace{}%
\AgdaSymbol{∀}\AgdaSpace{}%
\AgdaBound{x}\AgdaSpace{}%
\AgdaSymbol{→}\AgdaSpace{}%
\AgdaBound{x}\AgdaSpace{}%
\AgdaOperator{\AgdaFunction{+}}\AgdaSpace{}%
\AgdaNumber{0}\AgdaSpace{}%
\AgdaOperator{\AgdaDatatype{≡}}\AgdaSpace{}%
\AgdaBound{x}\<%
\\
\>[2]\AgdaFunction{plus-0}\AgdaSpace{}%
\AgdaInductiveConstructor{zero}%
\>[18]\AgdaSymbol{=}\AgdaSpace{}%
\AgdaInductiveConstructor{refl}\<%
\\
\>[2]\AgdaFunction{plus-0}\AgdaSpace{}%
\AgdaSymbol{(}\AgdaInductiveConstructor{suc}\AgdaSpace{}%
\AgdaBound{x}\AgdaSymbol{)}%
\>[18]\AgdaSymbol{=}\AgdaSpace{}%
\AgdaFunction{cong}\AgdaSpace{}%
\AgdaInductiveConstructor{suc}\AgdaSpace{}%
\AgdaOperator{\AgdaFunction{\$}}\AgdaSpace{}%
\AgdaFunction{plus-0}\AgdaSpace{}%
\AgdaBound{x}\<%
\\
\>[2]\AgdaSymbol{\{-\#}\AgdaSpace{}%
\AgdaKeyword{REWRITE}\AgdaSpace{}%
\AgdaFunction{plus-0}\AgdaSpace{}%
\AgdaSymbol{\#-\}}\<%
\end{code}
The definition of \AF{plus-0} proves the equivalence of the left- and
right-hand sides of the rule, and the \AK{REWRITE} pragma registers it as a
rewrite to be applied automatically during normalisation.  As another
example, we define the following rule that we were taught in
school:
\begin{code}[hide]%
\>[0]\AgdaKeyword{module}\AgdaSpace{}%
\AgdaModule{PlusZSolv}\AgdaSpace{}%
\AgdaKeyword{where}\<%
\\
\>[0][@{}l@{\AgdaIndent{0}}]%
\>[2]\AgdaKeyword{open}\AgdaSpace{}%
\AgdaKeyword{import}\AgdaSpace{}%
\AgdaModule{Data.Nat}\AgdaSpace{}%
\AgdaSymbol{as}\AgdaSpace{}%
\AgdaModule{ℕ}\AgdaSpace{}%
\AgdaKeyword{hiding}\AgdaSpace{}%
\AgdaSymbol{(}\AgdaOperator{\AgdaFunction{\AgdaUnderscore{}≟\AgdaUnderscore{}}}\AgdaSymbol{)}\<%
\\
\>[2]\AgdaKeyword{open}\AgdaSpace{}%
\AgdaKeyword{import}\AgdaSpace{}%
\AgdaModule{Relation.Binary.PropositionalEquality}\<%
\\
\>[2]\AgdaKeyword{open}\AgdaSpace{}%
\AgdaKeyword{import}\AgdaSpace{}%
\AgdaModule{Agda.Builtin.Equality.Rewrite}\<%
\end{code}
\begin{code}%
\>[2]\AgdaKeyword{open}\AgdaSpace{}%
\AgdaKeyword{import}\AgdaSpace{}%
\AgdaModule{Data.Nat.Solver}\AgdaSpace{}%
\AgdaSymbol{;}\AgdaSpace{}%
\AgdaKeyword{open}\AgdaSpace{}%
\AgdaModule{+-*-Solver}\<%
\\
\>[2]\AgdaFunction{sum-square}\AgdaSpace{}%
\AgdaSymbol{:}\AgdaSpace{}%
\AgdaSymbol{∀}\AgdaSpace{}%
\AgdaBound{x}\AgdaSpace{}%
\AgdaBound{y}\AgdaSpace{}%
\AgdaSymbol{→}\AgdaSpace{}%
\AgdaBound{x}\AgdaSpace{}%
\AgdaOperator{\AgdaPrimitive{*}}\AgdaSpace{}%
\AgdaBound{x}\AgdaSpace{}%
\AgdaOperator{\AgdaPrimitive{+}}\AgdaSpace{}%
\AgdaNumber{2}\AgdaSpace{}%
\AgdaOperator{\AgdaPrimitive{*}}\AgdaSpace{}%
\AgdaBound{x}\AgdaSpace{}%
\AgdaOperator{\AgdaPrimitive{*}}\AgdaSpace{}%
\AgdaBound{y}\AgdaSpace{}%
\AgdaOperator{\AgdaPrimitive{+}}\AgdaSpace{}%
\AgdaBound{y}\AgdaSpace{}%
\AgdaOperator{\AgdaPrimitive{*}}\AgdaSpace{}%
\AgdaBound{y}\AgdaSpace{}%
\AgdaOperator{\AgdaDatatype{≡}}\AgdaSpace{}%
\AgdaSymbol{(}\AgdaBound{x}\AgdaSpace{}%
\AgdaOperator{\AgdaPrimitive{+}}\AgdaSpace{}%
\AgdaBound{y}\AgdaSymbol{)}\AgdaSpace{}%
\AgdaOperator{\AgdaPrimitive{*}}\AgdaSpace{}%
\AgdaSymbol{(}\AgdaBound{x}\AgdaSpace{}%
\AgdaOperator{\AgdaPrimitive{+}}\AgdaSpace{}%
\AgdaBound{y}\AgdaSymbol{)}\<%
\\
\>[2]\AgdaFunction{sum-square}\AgdaSpace{}%
\AgdaSymbol{=}\AgdaSpace{}%
\AgdaFunction{solve}\AgdaSpace{}%
\AgdaNumber{2}\AgdaSpace{}%
\AgdaSymbol{(λ}\AgdaSpace{}%
\AgdaBound{x}\AgdaSpace{}%
\AgdaBound{y}\AgdaSpace{}%
\AgdaSymbol{→}\AgdaSpace{}%
\AgdaBound{x}\AgdaSpace{}%
\AgdaOperator{\AgdaFunction{:*}}\AgdaSpace{}%
\AgdaBound{x}\AgdaSpace{}%
\AgdaOperator{\AgdaFunction{:+}}\AgdaSpace{}%
\AgdaInductiveConstructor{con}\AgdaSpace{}%
\AgdaNumber{2}\AgdaSpace{}%
\AgdaOperator{\AgdaFunction{:*}}\AgdaSpace{}%
\AgdaBound{x}\AgdaSpace{}%
\AgdaOperator{\AgdaFunction{:*}}\AgdaSpace{}%
\AgdaBound{y}\AgdaSpace{}%
\AgdaOperator{\AgdaFunction{:+}}\AgdaSpace{}%
\AgdaBound{y}\AgdaSpace{}%
\AgdaOperator{\AgdaFunction{:*}}\AgdaSpace{}%
\AgdaBound{y}\AgdaSpace{}%
\AgdaOperator{\AgdaFunction{:=}}%
\>[72]\AgdaSymbol{(}\AgdaBound{x}\AgdaSpace{}%
\AgdaOperator{\AgdaFunction{:+}}\AgdaSpace{}%
\AgdaBound{y}\AgdaSymbol{)}\AgdaSpace{}%
\AgdaOperator{\AgdaFunction{:*}}\AgdaSpace{}%
\AgdaSymbol{(}\AgdaBound{x}\AgdaSpace{}%
\AgdaOperator{\AgdaFunction{:+}}\AgdaSpace{}%
\AgdaBound{y}\AgdaSymbol{))}\AgdaSpace{}%
\AgdaInductiveConstructor{refl}\<%
\\
\>[2]\AgdaSymbol{\{-\#}\AgdaSpace{}%
\AgdaKeyword{REWRITE}\AgdaSpace{}%
\AgdaFunction{sum-square}\AgdaSpace{}%
\AgdaSymbol{\#-\}}\<%
\end{code}

It might seem that such an example of rewriting expressions
over natural numbers are not very practical, but the benefit becomes more
obvious for more complex data structures.
For example, here is the famous fusion law for distributing \AF{map}s over
function composition \AF{\_∘\_}:
\begin{code}[hide]%
\>[2]\AgdaKeyword{open}\AgdaSpace{}%
\AgdaKeyword{import}\AgdaSpace{}%
\AgdaModule{Data.List}\AgdaSpace{}%
\AgdaKeyword{using}\AgdaSpace{}%
\AgdaSymbol{(}\AgdaFunction{map}\AgdaSymbol{)}\<%
\end{code}
\begin{code}%
\>[2]\AgdaFunction{map-∘}\AgdaSpace{}%
\AgdaSymbol{:}\AgdaSpace{}%
\AgdaSymbol{∀}\AgdaSpace{}%
\AgdaSymbol{\{}\AgdaBound{X}\AgdaSpace{}%
\AgdaBound{Y}\AgdaSpace{}%
\AgdaBound{Z}\AgdaSpace{}%
\AgdaSymbol{:}\AgdaSpace{}%
\AgdaPrimitive{Set}\AgdaSymbol{\}\{}\AgdaBound{g}\AgdaSpace{}%
\AgdaSymbol{:}\AgdaSpace{}%
\AgdaBound{X}\AgdaSpace{}%
\AgdaSymbol{→}\AgdaSpace{}%
\AgdaBound{Y}\AgdaSymbol{\}\{}\AgdaBound{f}\AgdaSpace{}%
\AgdaSymbol{:}\AgdaSpace{}%
\AgdaBound{Y}\AgdaSpace{}%
\AgdaSymbol{→}\AgdaSpace{}%
\AgdaBound{Z}\AgdaSymbol{\}}\AgdaSpace{}%
\AgdaSymbol{→}\AgdaSpace{}%
\AgdaSymbol{∀}\AgdaSpace{}%
\AgdaBound{xs}\AgdaSpace{}%
\AgdaSymbol{→}\AgdaSpace{}%
\AgdaSymbol{(}\AgdaFunction{map}\AgdaSpace{}%
\AgdaBound{f}\AgdaSpace{}%
\AgdaOperator{\AgdaFunction{∘}}\AgdaSpace{}%
\AgdaFunction{map}\AgdaSpace{}%
\AgdaBound{g}\AgdaSymbol{)}\AgdaSpace{}%
\AgdaBound{xs}\AgdaSpace{}%
\AgdaOperator{\AgdaDatatype{≡}}\AgdaSpace{}%
\AgdaFunction{map}\AgdaSpace{}%
\AgdaSymbol{(}\AgdaBound{f}\AgdaSpace{}%
\AgdaOperator{\AgdaFunction{∘}}\AgdaSpace{}%
\AgdaBound{g}\AgdaSymbol{)}\AgdaSpace{}%
\AgdaBound{xs}\<%
\\
\>[2]\AgdaFunction{map-∘}\AgdaSpace{}%
\AgdaInductiveConstructor{[]}%
\>[18]\AgdaSymbol{=}\AgdaSpace{}%
\AgdaInductiveConstructor{refl}\<%
\\
\>[2]\AgdaFunction{map-∘}\AgdaSpace{}%
\AgdaSymbol{(}\AgdaBound{x}\AgdaSpace{}%
\AgdaOperator{\AgdaInductiveConstructor{∷}}\AgdaSpace{}%
\AgdaBound{xs}\AgdaSymbol{)}%
\>[18]\AgdaSymbol{=}\AgdaSpace{}%
\AgdaFunction{cong}\AgdaSpace{}%
\AgdaSymbol{(\AgdaUnderscore{}}\AgdaSpace{}%
\AgdaOperator{\AgdaInductiveConstructor{∷\AgdaUnderscore{}}}\AgdaSymbol{)}\AgdaSpace{}%
\AgdaSymbol{(}\AgdaFunction{map-∘}\AgdaSpace{}%
\AgdaBound{xs}\AgdaSymbol{)}\<%
\\
\>[2]\AgdaSymbol{\{-\#}\AgdaSpace{}%
\AgdaKeyword{REWRITE}\AgdaSpace{}%
\AgdaFunction{map-∘}\AgdaSpace{}%
\AgdaSymbol{\#-\}}\<%
\end{code}
Instead of traversing all the elements
of \AB{xs} and then all the elements of the \AF{map} \AB{g} \AB{xs}, we can compute
the same result in a single traversal.  Generally, we often know a number of properties
about the data structures we are working with.  In a dependently-typed systems we can
make those properties explicit and formally verify them, but in rewriting-capable
systems we can selectively turn properties into optimisations.


One danger with rewrite rules is that we can get different result depending on the order of rule application.
The recently introduced confluence checker~\cite{10.1145/3434341}
helps to prevent this problem.  When it is turned on, it reports
when the registered set of rewrite rules is not
confluent.  For example, in case of \AF{plus-0} rule, the confluence checker
complains that
\AF{plus} (\AC{suc} \AB{x}) \AC{zero} can be rewritten to either
(\AC{suc} \AB{x}) or \AC{suc} (\AF{plus} \AB{x} \AC{zero}).  If
we add a new rewrite rule for \AF{plus} (\AC{suc} \AB{x}) \AC{zero} $\mapsto$
\AC{suc} \AB{x}, our rewrite system is again accepted.



\subsection{Monadic Workaround for Lets}
One of the unfortunate design choices of the Agda internal language is
the lack of a `let' construct.  All the lets we use in the code are eliminated
eagerly through substitution of the bound expression in the body.  While
this is semantically sound, it leads to unnecessary code duplication:
\begin{code}%
\>[2]\AgdaFunction{ex₈}\AgdaSpace{}%
\AgdaSymbol{:}\AgdaSpace{}%
\AgdaDatatype{ℕ}\AgdaSpace{}%
\AgdaSymbol{→}\AgdaSpace{}%
\AgdaDatatype{ℕ}\<%
\\
\>[2]\AgdaFunction{ex₈}\AgdaSpace{}%
\AgdaBound{x}\AgdaSpace{}%
\AgdaSymbol{=}\AgdaSpace{}%
\AgdaKeyword{let}\AgdaSpace{}%
\AgdaBound{a}\AgdaSpace{}%
\AgdaSymbol{=}\AgdaSpace{}%
\AgdaBound{x}\AgdaSpace{}%
\AgdaOperator{\AgdaPrimitive{*}}\AgdaSpace{}%
\AgdaBound{x}\AgdaSpace{}%
\AgdaOperator{\AgdaPrimitive{+}}\AgdaSpace{}%
\AgdaNumber{3}\AgdaSpace{}%
\AgdaOperator{\AgdaPrimitive{*}}\AgdaSpace{}%
\AgdaBound{x}\AgdaSpace{}%
\AgdaOperator{\AgdaPrimitive{+}}\AgdaSpace{}%
\AgdaNumber{5}\AgdaSpace{}%
\AgdaKeyword{in}\AgdaSpace{}%
\AgdaBound{a}\AgdaSpace{}%
\AgdaOperator{\AgdaPrimitive{+}}\AgdaSpace{}%
\AgdaBound{a}\AgdaSpace{}%
\AgdaComment{--\ ⇒\ (x\ *\ x\ +\ 3\ *\ x\ +\ 5)\ +\ (x\ *\ x\ +\ 3\ *\ x\ +\ 5)}\<%
\end{code}
While changing Agda itself to support `let' in the internal language
would be a major change, we can use the following
elegant workaround.  Agda's do-notation is a syntactic sugar that
expands to the monadic bind \AF{\_>>=\_}.
In particular, we can work in the identity monad by defining
$a~\AF{>>=}~f = f~a$ and adding it to our extractor,
\begin{code}[hide]%
\>[0]\AgdaKeyword{module}\AgdaSpace{}%
\AgdaModule{Monadic}\AgdaSpace{}%
\AgdaKeyword{where}\<%
\\
\>[0][@{}l@{\AgdaIndent{0}}]%
\>[2]\AgdaKeyword{open}\AgdaSpace{}%
\AgdaKeyword{import}\AgdaSpace{}%
\AgdaModule{Data.Nat}\AgdaSpace{}%
\AgdaSymbol{as}\AgdaSpace{}%
\AgdaModule{ℕ}\AgdaSpace{}%
\AgdaKeyword{hiding}\AgdaSpace{}%
\AgdaSymbol{(}\AgdaOperator{\AgdaFunction{\AgdaUnderscore{}≟\AgdaUnderscore{}}}\AgdaSymbol{)}\<%
\end{code}
\begin{code}[hide]%
\>[2]\AgdaOperator{\AgdaFunction{\AgdaUnderscore{}>>=\AgdaUnderscore{}}}\AgdaSpace{}%
\AgdaSymbol{:}\AgdaSpace{}%
\AgdaSymbol{∀}\AgdaSpace{}%
\AgdaSymbol{\{}\AgdaBound{ℓ₁}\AgdaSpace{}%
\AgdaBound{ℓ₂}\AgdaSymbol{\}\{}\AgdaBound{A}\AgdaSpace{}%
\AgdaSymbol{:}\AgdaSpace{}%
\AgdaPrimitive{Set}\AgdaSpace{}%
\AgdaBound{ℓ₁}\AgdaSymbol{\}\{}\AgdaBound{B}\AgdaSpace{}%
\AgdaSymbol{:}\AgdaSpace{}%
\AgdaPrimitive{Set}\AgdaSpace{}%
\AgdaBound{ℓ₂}\AgdaSymbol{\}}\AgdaSpace{}%
\AgdaSymbol{→}\AgdaSpace{}%
\AgdaBound{A}\AgdaSpace{}%
\AgdaSymbol{→}\AgdaSpace{}%
\AgdaSymbol{(}\AgdaBound{A}\AgdaSpace{}%
\AgdaSymbol{→}\AgdaSpace{}%
\AgdaBound{B}\AgdaSymbol{)}\AgdaSpace{}%
\AgdaSymbol{→}\AgdaSpace{}%
\AgdaBound{B}\<%
\\
\>[2]\AgdaBound{a}\AgdaSpace{}%
\AgdaOperator{\AgdaFunction{>>=}}\AgdaSpace{}%
\AgdaBound{f}\AgdaSpace{}%
\AgdaSymbol{=}\AgdaSpace{}%
\AgdaBound{f}\AgdaSpace{}%
\AgdaBound{a}\<%
\\
\>[2]\AgdaFunction{return}\AgdaSpace{}%
\AgdaSymbol{:}\AgdaSpace{}%
\AgdaSymbol{∀}\AgdaSpace{}%
\AgdaSymbol{\{}\AgdaBound{ℓ}\AgdaSymbol{\}\{}\AgdaBound{A}\AgdaSpace{}%
\AgdaSymbol{:}\AgdaSpace{}%
\AgdaPrimitive{Set}\AgdaSpace{}%
\AgdaBound{ℓ}\AgdaSymbol{\}}\AgdaSpace{}%
\AgdaSymbol{→}\AgdaSpace{}%
\AgdaBound{A}\AgdaSpace{}%
\AgdaSymbol{→}\AgdaSpace{}%
\AgdaBound{A}\<%
\\
\>[2]\AgdaFunction{return}\AgdaSpace{}%
\AgdaBound{a}\AgdaSpace{}%
\AgdaSymbol{=}\AgdaSpace{}%
\AgdaBound{a}\<%
\end{code}
allowing us to use do-notation instead of let:
\begin{code}%
\>[2]\AgdaFunction{ex₈′}\AgdaSpace{}%
\AgdaSymbol{:}\AgdaSpace{}%
\AgdaDatatype{ℕ}\AgdaSpace{}%
\AgdaSymbol{→}\AgdaSpace{}%
\AgdaDatatype{ℕ}\<%
\\
\>[2]\AgdaFunction{ex₈′}\AgdaSpace{}%
\AgdaBound{x}\AgdaSpace{}%
\AgdaSymbol{=}\AgdaSpace{}%
\AgdaKeyword{do}\AgdaSpace{}%
\AgdaBound{a}\AgdaSpace{}%
\AgdaOperator{\AgdaFunction{←}}\AgdaSpace{}%
\AgdaBound{x}\AgdaSpace{}%
\AgdaOperator{\AgdaPrimitive{*}}\AgdaSpace{}%
\AgdaBound{x}\AgdaSpace{}%
\AgdaOperator{\AgdaPrimitive{+}}\AgdaSpace{}%
\AgdaNumber{3}\AgdaSpace{}%
\AgdaOperator{\AgdaPrimitive{*}}\AgdaSpace{}%
\AgdaBound{x}\AgdaSpace{}%
\AgdaOperator{\AgdaPrimitive{+}}\AgdaSpace{}%
\AgdaNumber{5}\AgdaSymbol{;}\AgdaSpace{}%
\AgdaBound{a}\AgdaSpace{}%
\AgdaOperator{\AgdaPrimitive{+}}\AgdaSpace{}%
\AgdaBound{a}\<%
\end{code}


\section{\label{sec:array}Array Language}

In this section we switch to extraction of a different language
called Single Assignment C --- \sac{} for short.  We explain essence of the language,
our embedding for it in Agda, and the difference in the
extraction process when comparing to Kaleidoscope.

\subsection{\sac{} --- Single Assignment C}
\sac{} is a first-order array language that looks like C syntactically,
but nonetheless
is purely functional.  The main goal of \sac{} is to provide
a framework to efficiently operate with multi-dimensional arrays.
All types in \sac{} represent arrays with potentially unknown ranks
(number of dimensions) and shapes (extents along dimensions).
Its purely functional nature is achieved by ruling out
expressions that have side effects, undefined behaviour, pointers, and
other imperative constructions of C.  This allows the compiler
to use implicit memory management and to make decisions about
parallel execution of certain code parts without requiring
any explicit annotations.  The current compiler \texttt{sac2c}
supports efficient compilation to multicore and GPU
architectures~\cite{sac-nbody}.  We introduce the key aspects of
the language that are used in the extraction examples.  For
more information about \sac{} refer to~\cite{GrelSchoIJPP06}.

\paragraph{Type system}
The main distinctive features of \sac{} are its hierarchy of array types,
intersection types and the unified data-parallel array comprehensions.
In \sac{}, functions express rank-polymorphic computations.  That is, they
compute on arrays of arbitrary rank and shape.  The type system tracks
information about shapes by using explicit attributes.  For example,
arrays where elements are integers and the shape is statically known
are expressed as:
\begin{lstlisting}
  int[] scal;         int[42] vec;         int[10,10,10] ten;
\end{lstlisting}
The shape of an array at runtime is always given by a tuple of natural
numbers.  In types, the shape attribute is an approximation of the
runtime value.  In the example above array \texttt{scal} is of rank
zero, representing a scalar value. The \texttt{vec} is a 1-dimensional array
containing 42 elements, and \texttt{ten} is a 3-dimensional array of
shape $10\times10\times10$.

Arrays with static dimensions but without a static size can be specified
using a dot:
\begin{lstlisting}
  int[.] a;         int[.,.] b;         int[.,.,.] c;
\end{lstlisting}
The variables \texttt{a}, \texttt{b} and \texttt{c} are of ranks one,
two, and three respectively, with unknown shapes.
Finally, arrays can have a dynamic number of dimensions:
\begin{lstlisting}
  int[+] d;         int[*] e;
\end{lstlisting}
where \texttt{d} is an array of rank 1 or higher, and \texttt{e} is an array of any rank.

There is a natural partial order on type attributes according to the
precision with which they describe the rank and dimensions of an array:
\begin{lstlisting}
  [] $\le$ [*]      [42] $\le$ [.] $\le$ [+] $\le$ [*]     [2,2] $\le$ [.,.] $\le$ [+] $\le$ [*]
\end{lstlisting}
This shape hierarchy gives rise to function overloading based on the
shape of the arguments, where the compiler picks the most specific
instance in case of overlap.

A limitation of \sac{} is that there is no way to express
complex shape relations in case of statically unknown shapes.  For example,
it would be useful to specify matrix multiplication as:
\begin{lstlisting}
  int[m,n] matmul (int[m,k], int[k,n])
\end{lstlisting}
Unfortunately, this cannot be expressed as there is no notion of type-level variables.
This becomes even more problematic for functions like \texttt{take} or \texttt{drop} below:
\begin{lstlisting}
  int[.] take(int n, int[.] v)  // take(2, [1,2,3,4]) == [1,2]
  int[.] drop(int n, int[.] v)  // drop(2, [1,2,3,4]) == [3,4]
\end{lstlisting}
Annotating them with a precise size would require some form of
dependent types, which means that we would have to give up global type
inference.

The key language construct in \sac{} is the \texttt{with}-loop --- a
data-parallel array comprehension construct.  The programmer
specifies how index sets are to be mapped into element values
and whether the computed values form an array or are folded
into a single value.  This could be also thought of as a generalised
map/reduce construct.  Consider an example of matrix multiplication:
\begin{lstlisting}
  int[.,.] matmul (int[.,.] a, int[.,.] b) {
    M = shape(a)[0]; K = shape(a)[1]; N = shape(b)[1];
    return with {
      ([0,0] <= [i,j] < [M,N]): with {
         ([0] <= [k] < [K]): a[[i,k]]*b[[k,j]];
      }: fold (+, 0);
    }: genarray ([M,N], 0);
  }
\end{lstlisting}
First, we obtain the number of rows and columns in the matrices by
querying their shape (which is a 1-dimensional array) and selecting
its corresponding components.  The outer with-loop specifies
the index range from $[0,0]$ up to $[M,N]$ and tells that all the
computed values should be put into the array of shape $M\times N$.
The latter is specified with \texttt{genarray} at the end of the
with-loop, where the first argument is the shape of the result,
and the second one is the default element.  The default element serves
two purposes: i) providing a value for the array indices that were
not specified in the index ranges; ii) providing the shape of each
element.  The latter is important because the computed elements
must all have the same shape.
The shape of the result is a concatenation of the \texttt{genarray}
shape and the shape of the default element.
The inner with-loop computes the sum of point-wise multiplied
$i$-th row and $j$-th column, expressed by \texttt{fold (+, 0)}.
%
For more details on programming in \sac{} refer to~\cite{GrelckCEFP11}.

\subsection{Embedded Array Language} \label{sec:embedded-array-lang}
\begin{code}[hide]%
\>[0]\AgdaKeyword{postulate}\<%
\\
\>[0][@{}l@{\AgdaIndent{0}}]%
\>[2]\AgdaPostulate{⋯}\AgdaSpace{}%
\AgdaSymbol{:}\AgdaSpace{}%
\AgdaSymbol{∀}\AgdaSpace{}%
\AgdaSymbol{\{}\AgdaBound{a}\AgdaSymbol{\}\{}\AgdaBound{A}\AgdaSpace{}%
\AgdaSymbol{:}\AgdaSpace{}%
\AgdaPrimitive{Set}\AgdaSpace{}%
\AgdaBound{a}\AgdaSymbol{\}}\AgdaSpace{}%
\AgdaSymbol{→}\AgdaSpace{}%
\AgdaBound{A}\<%
\\
\>[0]\AgdaKeyword{module}\AgdaSpace{}%
\AgdaModule{\AgdaUnderscore{}}\AgdaSpace{}%
\AgdaKeyword{where}\<%
\\
\>[0]\AgdaKeyword{module}\AgdaSpace{}%
\AgdaModule{ArType}\AgdaSpace{}%
\AgdaKeyword{where}\<%
\\
\>[0][@{}l@{\AgdaIndent{0}}]%
\>[2]\AgdaKeyword{open}\AgdaSpace{}%
\AgdaKeyword{import}\AgdaSpace{}%
\AgdaModule{Data.Vec}\AgdaSpace{}%
\AgdaKeyword{using}\AgdaSpace{}%
\AgdaSymbol{(}\AgdaDatatype{Vec}\AgdaSymbol{;}\AgdaSpace{}%
\AgdaInductiveConstructor{[]}\AgdaSymbol{;}\AgdaSpace{}%
\AgdaOperator{\AgdaInductiveConstructor{\AgdaUnderscore{}∷\AgdaUnderscore{}}}\AgdaSymbol{)}\<%
\\
\>[2]\AgdaKeyword{open}\AgdaSpace{}%
\AgdaKeyword{import}\AgdaSpace{}%
\AgdaModule{Data.Nat}\<%
\\
\>[2]\AgdaKeyword{open}\AgdaSpace{}%
\AgdaKeyword{import}\AgdaSpace{}%
\AgdaModule{Data.Fin}\<%
\\
\>[2]\AgdaKeyword{open}\AgdaSpace{}%
\AgdaKeyword{import}\AgdaSpace{}%
\AgdaModule{Function}\<%
\\
\>[2]\AgdaKeyword{infixr}\AgdaSpace{}%
\AgdaNumber{5}\AgdaSpace{}%
\AgdaOperator{\AgdaInductiveConstructor{\AgdaUnderscore{}∷\AgdaUnderscore{}}}\<%
\end{code}

To embed \sac{}, we have to define a type of multi-dimensional
arrays, and three constructs: with-loops, shapes, and selections.  Our goal
is to express non-trivial shape relations
between the arguments of a function and to ensure in-bound array indexing
statically.  We achieve this with the following two Agda types:
\begin{code}%
\>[2]\AgdaKeyword{data}\AgdaSpace{}%
\AgdaDatatype{Ix}\AgdaSpace{}%
\AgdaSymbol{:}\AgdaSpace{}%
\AgdaSymbol{(}\AgdaBound{d}\AgdaSpace{}%
\AgdaSymbol{:}\AgdaSpace{}%
\AgdaDatatype{ℕ}\AgdaSymbol{)}\AgdaSpace{}%
\AgdaSymbol{→}\AgdaSpace{}%
\AgdaSymbol{(}\AgdaBound{s}\AgdaSpace{}%
\AgdaSymbol{:}\AgdaSpace{}%
\AgdaDatatype{Vec}\AgdaSpace{}%
\AgdaDatatype{ℕ}\AgdaSpace{}%
\AgdaBound{d}\AgdaSymbol{)}\AgdaSpace{}%
\AgdaSymbol{→}\AgdaSpace{}%
\AgdaPrimitive{Set}\AgdaSpace{}%
\AgdaKeyword{where}\<%
\\
\>[2][@{}l@{\AgdaIndent{0}}]%
\>[4]\AgdaInductiveConstructor{[]}%
\>[9]\AgdaSymbol{:}\AgdaSpace{}%
\AgdaDatatype{Ix}\AgdaSpace{}%
\AgdaNumber{0}\AgdaSpace{}%
\AgdaInductiveConstructor{[]}\<%
\\
\>[4]\AgdaOperator{\AgdaInductiveConstructor{\AgdaUnderscore{}∷\AgdaUnderscore{}}}%
\>[9]\AgdaSymbol{:}\AgdaSpace{}%
\AgdaSymbol{∀}\AgdaSpace{}%
\AgdaSymbol{\{}\AgdaBound{d}\AgdaSpace{}%
\AgdaBound{s}\AgdaSpace{}%
\AgdaBound{x}\AgdaSymbol{\}}\AgdaSpace{}%
\AgdaSymbol{→}\AgdaSpace{}%
\AgdaDatatype{Fin}\AgdaSpace{}%
\AgdaBound{x}\AgdaSpace{}%
\AgdaSymbol{→}\AgdaSpace{}%
\AgdaSymbol{(}\AgdaBound{ix}\AgdaSpace{}%
\AgdaSymbol{:}\AgdaSpace{}%
\AgdaDatatype{Ix}\AgdaSpace{}%
\AgdaBound{d}\AgdaSpace{}%
\AgdaBound{s}\AgdaSymbol{)}\AgdaSpace{}%
\AgdaSymbol{→}\AgdaSpace{}%
\AgdaDatatype{Ix}\AgdaSpace{}%
\AgdaSymbol{(}\AgdaInductiveConstructor{suc}\AgdaSpace{}%
\AgdaBound{d}\AgdaSymbol{)}\AgdaSpace{}%
\AgdaSymbol{(}\AgdaBound{x}\AgdaSpace{}%
\AgdaOperator{\AgdaInductiveConstructor{∷}}\AgdaSpace{}%
\AgdaBound{s}\AgdaSymbol{)}\<%
\\
\\[\AgdaEmptyExtraSkip]%
\>[2]\AgdaKeyword{record}\AgdaSpace{}%
\AgdaRecord{Ar}\AgdaSpace{}%
\AgdaSymbol{\{}\AgdaBound{a}\AgdaSymbol{\}}\AgdaSpace{}%
\AgdaSymbol{(}\AgdaBound{X}\AgdaSpace{}%
\AgdaSymbol{:}\AgdaSpace{}%
\AgdaPrimitive{Set}\AgdaSpace{}%
\AgdaBound{a}\AgdaSymbol{)}\AgdaSpace{}%
\AgdaSymbol{(}\AgdaBound{d}\AgdaSpace{}%
\AgdaSymbol{:}\AgdaSpace{}%
\AgdaDatatype{ℕ}\AgdaSymbol{)}\AgdaSpace{}%
\AgdaSymbol{(}\AgdaBound{s}\AgdaSpace{}%
\AgdaSymbol{:}\AgdaSpace{}%
\AgdaDatatype{Vec}\AgdaSpace{}%
\AgdaDatatype{ℕ}\AgdaSpace{}%
\AgdaBound{d}\AgdaSymbol{)}\AgdaSpace{}%
\AgdaSymbol{:}\AgdaSpace{}%
\AgdaPrimitive{Set}\AgdaSpace{}%
\AgdaBound{a}\AgdaSpace{}%
\AgdaKeyword{where}\<%
\\
\>[2][@{}l@{\AgdaIndent{0}}]%
\>[4]\AgdaKeyword{constructor}\AgdaSpace{}%
\AgdaInductiveConstructor{imap}\<%
\\
\>[4]\AgdaKeyword{field}\AgdaSpace{}%
\AgdaField{sel}\AgdaSpace{}%
\AgdaSymbol{:}\AgdaSpace{}%
\AgdaDatatype{Ix}\AgdaSpace{}%
\AgdaBound{d}\AgdaSpace{}%
\AgdaBound{s}\AgdaSpace{}%
\AgdaSymbol{→}\AgdaSpace{}%
\AgdaBound{X}\<%
\end{code}
Both types are indexed by a shape \AB{s}, represented
as a \AD{Vec}tor of natural numbers.  The \AD{Ix} type is a type of valid
indices within the index-space generated by the shape \AD{s}.  A valid
index in such an index-space is a tuple of natural numbers that is component-wise
less than the shape \AB{s}.  Finally, the array with elements of type \AB{X}
is given by a function from valid indices to $X$.
In some sense \AD{Ar} and \AD{Ix} are second-order versions of \AD{Vec}
and \AD{Fin}.  This could be also thought of as a
computational interpretation of the Mathematics of Arrays~\cite{LMRMullin:moa}
(where $\Psi$ becomes an array constructor), or as a generalisation of
pull arrays~\cite{pushpull}, or as simple containers~\cite{DBLP:journals/jfp/AltenkirchGHMM15}.

This encoding intrinsically guarantees that all the array accesses are within
bounds.
As for \texttt{fold}
\texttt{with}-loops, there is no need for a special construct:
we can define
a recursive function (analogous to reduce on \AD{Vec}), and let the extractor
translate its applications into the corresponding \texttt{fold} \texttt{with}-loop.

Consider now the matrix multiplication example expressed in the embedded
language:
\begin{code}[hide]%
\>[2]\AgdaKeyword{open}\AgdaSpace{}%
\AgdaModule{Ar}\AgdaSpace{}%
\AgdaKeyword{public}\<%
\\
\>[2]\AgdaKeyword{private}\<%
\\
\>[2][@{}l@{\AgdaIndent{0}}]%
\>[3]\AgdaKeyword{postulate}\<%
\\
\>[3][@{}l@{\AgdaIndent{0}}]%
\>[4]\AgdaPostulate{sum}\AgdaSpace{}%
\AgdaSymbol{:}\AgdaSpace{}%
\AgdaSymbol{∀}\AgdaSpace{}%
\AgdaSymbol{\{}\AgdaBound{n}\AgdaSymbol{\}}%
\>[17]\AgdaSymbol{→}\AgdaSpace{}%
\AgdaRecord{Ar}\AgdaSpace{}%
\AgdaDatatype{ℕ}\AgdaSpace{}%
\AgdaNumber{1}\AgdaSpace{}%
\AgdaSymbol{(}\AgdaBound{n}\AgdaSpace{}%
\AgdaOperator{\AgdaInductiveConstructor{∷}}\AgdaSpace{}%
\AgdaInductiveConstructor{[]}\AgdaSymbol{)}\AgdaSpace{}%
\AgdaSymbol{→}\AgdaSpace{}%
\AgdaDatatype{ℕ}\<%
\end{code}
\begin{code}%
\>[2]\AgdaFunction{mm}\AgdaSpace{}%
\AgdaSymbol{:}\AgdaSpace{}%
\AgdaSymbol{∀}\AgdaSpace{}%
\AgdaSymbol{\{}\AgdaBound{a}\AgdaSpace{}%
\AgdaBound{b}\AgdaSpace{}%
\AgdaBound{c}\AgdaSymbol{\}}\AgdaSpace{}%
\AgdaSymbol{→}\AgdaSpace{}%
\AgdaKeyword{let}\AgdaSpace{}%
\AgdaBound{Mat}\AgdaSpace{}%
\AgdaBound{x}\AgdaSpace{}%
\AgdaBound{y}\AgdaSpace{}%
\AgdaSymbol{=}\AgdaSpace{}%
\AgdaRecord{Ar}\AgdaSpace{}%
\AgdaDatatype{ℕ}\AgdaSpace{}%
\AgdaNumber{2}\AgdaSpace{}%
\AgdaSymbol{(}\AgdaBound{x}\AgdaSpace{}%
\AgdaOperator{\AgdaInductiveConstructor{∷}}\AgdaSpace{}%
\AgdaBound{y}\AgdaSpace{}%
\AgdaOperator{\AgdaInductiveConstructor{∷}}\AgdaSpace{}%
\AgdaInductiveConstructor{[]}\AgdaSymbol{)}\AgdaSpace{}%
\AgdaKeyword{in}\AgdaSpace{}%
\AgdaBound{Mat}\AgdaSpace{}%
\AgdaBound{a}\AgdaSpace{}%
\AgdaBound{b}\AgdaSpace{}%
\AgdaSymbol{→}\AgdaSpace{}%
\AgdaBound{Mat}\AgdaSpace{}%
\AgdaBound{b}\AgdaSpace{}%
\AgdaBound{c}\AgdaSpace{}%
\AgdaSymbol{→}\AgdaSpace{}%
\AgdaBound{Mat}\AgdaSpace{}%
\AgdaBound{a}\AgdaSpace{}%
\AgdaBound{c}\<%
\\
\>[2]\AgdaFunction{mm}%
\>[133I]\AgdaSymbol{(}\AgdaInductiveConstructor{imap}\AgdaSpace{}%
\AgdaBound{a}\AgdaSymbol{)}\AgdaSpace{}%
\AgdaSymbol{(}\AgdaInductiveConstructor{imap}\AgdaSpace{}%
\AgdaBound{b}\AgdaSymbol{)}\AgdaSpace{}%
\AgdaSymbol{=}\AgdaSpace{}%
\AgdaInductiveConstructor{imap}\AgdaSpace{}%
\AgdaFunction{body}\AgdaSpace{}%
\AgdaKeyword{where}\<%
\\
\>[.][@{}l@{}]\<[133I]%
\>[5]\AgdaFunction{body}\AgdaSpace{}%
\AgdaSymbol{:}\AgdaSpace{}%
\AgdaSymbol{\AgdaUnderscore{}}\<%
\\
\>[5]\AgdaFunction{body}\AgdaSpace{}%
\AgdaSymbol{(}\AgdaBound{i}\AgdaSpace{}%
\AgdaOperator{\AgdaInductiveConstructor{∷}}\AgdaSpace{}%
\AgdaBound{j}\AgdaSpace{}%
\AgdaOperator{\AgdaInductiveConstructor{∷}}\AgdaSpace{}%
\AgdaInductiveConstructor{[]}\AgdaSymbol{)}\AgdaSpace{}%
\AgdaSymbol{=}\AgdaSpace{}%
\AgdaPostulate{sum}\AgdaSpace{}%
\AgdaOperator{\AgdaFunction{\$}}\AgdaSpace{}%
\AgdaInductiveConstructor{imap}\AgdaSpace{}%
\AgdaSymbol{λ}\AgdaSpace{}%
\AgdaKeyword{where}\AgdaSpace{}%
\AgdaSymbol{(}\AgdaBound{k}\AgdaSpace{}%
\AgdaOperator{\AgdaInductiveConstructor{∷}}\AgdaSpace{}%
\AgdaInductiveConstructor{[]}\AgdaSymbol{)}\AgdaSpace{}%
\AgdaSymbol{→}\AgdaSpace{}%
\AgdaBound{a}\AgdaSpace{}%
\AgdaSymbol{(}\AgdaBound{i}\AgdaSpace{}%
\AgdaOperator{\AgdaInductiveConstructor{∷}}\AgdaSpace{}%
\AgdaBound{k}\AgdaSpace{}%
\AgdaOperator{\AgdaInductiveConstructor{∷}}\AgdaSpace{}%
\AgdaInductiveConstructor{[]}\AgdaSymbol{)}\AgdaSpace{}%
\AgdaOperator{\AgdaPrimitive{*}}\AgdaSpace{}%
\AgdaBound{b}\AgdaSpace{}%
\AgdaSymbol{(}\AgdaBound{k}\AgdaSpace{}%
\AgdaOperator{\AgdaInductiveConstructor{∷}}\AgdaSpace{}%
\AgdaBound{j}\AgdaSpace{}%
\AgdaOperator{\AgdaInductiveConstructor{∷}}\AgdaSpace{}%
\AgdaInductiveConstructor{[]}\AgdaSymbol{)}\<%
\end{code}
With a similar level of expressiveness, the
implementation encodes the correct shape relation between the arguments and
guarantees in-bound indexing without any explicit proofs.

Our definition of \AD{Ar} satisfies the useful property that
any composition of operations on arrays normalises to a single
\AC{imap}.  Consider an example:
\begin{code}[hide]%
\>[0]\AgdaKeyword{module}\AgdaSpace{}%
\AgdaModule{ExFuse}\AgdaSpace{}%
\AgdaKeyword{where}\<%
\\
\>[0][@{}l@{\AgdaIndent{0}}]%
\>[2]\AgdaKeyword{open}\AgdaSpace{}%
\AgdaModule{ArType}\<%
\\
\>[2]\AgdaKeyword{open}\AgdaSpace{}%
\AgdaKeyword{import}\AgdaSpace{}%
\AgdaModule{Data.Nat}\AgdaSpace{}%
\AgdaSymbol{as}\AgdaSpace{}%
\AgdaModule{ℕ}\AgdaSpace{}%
\AgdaKeyword{using}\AgdaSpace{}%
\AgdaSymbol{(}\AgdaDatatype{ℕ}\AgdaSymbol{;}\AgdaSpace{}%
\AgdaInductiveConstructor{zero}\AgdaSymbol{;}\AgdaSpace{}%
\AgdaInductiveConstructor{suc}\AgdaSymbol{)}\<%
\\
\>[2]\AgdaKeyword{open}\AgdaSpace{}%
\AgdaKeyword{import}\AgdaSpace{}%
\AgdaModule{Data.Vec}\AgdaSpace{}%
\AgdaKeyword{using}\AgdaSpace{}%
\AgdaSymbol{(}\AgdaDatatype{Vec}\AgdaSymbol{;}\AgdaSpace{}%
\AgdaInductiveConstructor{[]}\AgdaSymbol{;}%
\>[40]\AgdaOperator{\AgdaInductiveConstructor{\AgdaUnderscore{}∷\AgdaUnderscore{}}}\AgdaSymbol{;}\AgdaSpace{}%
\AgdaFunction{reverse}\AgdaSymbol{;}\AgdaSpace{}%
\AgdaOperator{\AgdaFunction{\AgdaUnderscore{}++\AgdaUnderscore{}}}\AgdaSymbol{;}\AgdaSpace{}%
\AgdaFunction{splitAt}\AgdaSymbol{)}\<%
\\
\>[2]\AgdaKeyword{open}\AgdaSpace{}%
\AgdaKeyword{import}\AgdaSpace{}%
\AgdaModule{Relation.Binary.PropositionalEquality}\<%
\\
\>[2]\AgdaKeyword{open}\AgdaSpace{}%
\AgdaKeyword{import}\AgdaSpace{}%
\AgdaModule{Data.Nat.Properties}\<%
\\
\>[2]\AgdaKeyword{open}\AgdaSpace{}%
\AgdaKeyword{import}\AgdaSpace{}%
\AgdaModule{Data.Product}\<%
\\
\>[2]\AgdaKeyword{open}\AgdaSpace{}%
\AgdaKeyword{import}\AgdaSpace{}%
\AgdaModule{Reflection}\<%
\\
\>[2]\AgdaKeyword{open}\AgdaSpace{}%
\AgdaKeyword{import}\AgdaSpace{}%
\AgdaModule{Data.List}\AgdaSpace{}%
\AgdaKeyword{using}\AgdaSpace{}%
\AgdaSymbol{(}\AgdaDatatype{List}\AgdaSymbol{;}\AgdaSpace{}%
\AgdaInductiveConstructor{[]}\AgdaSymbol{;}\AgdaSpace{}%
\AgdaOperator{\AgdaInductiveConstructor{\AgdaUnderscore{}∷\AgdaUnderscore{}}}\AgdaSymbol{)}\<%
\\
\>[2]\AgdaKeyword{open}\AgdaSpace{}%
\AgdaKeyword{import}\AgdaSpace{}%
\AgdaModule{Function}\AgdaSpace{}%
\AgdaKeyword{using}\AgdaSpace{}%
\AgdaSymbol{(}\AgdaOperator{\AgdaFunction{\AgdaUnderscore{}\$\AgdaUnderscore{}}}\AgdaSymbol{)}\<%
\\
\>[2]\AgdaKeyword{open}\AgdaSpace{}%
\AgdaKeyword{import}\AgdaSpace{}%
\AgdaModule{Data.Fin}\AgdaSpace{}%
\AgdaKeyword{using}\AgdaSpace{}%
\AgdaSymbol{(}\AgdaDatatype{Fin}\AgdaSymbol{;}\AgdaSpace{}%
\AgdaInductiveConstructor{zero}\AgdaSymbol{;}\AgdaSpace{}%
\AgdaInductiveConstructor{suc}\AgdaSymbol{)}\<%
\\
\>[2]\AgdaKeyword{infixl}\AgdaSpace{}%
\AgdaNumber{7}\AgdaSpace{}%
\AgdaOperator{\AgdaFunction{\AgdaUnderscore{}*\AgdaUnderscore{}}}\<%
\\
\>[2]\AgdaKeyword{infixl}\AgdaSpace{}%
\AgdaNumber{6}\AgdaSpace{}%
\AgdaOperator{\AgdaFunction{\AgdaUnderscore{}+\AgdaUnderscore{}}}\<%
\\
\>[2]\AgdaKeyword{postulate}\<%
\\
\>[2][@{}l@{\AgdaIndent{0}}]%
\>[4]\AgdaPostulate{sum}\AgdaSpace{}%
\AgdaSymbol{:}\AgdaSpace{}%
\AgdaSymbol{∀}\AgdaSpace{}%
\AgdaSymbol{\{}\AgdaBound{d}\AgdaSpace{}%
\AgdaBound{s}\AgdaSymbol{\}}\AgdaSpace{}%
\AgdaSymbol{→}\AgdaSpace{}%
\AgdaRecord{Ar}\AgdaSpace{}%
\AgdaDatatype{ℕ}\AgdaSpace{}%
\AgdaBound{d}\AgdaSpace{}%
\AgdaBound{s}\AgdaSpace{}%
\AgdaSymbol{→}\AgdaSpace{}%
\AgdaDatatype{ℕ}\<%
\\
\>[4]\AgdaPostulate{reverse-inv}\AgdaSpace{}%
\AgdaSymbol{:}\AgdaSpace{}%
\AgdaSymbol{∀}\AgdaSpace{}%
\AgdaSymbol{\{}\AgdaBound{X}\AgdaSpace{}%
\AgdaSymbol{:}\AgdaSpace{}%
\AgdaPrimitive{Set}\AgdaSymbol{\}\{}\AgdaBound{n}\AgdaSymbol{\}}\AgdaSpace{}%
\AgdaSymbol{(}\AgdaBound{xs}\AgdaSpace{}%
\AgdaSymbol{:}\AgdaSpace{}%
\AgdaDatatype{Vec}\AgdaSpace{}%
\AgdaBound{X}\AgdaSpace{}%
\AgdaBound{n}\AgdaSymbol{)}\AgdaSpace{}%
\AgdaSymbol{→}\AgdaSpace{}%
\AgdaFunction{reverse}\AgdaSpace{}%
\AgdaSymbol{(}\AgdaFunction{reverse}\AgdaSpace{}%
\AgdaBound{xs}\AgdaSymbol{)}\AgdaSpace{}%
\AgdaOperator{\AgdaDatatype{≡}}\AgdaSpace{}%
\AgdaBound{xs}\<%
\\
\>[4]\AgdaPostulate{ix-reverse}\AgdaSpace{}%
\AgdaSymbol{:}\AgdaSpace{}%
\AgdaSymbol{∀}\AgdaSpace{}%
\AgdaSymbol{\{}\AgdaBound{d}\AgdaSpace{}%
\AgdaBound{s}\AgdaSymbol{\}}\AgdaSpace{}%
\AgdaSymbol{→}\AgdaSpace{}%
\AgdaDatatype{Ix}\AgdaSpace{}%
\AgdaBound{d}\AgdaSpace{}%
\AgdaBound{s}\AgdaSpace{}%
\AgdaSymbol{→}\AgdaSpace{}%
\AgdaDatatype{Ix}\AgdaSpace{}%
\AgdaBound{d}\AgdaSpace{}%
\AgdaSymbol{(}\AgdaFunction{reverse}\AgdaSpace{}%
\AgdaBound{s}\AgdaSymbol{)}\<%
\end{code}
\begin{code}%
\>[2]\AgdaOperator{\AgdaFunction{\AgdaUnderscore{}*\AgdaUnderscore{}}}\AgdaSpace{}%
\AgdaOperator{\AgdaFunction{\AgdaUnderscore{}+\AgdaUnderscore{}}}\AgdaSpace{}%
\AgdaSymbol{:}\AgdaSpace{}%
\AgdaSymbol{∀}\AgdaSpace{}%
\AgdaSymbol{\{}\AgdaBound{d}\AgdaSpace{}%
\AgdaBound{s}\AgdaSymbol{\}}\AgdaSpace{}%
\AgdaSymbol{→}\AgdaSpace{}%
\AgdaSymbol{(}\AgdaBound{a}\AgdaSpace{}%
\AgdaBound{b}\AgdaSpace{}%
\AgdaSymbol{:}\AgdaSpace{}%
\AgdaRecord{Ar}\AgdaSpace{}%
\AgdaDatatype{ℕ}\AgdaSpace{}%
\AgdaBound{d}\AgdaSpace{}%
\AgdaBound{s}\AgdaSymbol{)}\AgdaSpace{}%
\AgdaSymbol{→}\AgdaSpace{}%
\AgdaRecord{Ar}\AgdaSpace{}%
\AgdaDatatype{ℕ}\AgdaSpace{}%
\AgdaBound{d}\AgdaSpace{}%
\AgdaBound{s}\<%
\\
\>[2]\AgdaOperator{\AgdaFunction{\AgdaUnderscore{}+\AgdaUnderscore{}}}\AgdaSpace{}%
\AgdaBound{a}\AgdaSpace{}%
\AgdaBound{b}\AgdaSpace{}%
\AgdaSymbol{=}\AgdaSpace{}%
\AgdaInductiveConstructor{imap}\AgdaSpace{}%
\AgdaSymbol{λ}\AgdaSpace{}%
\AgdaBound{iv}\AgdaSpace{}%
\AgdaSymbol{→}\AgdaSpace{}%
\AgdaField{sel}\AgdaSpace{}%
\AgdaBound{a}\AgdaSpace{}%
\AgdaBound{iv}\AgdaSpace{}%
\AgdaOperator{\AgdaPrimitive{ℕ.+}}\AgdaSpace{}%
\AgdaField{sel}\AgdaSpace{}%
\AgdaBound{b}\AgdaSpace{}%
\AgdaBound{iv}\<%
\\
\>[2]\AgdaOperator{\AgdaFunction{\AgdaUnderscore{}*\AgdaUnderscore{}}}\AgdaSpace{}%
\AgdaBound{a}\AgdaSpace{}%
\AgdaBound{b}\AgdaSpace{}%
\AgdaSymbol{=}\AgdaSpace{}%
\AgdaInductiveConstructor{imap}\AgdaSpace{}%
\AgdaSymbol{λ}\AgdaSpace{}%
\AgdaBound{iv}\AgdaSpace{}%
\AgdaSymbol{→}\AgdaSpace{}%
\AgdaField{sel}\AgdaSpace{}%
\AgdaBound{a}\AgdaSpace{}%
\AgdaBound{iv}\AgdaSpace{}%
\AgdaOperator{\AgdaPrimitive{ℕ.*}}\AgdaSpace{}%
\AgdaField{sel}\AgdaSpace{}%
\AgdaBound{b}\AgdaSpace{}%
\AgdaBound{iv}\<%
\\
\\[\AgdaEmptyExtraSkip]%
\>[2]\AgdaOperator{\AgdaFunction{\AgdaUnderscore{}ᵀ}}\AgdaSpace{}%
\AgdaSymbol{:}\AgdaSpace{}%
\AgdaSymbol{∀}\AgdaSpace{}%
\AgdaSymbol{\{}\AgdaBound{X}\AgdaSpace{}%
\AgdaSymbol{:}\AgdaSpace{}%
\AgdaPrimitive{Set}\AgdaSymbol{\}\{}\AgdaBound{d}\AgdaSpace{}%
\AgdaBound{s}\AgdaSymbol{\}}\AgdaSpace{}%
\AgdaSymbol{→}\AgdaSpace{}%
\AgdaRecord{Ar}\AgdaSpace{}%
\AgdaBound{X}\AgdaSpace{}%
\AgdaBound{d}\AgdaSpace{}%
\AgdaBound{s}\AgdaSpace{}%
\AgdaSymbol{→}\AgdaSpace{}%
\AgdaRecord{Ar}\AgdaSpace{}%
\AgdaBound{X}\AgdaSpace{}%
\AgdaBound{d}\AgdaSpace{}%
\AgdaSymbol{(}\AgdaFunction{reverse}\AgdaSpace{}%
\AgdaBound{s}\AgdaSymbol{)}\<%
\\
\>[2]\AgdaOperator{\AgdaFunction{\AgdaUnderscore{}ᵀ}}\AgdaSpace{}%
\AgdaBound{a}\AgdaSpace{}%
\AgdaSymbol{=}\AgdaSpace{}%
\AgdaInductiveConstructor{imap}\AgdaSpace{}%
\AgdaSymbol{λ}\AgdaSpace{}%
\AgdaBound{iv}\AgdaSpace{}%
\AgdaSymbol{→}\AgdaSpace{}%
\AgdaField{sel}\AgdaSpace{}%
\AgdaBound{a}\AgdaSpace{}%
\AgdaSymbol{(}\AgdaFunction{subst}\AgdaSpace{}%
\AgdaSymbol{(}\AgdaDatatype{Ix}\AgdaSpace{}%
\AgdaSymbol{\AgdaUnderscore{})}\AgdaSpace{}%
\AgdaSymbol{(}\AgdaPostulate{reverse-inv}\AgdaSpace{}%
\AgdaSymbol{\AgdaUnderscore{})}\AgdaSpace{}%
\AgdaSymbol{(}\AgdaPostulate{ix-reverse}\AgdaSpace{}%
\AgdaBound{iv}\AgdaSymbol{))}\<%
\\
\\[\AgdaEmptyExtraSkip]%
\>[2]\AgdaFunction{ex}\AgdaSpace{}%
\AgdaSymbol{:}\AgdaSpace{}%
\AgdaSymbol{∀}\AgdaSpace{}%
\AgdaSymbol{\{}\AgdaBound{m}\AgdaSpace{}%
\AgdaBound{n}\AgdaSymbol{\}}\AgdaSpace{}%
\AgdaSymbol{→}\AgdaSpace{}%
\AgdaRecord{Ar}\AgdaSpace{}%
\AgdaDatatype{ℕ}\AgdaSpace{}%
\AgdaNumber{2}\AgdaSpace{}%
\AgdaSymbol{(}\AgdaBound{n}\AgdaSpace{}%
\AgdaOperator{\AgdaInductiveConstructor{∷}}\AgdaSpace{}%
\AgdaBound{m}\AgdaSpace{}%
\AgdaOperator{\AgdaInductiveConstructor{∷}}\AgdaSpace{}%
\AgdaInductiveConstructor{[]}\AgdaSymbol{)}\AgdaSpace{}%
\AgdaSymbol{→}\AgdaSpace{}%
\AgdaRecord{Ar}\AgdaSpace{}%
\AgdaDatatype{ℕ}\AgdaSpace{}%
\AgdaNumber{2}\AgdaSpace{}%
\AgdaSymbol{(}\AgdaBound{m}\AgdaSpace{}%
\AgdaOperator{\AgdaInductiveConstructor{∷}}\AgdaSpace{}%
\AgdaBound{n}\AgdaSpace{}%
\AgdaOperator{\AgdaInductiveConstructor{∷}}\AgdaSpace{}%
\AgdaInductiveConstructor{[]}\AgdaSymbol{)}\AgdaSpace{}%
\AgdaSymbol{→}\AgdaSpace{}%
\AgdaRecord{Ar}\AgdaSpace{}%
\AgdaDatatype{ℕ}\AgdaSpace{}%
\AgdaNumber{2}\AgdaSpace{}%
\AgdaSymbol{(}\AgdaBound{m}\AgdaSpace{}%
\AgdaOperator{\AgdaInductiveConstructor{∷}}\AgdaSpace{}%
\AgdaBound{n}\AgdaSpace{}%
\AgdaOperator{\AgdaInductiveConstructor{∷}}\AgdaSpace{}%
\AgdaInductiveConstructor{[]}\AgdaSymbol{)}\<%
\\
\>[2]\AgdaFunction{ex}\AgdaSpace{}%
\AgdaBound{a}\AgdaSpace{}%
\AgdaBound{b}\AgdaSpace{}%
\AgdaSymbol{=}\AgdaSpace{}%
\AgdaBound{a}\AgdaSpace{}%
\AgdaOperator{\AgdaFunction{ᵀ}}\AgdaSpace{}%
\AgdaOperator{\AgdaFunction{+}}\AgdaSpace{}%
\AgdaSymbol{(}\AgdaBound{b}\AgdaSpace{}%
\AgdaOperator{\AgdaFunction{*}}\AgdaSpace{}%
\AgdaBound{b}\AgdaSymbol{)}\<%
\end{code}
Here we defined \AF{\_+\_} and \AF{\_*\_} as element-wise operations
on the array elements.  The \AF{\_ᵀ} is a transposition of the matrix,
which reverses the order of the dimensions.  Note that all
of these are defined in a rank-polymorphic style.  For transposition
we had to apply a proof (\AF{reverse-inv}) that reversing an
index is involutive.  The body of \AF{ex} is given as four
operations on the entire arrays, conceptually creating a new copy
of an array at every application.  Due to our encoding, the
body of \AF{ex} normalises into a single \AC{imap}.
This is largely possible because we defined \AD{Ar} as a record,
and these are guaranteed to preserve $\eta$-equality.  That is,
every \AB{x} : \AD{Ar} \AB{d} \AB{s} is \emph{definitionally} equal
to \AC{imap} (\AR{sel} \AB{x}).

During the implementation of our extractor for \sac{} in Agda, we
encountered an unexpected challenge related to the definition of
\AD{Ar}.  By defining it as a record type, the elaborator of Agda
decides to erase the (implicit) arguments $a$, $X$, $d$, and $s$ of
the constructor \AC{imap}, replacing them by the constructor
\AC{unknown} in the reflected syntax.
The reason why Agda does this is because these parameters can always
be reconstructed from the type of the array.
However, inferring them is far from trivial as \AC{imap} may appear in
arbitrary contexts.
To work around this issue, we extended the reflection API of Agda with
a new primitive \AF{withReconstructed} that instructs all the further
calls to \AF{getDefinition}, \AF{normalise}, \AF{reduce}, \etc{} to
reconstruct the parameters that are normally marked as \AC{unknown}.
We use this function when \AF{kompile} obtains the representation of
each definition. For more details on this new feature, see \url{https://github.com/agda/agda/pull/5022}.

\subsection{Validating Types}
One of the major differences between extracting into Kaleidoscope and
into \sac{} is the presence of the non-trivial type system in the latter.
This requires us to choose what Agda types are going to be supported
and how to translate them into the target language.

\sac{} lacks support for heterogeneously nested arrays: all the
elements in the array must be of the same shape.  Therefore, there
is no way to construct the following types:
\begin{lstlisting}
  (int[.])[5]       (int[.])[.]     (int[*])[.]    (int[*])[*]
\end{lstlisting}
Furthermore, syntactically, there is no way to express a nested array
type.  However, one can deal with homogeneous nesting by flattening as
follows:
\begin{lstlisting}
  (int[5])[6] => int[6,5]        (int[$\tau$])[$\sigma$] => int([$\sigma$] ++ [$\tau$])
\end{lstlisting}
Also, the \texttt{with}-loop construct makes it possible to express
the computation in a nested style, but the resulting array type
is flattened according to the scheme above.   Consider an example:
\begin{lstlisting}
  int[5] foo (int[1]);  // some function that produces 5-element vectors.
  int[6] gen (int[6,5] a) {
    return with {
      ([0] <= iv < [6]): foo (iv);
    }: genarray ([6], with{}: genarray ([5], 0));
  }
\end{lstlisting}
The function \texttt{gen} computes a two-dimensional array.  The
\texttt{with}-loop that this array generates has a 1-dimensional
index-space (specified by the shape [6]), and non-scalar elements.
The latter is given by the shape of the default element, which is a
vector of 5 zeroes.  As a result we get an array of shape [6,5].

This suggests that nested \AC{imap}s can be mapped directly to
with-loops, translating nested array types in Agda into flattened
types in \sac{}.  However, while \AC{imap} is a constructor for \AD{Ar},
there is also the projection \AR{sel}.  Selecting into a nested array
would result in selection on a partial index:
\begin{code}%
\>[2]\AgdaFunction{partial-sel}\AgdaSpace{}%
\AgdaSymbol{:}\AgdaSpace{}%
\AgdaRecord{Ar}\AgdaSpace{}%
\AgdaSymbol{(}\AgdaRecord{Ar}\AgdaSpace{}%
\AgdaDatatype{ℕ}\AgdaSpace{}%
\AgdaNumber{1}\AgdaSpace{}%
\AgdaSymbol{(}\AgdaNumber{5}\AgdaSpace{}%
\AgdaOperator{\AgdaInductiveConstructor{∷}}\AgdaSpace{}%
\AgdaInductiveConstructor{[]}\AgdaSymbol{))}\AgdaSpace{}%
\AgdaNumber{1}\AgdaSpace{}%
\AgdaSymbol{(}\AgdaNumber{6}\AgdaSpace{}%
\AgdaOperator{\AgdaInductiveConstructor{∷}}\AgdaSpace{}%
\AgdaInductiveConstructor{[]}\AgdaSymbol{)}\AgdaSpace{}%
\AgdaSymbol{→}\AgdaSpace{}%
\AgdaRecord{Ar}\AgdaSpace{}%
\AgdaDatatype{ℕ}\AgdaSpace{}%
\AgdaNumber{1}\AgdaSpace{}%
\AgdaSymbol{(}\AgdaNumber{5}\AgdaSpace{}%
\AgdaOperator{\AgdaInductiveConstructor{∷}}\AgdaSpace{}%
\AgdaInductiveConstructor{[]}\AgdaSymbol{)}\<%
\\
\>[2]\AgdaFunction{partial-sel}\AgdaSpace{}%
\AgdaBound{x}\AgdaSpace{}%
\AgdaSymbol{=}\AgdaSpace{}%
\AgdaField{sel}\AgdaSpace{}%
\AgdaBound{x}\AgdaSpace{}%
\AgdaSymbol{(}\AgdaInductiveConstructor{zero}\AgdaSpace{}%
\AgdaOperator{\AgdaInductiveConstructor{∷}}\AgdaSpace{}%
\AgdaInductiveConstructor{[]}\AgdaSymbol{)}\<%
\\
\>[2]\AgdaComment{--\ int[5]\ partial\AgdaUnderscore{}sel\ (int[5,6]\ a)\ \{\ return\ ??\ \}}\<%
\end{code}
The argument to \AF{partial-sel} is a nested array of shape [6],
with inner elements of shape [5].  We can represent it in \sac{} as
an array of shape [5,6].  Selections into such an array require
two-element indices, but in the above code, selection happens on
a 1-element index.  Fortunately, we can generalise \sac{} selections
as follows:
\begin{lstlisting}
  int[*] sel(int[.] idx, int[*] a) {
    sh_inner = drop (shape (idx), shape (a));
    return with {
      (0*sh_inner <= iv < sh_inner): a[idx ++ iv];
    }: genarray (sh_inner, 0);
  }
\end{lstlisting}
When selecting an array \texttt{a} at index \texttt{idx},
the shape of the result is computed by dropping
\texttt{idx}-many elements from the shape of the argument.  The
content of the result is given by a mapping \texttt{iv} $\mapsto$
\texttt{a}[\texttt{idx} \texttt{++} \texttt{iv}], where \texttt{iv}
iterates over the index space of the resulting array.  Essentially,
we partially apply the selection operation to \texttt{idx}.  Partial
selection is a well-known pattern and it is defined in the standard
library for all the supported base types such as int, float, \etc{}

Array shapes in Agda are represented by the \AD{Vec} type, whereas
\sac{} shapes are 1-dimensional arrays.  Mapping a vector type is
straight-forward, as we only need to implement nil/cons to construct
vectors and head/tail to eliminate them\footnote{Here we show
  1-dimensional versions of the functions, but in reality we
  implement rank-polymorphic cons/hd/tail in a way
  similar to \texttt{sel} from above.}:
\begin{lstlisting}
int[.] cons (int x, int[.] xs) {          int[.] tl (int[.] xs) {
  return with {                             return with {
    ([0] <= iv <= [0]): x;                    (. <= iv <= .): xs[iv+1];
    ([1] <= iv <= .): xs[iv - 1];           }: genarray (shape (xs) - 1); }
  }: genarray (1 + shape (xs));
}                                         int hd (int[.] xs) { return xs[0]; }
\end{lstlisting}
Finally, note that if we can extract \AD{Vec}s, we can extract
\AD{List}s into exactly the same target language constructs.
The only difference lies in
the analysis of the nesting of the type.  \AD{Ar} of \AD{Vec}
and \AD{Vec} of \AD{Ar} are always homogeneous as long as the leaf
element is some base type like \AD{ℕ}.  \AD{List}s of base types
or lists of homogeneous arrays are also homogeneous.  However, whenever
\AD{List} shows up on any inner level of the nesting, we loose
homogeneity, \eg{} \AD{List}\AF{∘}\AD{List} is inhomogeneous, because
the inner elements may be of different sizes.  We implement this analysis
in the extractor, therefore allowing for the combination of nested
\AD{Ar}, \AD{Vec} and \AD{List} over the base types \AD{ℕ} and \AD{Float}.

\begin{code}[hide]%
\>[0]\AgdaKeyword{open}\AgdaSpace{}%
\AgdaKeyword{import}\AgdaSpace{}%
\AgdaModule{Data.Vec}\AgdaSpace{}%
\AgdaSymbol{as}\AgdaSpace{}%
\AgdaModule{V}\AgdaSpace{}%
\AgdaKeyword{using}\AgdaSpace{}%
\AgdaSymbol{(}\AgdaDatatype{Vec}\AgdaSymbol{;}\AgdaSpace{}%
\AgdaInductiveConstructor{[]}\AgdaSymbol{;}\AgdaSpace{}%
\AgdaOperator{\AgdaInductiveConstructor{\AgdaUnderscore{}∷\AgdaUnderscore{}}}\AgdaSymbol{)}\<%
\\
\>[0]\AgdaKeyword{open}\AgdaSpace{}%
\AgdaKeyword{import}\AgdaSpace{}%
\AgdaModule{Data.Nat}\AgdaSpace{}%
\AgdaSymbol{as}\AgdaSpace{}%
\AgdaModule{ℕ}\AgdaSpace{}%
\AgdaKeyword{using}\AgdaSpace{}%
\AgdaSymbol{(}\AgdaDatatype{ℕ}\AgdaSymbol{;}\AgdaSpace{}%
\AgdaInductiveConstructor{zero}\AgdaSymbol{;}\AgdaSpace{}%
\AgdaInductiveConstructor{suc}\AgdaSymbol{)}\<%
\\
\>[0]\AgdaKeyword{open}\AgdaSpace{}%
\AgdaKeyword{import}\AgdaSpace{}%
\AgdaModule{Data.Unit}\<%
\\
\>[0]\AgdaKeyword{open}\AgdaSpace{}%
\AgdaKeyword{import}\AgdaSpace{}%
\AgdaModule{Data.Fin}\AgdaSpace{}%
\AgdaKeyword{using}\AgdaSpace{}%
\AgdaSymbol{(}\AgdaDatatype{Fin}\AgdaSymbol{;}\AgdaSpace{}%
\AgdaInductiveConstructor{zero}\AgdaSymbol{;}\AgdaSpace{}%
\AgdaInductiveConstructor{suc}\AgdaSymbol{)}\<%
\\
\>[0]\AgdaKeyword{open}\AgdaSpace{}%
\AgdaKeyword{import}\AgdaSpace{}%
\AgdaModule{Data.List}\<%
\\
\>[0]\AgdaKeyword{open}\AgdaSpace{}%
\AgdaKeyword{import}\AgdaSpace{}%
\AgdaModule{Function}\<%
\\
\>[0]\AgdaKeyword{open}\AgdaSpace{}%
\AgdaKeyword{import}\AgdaSpace{}%
\AgdaModule{Reflection}\<%
\\
\>[0]\AgdaKeyword{module}\AgdaSpace{}%
\AgdaModule{\AgdaUnderscore{}}\AgdaSpace{}%
\AgdaKeyword{where}\<%
\\
\>[0]\AgdaKeyword{module}\AgdaSpace{}%
\AgdaModule{APL}\AgdaSpace{}%
\AgdaKeyword{where}\<%
\\
\>[0][@{}l@{\AgdaIndent{0}}]%
\>[2]\AgdaKeyword{data}\AgdaSpace{}%
\AgdaDatatype{Ix}\AgdaSpace{}%
\AgdaSymbol{:}\AgdaSpace{}%
\AgdaSymbol{(}\AgdaBound{d}\AgdaSpace{}%
\AgdaSymbol{:}\AgdaSpace{}%
\AgdaDatatype{ℕ}\AgdaSymbol{)}\AgdaSpace{}%
\AgdaSymbol{→}\AgdaSpace{}%
\AgdaSymbol{(}\AgdaBound{s}\AgdaSpace{}%
\AgdaSymbol{:}\AgdaSpace{}%
\AgdaDatatype{Vec}\AgdaSpace{}%
\AgdaDatatype{ℕ}\AgdaSpace{}%
\AgdaBound{d}\AgdaSymbol{)}\AgdaSpace{}%
\AgdaSymbol{→}\AgdaSpace{}%
\AgdaPrimitive{Set}\AgdaSpace{}%
\AgdaKeyword{where}\<%
\\
\>[2][@{}l@{\AgdaIndent{0}}]%
\>[4]\AgdaInductiveConstructor{[]}%
\>[9]\AgdaSymbol{:}\AgdaSpace{}%
\AgdaDatatype{Ix}\AgdaSpace{}%
\AgdaNumber{0}\AgdaSpace{}%
\AgdaInductiveConstructor{[]}\<%
\\
\>[4]\AgdaOperator{\AgdaInductiveConstructor{\AgdaUnderscore{}∷\AgdaUnderscore{}}}%
\>[9]\AgdaSymbol{:}\AgdaSpace{}%
\AgdaSymbol{∀}\AgdaSpace{}%
\AgdaSymbol{\{}\AgdaBound{d}\AgdaSpace{}%
\AgdaBound{s}\AgdaSpace{}%
\AgdaBound{x}\AgdaSymbol{\}}\AgdaSpace{}%
\AgdaSymbol{→}\AgdaSpace{}%
\AgdaDatatype{Fin}\AgdaSpace{}%
\AgdaBound{x}\AgdaSpace{}%
\AgdaSymbol{→}\AgdaSpace{}%
\AgdaSymbol{(}\AgdaBound{ix}\AgdaSpace{}%
\AgdaSymbol{:}\AgdaSpace{}%
\AgdaDatatype{Ix}\AgdaSpace{}%
\AgdaBound{d}\AgdaSpace{}%
\AgdaBound{s}\AgdaSymbol{)}\AgdaSpace{}%
\AgdaSymbol{→}\AgdaSpace{}%
\AgdaDatatype{Ix}\AgdaSpace{}%
\AgdaSymbol{(}\AgdaInductiveConstructor{suc}\AgdaSpace{}%
\AgdaBound{d}\AgdaSymbol{)}\AgdaSpace{}%
\AgdaSymbol{(}\AgdaBound{x}\AgdaSpace{}%
\AgdaOperator{\AgdaInductiveConstructor{∷}}\AgdaSpace{}%
\AgdaBound{s}\AgdaSymbol{)}\<%
\\
\\[\AgdaEmptyExtraSkip]%
\>[2]\AgdaKeyword{record}\AgdaSpace{}%
\AgdaRecord{Ar}\AgdaSpace{}%
\AgdaSymbol{\{}\AgdaBound{a}\AgdaSymbol{\}}\AgdaSpace{}%
\AgdaSymbol{(}\AgdaBound{X}\AgdaSpace{}%
\AgdaSymbol{:}\AgdaSpace{}%
\AgdaPrimitive{Set}\AgdaSpace{}%
\AgdaBound{a}\AgdaSymbol{)}\AgdaSpace{}%
\AgdaSymbol{(}\AgdaBound{d}\AgdaSpace{}%
\AgdaSymbol{:}\AgdaSpace{}%
\AgdaDatatype{ℕ}\AgdaSymbol{)}\AgdaSpace{}%
\AgdaSymbol{(}\AgdaBound{s}\AgdaSpace{}%
\AgdaSymbol{:}\AgdaSpace{}%
\AgdaDatatype{Vec}\AgdaSpace{}%
\AgdaDatatype{ℕ}\AgdaSpace{}%
\AgdaBound{d}\AgdaSymbol{)}\AgdaSpace{}%
\AgdaSymbol{:}\AgdaSpace{}%
\AgdaPrimitive{Set}\AgdaSpace{}%
\AgdaBound{a}\AgdaSpace{}%
\AgdaKeyword{where}\<%
\\
\>[2][@{}l@{\AgdaIndent{0}}]%
\>[4]\AgdaKeyword{constructor}\AgdaSpace{}%
\AgdaInductiveConstructor{imap}\<%
\\
\>[4]\AgdaKeyword{field}\AgdaSpace{}%
\AgdaField{sel}\AgdaSpace{}%
\AgdaSymbol{:}\AgdaSpace{}%
\AgdaDatatype{Ix}\AgdaSpace{}%
\AgdaBound{d}\AgdaSpace{}%
\AgdaBound{s}\AgdaSpace{}%
\AgdaSymbol{→}\AgdaSpace{}%
\AgdaBound{X}\<%
\end{code}
\section{\label{sec:apl}APL and CNN}

In this section we consider the embedding of an
APL subset that is large enough to port the implementation
of a convolutional neural network~\cite{cnninapl}.  APL presents an interesting case
for our approach as it introduces the notions that are
difficult to express in Agda, and presumably any other existing theorem
prover.

APL is a language that pioneered the concept of rank- and shape-polymorphic
programming.  Expressions in APL are written in index-free combinator style
with few syntactic rules.  The language is dynamically typed, and
each combinator is an operation on (multi-dimensional) arrays.  Consider
the following (valid) APL expression:
\begin{flushleft}
  \qquad\apl{b ← 2 ÷⍨ (1 ⌽ a) + ¯1 ⌽ a}\qquad\qquad
  $b_i = \frac{1}{2}\left(a_{(i-1)\%n} + a_{(i+1)\%n}\right)$
\end{flushleft}
%
It computes a two-point convolution of the array {\apl{a}} using cyclic
boundaries.  This is done by
first rotating vectors along the last axis of {\apl{a}} one element to the
left ({\apl{¯1 ⌽ a}}), then one element to the right
({\apl{1 ⌽ a}}), then adding these results element-wise
({\apl{+}}), and then dividing each element by two ({\apl{2 ÷⍨}}).
APL expressions such as this one are applicable to \apl{a} of \emph{any} rank, including
zero-dimensional arrays.
Not only has the initial set of APL combinators been found useful in practice,
but it also gives rise to the number of universal equalities such as
\apl{(-x) ⌽ x ⌽ a ≡ a}, which says: if we first rotate vectors in the last
axis of \apl{a} by \apl{x} elements in one direction and then rotate by \apl{x}
elements in the opposite direction, we will always get back the same array.
These universal equalities are based on simple
arithmetic facts, yet they give a powerful reasoning technique and they can
be used as rewrite rules for automatic program transformations.

\subsection{Embedding of APL}
The semantics of each APL operator is heavily
overloaded: the same symbol has different meanings
depending on how many arguments are being passed and what these arguments
are, \ie{} their shapes, sign, \etc{}  For example, consider the
\apl{/} (slash) symbol that can be used as follows:
\[
\begin{tabular}{ll}
  \apl{ +/a    } & sum array elements, \apl{+} is an argument   \\
  \apl{2+/a    } & sum in groups of 2, \apl{+} and \apl{2} are arguments \\
  \apl{ 2/a    } & replicate each element 2 times \\
  \apl{ +/[k]a } & sum over the $k$-th axis, \apl{[k]} is an optional axis specification
\end{tabular}
\]

While the embedding does not have to match the original syntax one-to-one,
we would like to preserve one behaviour of the operators
that is used incredibly often --- the automatic cast between scalars,
vectors, and multi-dimensional arrays.  In APL every object is an
array, therefore vectors and scalars can be simply used as arguments
to the functions that expect arrays.  Shapes of arrays are 1-dimensional
arrays themselves.  Replicating such a behaviour in Agda would lead
to infinite recursion: we would have to index \AD{Ar} type with
\AD{Ar}, which is not possible.  Furthermore, binary operations
in APL have the following casting behaviour:
\[
\begin{tabular}{ll}
  \apl{1 2 3   + 1    } & computes to \apl{2 3 4} \\
  \apl{1       + 1 2 3} & computes to \apl{2 3 4} \\
  \apl{1 2 3   + 1 2 3} & computes to \apl{2 4 6} \\
  \apl{1 2 3 4 + 1 2 3} & runtime error
\end{tabular}
\]
If one of the arguments to the binary operation is a singleton array,
it is automatically replicated to match the shape of the other element.

Normally, overloading in Agda is solved by using instance arguments.
These are special kind of implicit arguments that are resolved using
instance resolution, achieving a similar effect as
classes and instances in Haskell.
In our case, we define a relation \AD{dy-args}
between the ranks and shapes of the arguments of
the binary operation:
\begin{code}%
\>[2]\AgdaKeyword{data}\AgdaSpace{}%
\AgdaDatatype{dy-args}\AgdaSpace{}%
\AgdaSymbol{:}\AgdaSpace{}%
\AgdaSymbol{∀}\AgdaSpace{}%
\AgdaBound{m}\AgdaSpace{}%
\AgdaBound{n}\AgdaSpace{}%
\AgdaSymbol{→}\AgdaSpace{}%
\AgdaDatatype{Vec}\AgdaSpace{}%
\AgdaDatatype{ℕ}\AgdaSpace{}%
\AgdaBound{m}\AgdaSpace{}%
\AgdaSymbol{→}\AgdaSpace{}%
\AgdaDatatype{Vec}\AgdaSpace{}%
\AgdaDatatype{ℕ}\AgdaSpace{}%
\AgdaBound{n}\AgdaSpace{}%
\AgdaSymbol{→}\AgdaSpace{}%
\AgdaPrimitive{Set}\AgdaSpace{}%
\AgdaKeyword{where}\<%
\\
\>[2][@{}l@{\AgdaIndent{0}}]%
\>[4]\AgdaInductiveConstructor{n-n}\AgdaSpace{}%
\AgdaSymbol{:}\AgdaSpace{}%
\AgdaSymbol{∀}\AgdaSpace{}%
\AgdaSymbol{\{}\AgdaBound{n}\AgdaSpace{}%
\AgdaBound{s}\AgdaSymbol{\}}\AgdaSpace{}%
\AgdaSymbol{→}\AgdaSpace{}%
\AgdaDatatype{dy-args}\AgdaSpace{}%
\AgdaBound{n}\AgdaSpace{}%
\AgdaBound{n}\AgdaSpace{}%
\AgdaBound{s}%
\>[35]\AgdaBound{s}\<%
\\
\>[4]\AgdaInductiveConstructor{n-0}\AgdaSpace{}%
\AgdaSymbol{:}\AgdaSpace{}%
\AgdaSymbol{∀}\AgdaSpace{}%
\AgdaSymbol{\{}\AgdaBound{n}\AgdaSpace{}%
\AgdaBound{s}\AgdaSymbol{\}}\AgdaSpace{}%
\AgdaSymbol{→}\AgdaSpace{}%
\AgdaDatatype{dy-args}\AgdaSpace{}%
\AgdaBound{n}\AgdaSpace{}%
\AgdaNumber{0}\AgdaSpace{}%
\AgdaBound{s}%
\>[35]\AgdaInductiveConstructor{[]}\<%
\\
\>[4]\AgdaInductiveConstructor{0-n}\AgdaSpace{}%
\AgdaSymbol{:}\AgdaSpace{}%
\AgdaSymbol{∀}\AgdaSpace{}%
\AgdaSymbol{\{}\AgdaBound{n}\AgdaSpace{}%
\AgdaBound{s}\AgdaSymbol{\}}\AgdaSpace{}%
\AgdaSymbol{→}\AgdaSpace{}%
\AgdaDatatype{dy-args}\AgdaSpace{}%
\AgdaNumber{0}\AgdaSpace{}%
\AgdaBound{n}\AgdaSpace{}%
\AgdaInductiveConstructor{[]}\AgdaSpace{}%
\AgdaBound{s}\<%
\end{code}
The constructors of \AD{dy-args} specify valid ways of calling
a binary operation: either the shapes are identical, or one of
them is a scalar (rank zero, shape empty).
However, when we register these constructors as instances, Agda fails
to resolve them when two zero-dimensional arrays are supplied as
arguments.  In this case all three instances fit, but Agda can only
accept a unique solution.  Ironically, in this case, all the three
instances would lead to the same correct result.

We solve this problem by using metaprogramming:
we define a macro and use it to resolve a given
hidden argument.  Within the macro, we are free to make arbitrary
choices in case of non-unique solutions. Concretely, we
define a macro \AF{dy-args-ok?} that tries to construct an element
of type $\AD{dy-args}~m~n~\AB{sx}~\AB{sy}$.
We then define a lifting function for binary operations as follows:
\begin{code}%
\>[2]\AgdaFunction{dy-args-dim}\AgdaSpace{}%
\AgdaSymbol{:}\AgdaSpace{}%
\AgdaSymbol{∀}\AgdaSpace{}%
\AgdaSymbol{\{}\AgdaBound{m}\AgdaSpace{}%
\AgdaBound{n}\AgdaSpace{}%
\AgdaBound{sx}\AgdaSpace{}%
\AgdaBound{sy}\AgdaSymbol{\}}\AgdaSpace{}%
\AgdaSymbol{→}\AgdaSpace{}%
\AgdaDatatype{dy-args}\AgdaSpace{}%
\AgdaBound{m}\AgdaSpace{}%
\AgdaBound{n}\AgdaSpace{}%
\AgdaBound{sx}\AgdaSpace{}%
\AgdaBound{sy}\AgdaSpace{}%
\AgdaSymbol{→}\AgdaSpace{}%
\AgdaDatatype{ℕ}%
\>[55]\AgdaComment{--\ pick\ the\ largest\ rank}\<%
\\
\>[2]\AgdaFunction{dy-args-shp}\AgdaSpace{}%
\AgdaSymbol{:}\AgdaSpace{}%
\AgdaSymbol{∀}\AgdaSpace{}%
\AgdaSymbol{\{}\AgdaBound{m}\AgdaSpace{}%
\AgdaBound{n}\AgdaSpace{}%
\AgdaBound{sx}\AgdaSpace{}%
\AgdaBound{sy}\AgdaSymbol{\}}\AgdaSpace{}%
\AgdaSymbol{→}\AgdaSpace{}%
\AgdaSymbol{(}\AgdaBound{dy}\AgdaSpace{}%
\AgdaSymbol{:}\AgdaSpace{}%
\AgdaDatatype{dy-args}\AgdaSpace{}%
\AgdaBound{m}\AgdaSpace{}%
\AgdaBound{n}\AgdaSpace{}%
\AgdaBound{sx}\AgdaSpace{}%
\AgdaBound{sy}\AgdaSymbol{)}\AgdaSpace{}%
\AgdaSymbol{→}\AgdaSpace{}%
\AgdaDatatype{Vec}\AgdaSpace{}%
\AgdaDatatype{ℕ}\AgdaSpace{}%
\AgdaSymbol{(}\AgdaFunction{dy-args-dim}\AgdaSpace{}%
\AgdaBound{dy}\AgdaSymbol{)}\<%
\end{code}
\begin{code}[hide]%
\>[2]\AgdaFunction{dy-args-ok?}\AgdaSpace{}%
\AgdaSymbol{:}\AgdaSpace{}%
\AgdaDatatype{Term}\AgdaSpace{}%
\AgdaSymbol{→}\AgdaSpace{}%
\AgdaPostulate{TC}\AgdaSpace{}%
\AgdaRecord{⊤}\AgdaSpace{}%
\AgdaComment{--\ macro\ that\ resolves\ the\ instances}\<%
\\
\>[2]\AgdaFunction{dy-args-dim}\AgdaSpace{}%
\AgdaSymbol{\{}\AgdaBound{m}\AgdaSymbol{\}}%
\>[21]\AgdaInductiveConstructor{n-n}\AgdaSpace{}%
\AgdaSymbol{=}\AgdaSpace{}%
\AgdaBound{m}\<%
\\
\>[2]\AgdaFunction{dy-args-dim}\AgdaSpace{}%
\AgdaSymbol{\{}\AgdaBound{m}\AgdaSymbol{\}}%
\>[21]\AgdaInductiveConstructor{n-0}\AgdaSpace{}%
\AgdaSymbol{=}\AgdaSpace{}%
\AgdaBound{m}\<%
\\
\>[2]\AgdaFunction{dy-args-dim}\AgdaSpace{}%
\AgdaSymbol{\{}\AgdaBound{m}\AgdaSymbol{\}\{}\AgdaBound{n}\AgdaSymbol{\}}\AgdaSpace{}%
\AgdaInductiveConstructor{0-n}\AgdaSpace{}%
\AgdaSymbol{=}\AgdaSpace{}%
\AgdaBound{n}\<%
\\
\\[\AgdaEmptyExtraSkip]%
\>[2]\AgdaFunction{dy-args-shp}\AgdaSpace{}%
\AgdaSymbol{\{}\AgdaBound{m}\AgdaSymbol{\}\{}\AgdaBound{n}\AgdaSymbol{\}\{}\AgdaBound{sx}\AgdaSymbol{\}}%
\>[29]\AgdaInductiveConstructor{n-n}\AgdaSpace{}%
\AgdaSymbol{=}\AgdaSpace{}%
\AgdaBound{sx}\<%
\\
\>[2]\AgdaFunction{dy-args-shp}\AgdaSpace{}%
\AgdaSymbol{\{}\AgdaBound{m}\AgdaSymbol{\}\{}\AgdaBound{n}\AgdaSymbol{\}\{}\AgdaBound{sx}\AgdaSymbol{\}}%
\>[29]\AgdaInductiveConstructor{n-0}\AgdaSpace{}%
\AgdaSymbol{=}\AgdaSpace{}%
\AgdaBound{sx}\<%
\\
\>[2]\AgdaFunction{dy-args-shp}\AgdaSpace{}%
\AgdaSymbol{\{}\AgdaBound{m}\AgdaSymbol{\}\{}\AgdaBound{n}\AgdaSymbol{\}\{}\AgdaBound{sx}\AgdaSymbol{\}\{}\AgdaBound{sy}\AgdaSymbol{\}}\AgdaSpace{}%
\AgdaInductiveConstructor{0-n}\AgdaSpace{}%
\AgdaSymbol{=}\AgdaSpace{}%
\AgdaBound{sy}\<%
\\
\\[\AgdaEmptyExtraSkip]%
\>[2]\AgdaFunction{dy-args-ok?}\AgdaSpace{}%
\AgdaBound{hole}\AgdaSpace{}%
\AgdaSymbol{=}\AgdaSpace{}%
\AgdaKeyword{do}\<%
\\
\>[2][@{}l@{\AgdaIndent{0}}]%
\>[4]\AgdaBound{goal}\AgdaSpace{}%
\AgdaOperator{\AgdaFunction{←}}\AgdaSpace{}%
\AgdaPostulate{inferType}\AgdaSpace{}%
\AgdaBound{hole}\<%
\\
\>[4]\AgdaInductiveConstructor{def}\AgdaSpace{}%
\AgdaSymbol{(}\AgdaKeyword{quote}\AgdaSpace{}%
\AgdaDatatype{dy-args}\AgdaSymbol{)}\AgdaSpace{}%
\AgdaSymbol{(}\AgdaInductiveConstructor{vArg}\AgdaSpace{}%
\AgdaBound{m}\AgdaSpace{}%
\AgdaOperator{\AgdaInductiveConstructor{∷}}\AgdaSpace{}%
\AgdaInductiveConstructor{vArg}\AgdaSpace{}%
\AgdaBound{n}\AgdaSpace{}%
\AgdaOperator{\AgdaInductiveConstructor{∷}}\AgdaSpace{}%
\AgdaInductiveConstructor{vArg}\AgdaSpace{}%
\AgdaBound{sx}\AgdaSpace{}%
\AgdaOperator{\AgdaInductiveConstructor{∷}}\AgdaSpace{}%
\AgdaInductiveConstructor{vArg}\AgdaSpace{}%
\AgdaBound{sy}\AgdaSpace{}%
\AgdaOperator{\AgdaInductiveConstructor{∷}}\AgdaSpace{}%
\AgdaInductiveConstructor{[]}\AgdaSymbol{)}\AgdaSpace{}%
\AgdaOperator{\AgdaFunction{←}}\AgdaSpace{}%
\AgdaPostulate{reduce}\AgdaSpace{}%
\AgdaBound{goal}\<%
\\
\>[4][@{}l@{\AgdaIndent{0}}]%
\>[6]\AgdaKeyword{where}\AgdaSpace{}%
\AgdaCatchallClause{\AgdaSymbol{\AgdaUnderscore{}}}\AgdaSpace{}%
\AgdaSymbol{→}\AgdaSpace{}%
\AgdaPostulate{typeError}\AgdaSpace{}%
\AgdaSymbol{(}\AgdaInductiveConstructor{strErr}\AgdaSpace{}%
\AgdaString{"expected\ dy-args\ expression\ in\ goal,\ found:"}\AgdaSpace{}%
\AgdaOperator{\AgdaInductiveConstructor{∷}}\AgdaSpace{}%
\AgdaInductiveConstructor{termErr}\AgdaSpace{}%
\AgdaBound{goal}\AgdaSpace{}%
\AgdaOperator{\AgdaInductiveConstructor{∷}}\AgdaSpace{}%
\AgdaInductiveConstructor{[]}\AgdaSymbol{)}\<%
\\
\>[4]\AgdaPostulate{reduce}\AgdaSpace{}%
\AgdaBound{m}\AgdaSpace{}%
\AgdaOperator{\AgdaFunction{>>=}}\AgdaSpace{}%
\AgdaSymbol{λ}\AgdaSpace{}%
\AgdaKeyword{where}\<%
\\
\>[4][@{}l@{\AgdaIndent{0}}]%
\>[6]\AgdaSymbol{(}\AgdaInductiveConstructor{lit}\AgdaSpace{}%
\AgdaSymbol{(}\AgdaInductiveConstructor{nat}\AgdaSpace{}%
\AgdaNumber{0}\AgdaSymbol{))}\AgdaSpace{}%
\AgdaSymbol{→}\AgdaSpace{}%
\AgdaPostulate{unify}\AgdaSpace{}%
\AgdaBound{hole}\AgdaSpace{}%
\AgdaSymbol{(}\AgdaInductiveConstructor{con}\AgdaSpace{}%
\AgdaSymbol{(}\AgdaKeyword{quote}\AgdaSpace{}%
\AgdaInductiveConstructor{0-n}\AgdaSymbol{)}\AgdaSpace{}%
\AgdaInductiveConstructor{[]}\AgdaSymbol{)}\<%
\\
\>[6]\AgdaSymbol{(}\AgdaInductiveConstructor{meta}\AgdaSpace{}%
\AgdaBound{id}\AgdaSpace{}%
\AgdaSymbol{\AgdaUnderscore{})}\AgdaSpace{}%
\AgdaSymbol{→}\AgdaSpace{}%
\AgdaPostulate{blockOnMeta}\AgdaSpace{}%
\AgdaBound{id}\<%
\\
\>[6]\AgdaCatchallClause{\AgdaBound{m}}%
\>[252I]\AgdaSymbol{→}\AgdaSpace{}%
\AgdaPostulate{reduce}\AgdaSpace{}%
\AgdaBound{n}\AgdaSpace{}%
\AgdaOperator{\AgdaFunction{>>=}}\AgdaSpace{}%
\AgdaSymbol{λ}\AgdaSpace{}%
\AgdaKeyword{where}\<%
\\
\>[.][@{}l@{}]\<[252I]%
\>[8]\AgdaSymbol{(}\AgdaInductiveConstructor{lit}\AgdaSpace{}%
\AgdaSymbol{(}\AgdaInductiveConstructor{nat}\AgdaSpace{}%
\AgdaNumber{0}\AgdaSymbol{))}\AgdaSpace{}%
\AgdaSymbol{→}\AgdaSpace{}%
\AgdaPostulate{unify}\AgdaSpace{}%
\AgdaBound{hole}\AgdaSpace{}%
\AgdaSymbol{(}\AgdaInductiveConstructor{con}\AgdaSpace{}%
\AgdaSymbol{(}\AgdaKeyword{quote}\AgdaSpace{}%
\AgdaInductiveConstructor{n-0}\AgdaSymbol{)}\AgdaSpace{}%
\AgdaInductiveConstructor{[]}\AgdaSymbol{)}\<%
\\
\>[8]\AgdaSymbol{(}\AgdaInductiveConstructor{meta}\AgdaSpace{}%
\AgdaBound{id}\AgdaSpace{}%
\AgdaSymbol{\AgdaUnderscore{})}\AgdaSpace{}%
\AgdaSymbol{→}\AgdaSpace{}%
\AgdaPostulate{blockOnMeta}\AgdaSpace{}%
\AgdaBound{id}\<%
\\
\>[8]\AgdaCatchallClause{\AgdaBound{n}}%
\>[272I]\AgdaSymbol{→}\AgdaSpace{}%
\AgdaKeyword{do}\<%
\\
\>[.][@{}l@{}]\<[272I]%
\>[10]\AgdaPostulate{catchTC}\<%
\\
\>[10][@{}l@{\AgdaIndent{0}}]%
\>[12]\AgdaSymbol{(}\AgdaPostulate{unify}\AgdaSpace{}%
\AgdaBound{m}\AgdaSpace{}%
\AgdaBound{n}\AgdaSymbol{)}\<%
\\
\>[12]\AgdaSymbol{(}\AgdaPostulate{typeError}\AgdaSpace{}%
\AgdaSymbol{(}\AgdaInductiveConstructor{strErr}\AgdaSpace{}%
\AgdaString{"no\ valid\ dy-args\ found\ for\ goal:\ "}\AgdaSpace{}%
\AgdaOperator{\AgdaInductiveConstructor{∷}}\AgdaSpace{}%
\AgdaInductiveConstructor{termErr}\AgdaSpace{}%
\AgdaBound{goal}\AgdaSpace{}%
\AgdaOperator{\AgdaInductiveConstructor{∷}}\AgdaSpace{}%
\AgdaInductiveConstructor{[]}\AgdaSymbol{))}\<%
\\
\>[10]\AgdaPostulate{unify}\AgdaSpace{}%
\AgdaBound{hole}\AgdaSpace{}%
\AgdaSymbol{(}\AgdaInductiveConstructor{con}\AgdaSpace{}%
\AgdaSymbol{(}\AgdaKeyword{quote}\AgdaSpace{}%
\AgdaInductiveConstructor{n-n}\AgdaSymbol{)}\AgdaSpace{}%
\AgdaInductiveConstructor{[]}\AgdaSymbol{)}\<%
\end{code}
\begin{code}%
\>[2]\AgdaFunction{dy-type}\AgdaSpace{}%
\AgdaSymbol{:}\AgdaSpace{}%
\AgdaSymbol{∀}\AgdaSpace{}%
\AgdaBound{a}\AgdaSpace{}%
\AgdaSymbol{→}\AgdaSpace{}%
\AgdaPrimitive{Set}\AgdaSpace{}%
\AgdaBound{a}\AgdaSpace{}%
\AgdaSymbol{→}\AgdaSpace{}%
\AgdaPrimitive{Set}\AgdaSpace{}%
\AgdaBound{a}\<%
\\
\>[2]\AgdaFunction{dy-type}\AgdaSpace{}%
\AgdaBound{a}\AgdaSpace{}%
\AgdaBound{X}\AgdaSpace{}%
\AgdaSymbol{=}%
\>[300I]\AgdaSymbol{∀}\AgdaSpace{}%
\AgdaSymbol{\{}\AgdaBound{m}\AgdaSpace{}%
\AgdaBound{n}\AgdaSpace{}%
\AgdaBound{sx}\AgdaSpace{}%
\AgdaBound{sy}\AgdaSymbol{\}}\AgdaSpace{}%
\AgdaSymbol{\{@(}\AgdaKeyword{tactic}\AgdaSpace{}%
\AgdaFunction{dy-args-ok?}\AgdaSymbol{)}\AgdaSpace{}%
\AgdaBound{args}\AgdaSpace{}%
\AgdaSymbol{:}\AgdaSpace{}%
\AgdaDatatype{dy-args}\AgdaSpace{}%
\AgdaBound{m}\AgdaSpace{}%
\AgdaBound{n}\AgdaSpace{}%
\AgdaBound{sx}\AgdaSpace{}%
\AgdaBound{sy}\AgdaSymbol{\}}\<%
\\
\>[.][@{}l@{}]\<[300I]%
\>[16]\AgdaSymbol{→}\AgdaSpace{}%
\AgdaRecord{Ar}\AgdaSpace{}%
\AgdaBound{X}\AgdaSpace{}%
\AgdaBound{m}\AgdaSpace{}%
\AgdaBound{sx}\AgdaSpace{}%
\AgdaSymbol{→}\AgdaSpace{}%
\AgdaRecord{Ar}\AgdaSpace{}%
\AgdaBound{X}\AgdaSpace{}%
\AgdaBound{n}\AgdaSpace{}%
\AgdaBound{sy}\AgdaSpace{}%
\AgdaSymbol{→}\AgdaSpace{}%
\AgdaRecord{Ar}\AgdaSpace{}%
\AgdaBound{X}\AgdaSpace{}%
\AgdaSymbol{\AgdaUnderscore{}}\AgdaSpace{}%
\AgdaSymbol{(}\AgdaFunction{dy-args-shp}\AgdaSpace{}%
\AgdaBound{args}\AgdaSymbol{)}\<%
\\
\\[\AgdaEmptyExtraSkip]%
\>[2]\AgdaFunction{lift′}\AgdaSpace{}%
\AgdaSymbol{:}\AgdaSpace{}%
\AgdaSymbol{∀}\AgdaSpace{}%
\AgdaSymbol{\{}\AgdaBound{a}\AgdaSymbol{\}\{}\AgdaBound{X}\AgdaSpace{}%
\AgdaSymbol{:}\AgdaSpace{}%
\AgdaPrimitive{Set}\AgdaSpace{}%
\AgdaBound{a}\AgdaSymbol{\}}\AgdaSpace{}%
\AgdaSymbol{→}\AgdaSpace{}%
\AgdaSymbol{(}\AgdaOperator{\AgdaBound{\AgdaUnderscore{}⊕\AgdaUnderscore{}}}\AgdaSpace{}%
\AgdaSymbol{:}\AgdaSpace{}%
\AgdaBound{X}\AgdaSpace{}%
\AgdaSymbol{→}\AgdaSpace{}%
\AgdaBound{X}\AgdaSpace{}%
\AgdaSymbol{→}\AgdaSpace{}%
\AgdaBound{X}\AgdaSymbol{)}\AgdaSpace{}%
\AgdaSymbol{→}\AgdaSpace{}%
\AgdaFunction{dy-type}\AgdaSpace{}%
\AgdaBound{a}\AgdaSpace{}%
\AgdaBound{X}\<%
\\
\>[2]\AgdaFunction{lift′}\AgdaSpace{}%
\AgdaOperator{\AgdaBound{\AgdaUnderscore{}⊕\AgdaUnderscore{}}}\AgdaSpace{}%
\AgdaSymbol{\{}\AgdaArgument{args}\AgdaSpace{}%
\AgdaSymbol{=}\AgdaSpace{}%
\AgdaInductiveConstructor{n-n}\AgdaSymbol{\}}\AgdaSpace{}%
\AgdaSymbol{(}\AgdaInductiveConstructor{imap}\AgdaSpace{}%
\AgdaBound{a}\AgdaSymbol{)}\AgdaSpace{}%
\AgdaSymbol{(}\AgdaInductiveConstructor{imap}\AgdaSpace{}%
\AgdaBound{b}\AgdaSymbol{)}\AgdaSpace{}%
\AgdaSymbol{=}\AgdaSpace{}%
\AgdaInductiveConstructor{imap}\AgdaSpace{}%
\AgdaSymbol{λ}\AgdaSpace{}%
\AgdaBound{iv}\AgdaSpace{}%
\AgdaSymbol{→}\AgdaSpace{}%
\AgdaBound{a}\AgdaSpace{}%
\AgdaBound{iv}\AgdaSpace{}%
\AgdaOperator{\AgdaBound{⊕}}\AgdaSpace{}%
\AgdaBound{b}\AgdaSpace{}%
\AgdaBound{iv}\<%
\\
\>[2]\AgdaFunction{lift′}\AgdaSpace{}%
\AgdaOperator{\AgdaBound{\AgdaUnderscore{}⊕\AgdaUnderscore{}}}\AgdaSpace{}%
\AgdaSymbol{\{}\AgdaArgument{args}\AgdaSpace{}%
\AgdaSymbol{=}\AgdaSpace{}%
\AgdaInductiveConstructor{n-0}\AgdaSymbol{\}}\AgdaSpace{}%
\AgdaSymbol{(}\AgdaInductiveConstructor{imap}\AgdaSpace{}%
\AgdaBound{a}\AgdaSymbol{)}\AgdaSpace{}%
\AgdaSymbol{(}\AgdaInductiveConstructor{imap}\AgdaSpace{}%
\AgdaBound{b}\AgdaSymbol{)}\AgdaSpace{}%
\AgdaSymbol{=}\AgdaSpace{}%
\AgdaInductiveConstructor{imap}\AgdaSpace{}%
\AgdaSymbol{λ}\AgdaSpace{}%
\AgdaBound{iv}\AgdaSpace{}%
\AgdaSymbol{→}\AgdaSpace{}%
\AgdaBound{a}\AgdaSpace{}%
\AgdaBound{iv}\AgdaSpace{}%
\AgdaOperator{\AgdaBound{⊕}}\AgdaSpace{}%
\AgdaBound{b}\AgdaSpace{}%
\AgdaInductiveConstructor{[]}\<%
\\
\>[2]\AgdaFunction{lift′}\AgdaSpace{}%
\AgdaOperator{\AgdaBound{\AgdaUnderscore{}⊕\AgdaUnderscore{}}}\AgdaSpace{}%
\AgdaSymbol{\{}\AgdaArgument{args}\AgdaSpace{}%
\AgdaSymbol{=}\AgdaSpace{}%
\AgdaInductiveConstructor{0-n}\AgdaSymbol{\}}\AgdaSpace{}%
\AgdaSymbol{(}\AgdaInductiveConstructor{imap}\AgdaSpace{}%
\AgdaBound{a}\AgdaSymbol{)}\AgdaSpace{}%
\AgdaSymbol{(}\AgdaInductiveConstructor{imap}\AgdaSpace{}%
\AgdaBound{b}\AgdaSymbol{)}\AgdaSpace{}%
\AgdaSymbol{=}\AgdaSpace{}%
\AgdaInductiveConstructor{imap}\AgdaSpace{}%
\AgdaSymbol{λ}\AgdaSpace{}%
\AgdaBound{iv}\AgdaSpace{}%
\AgdaSymbol{→}\AgdaSpace{}%
\AgdaBound{a}\AgdaSpace{}%
\AgdaInductiveConstructor{[]}\AgdaSpace{}%
\AgdaOperator{\AgdaBound{⊕}}\AgdaSpace{}%
\AgdaBound{b}\AgdaSpace{}%
\AgdaBound{iv}\<%
\end{code}
We define the \AF{dy-args-dim} and \AF{dy-args-shp} to pick the largest
rank and shape from the arguments that are related by \AF{dy-args}.
The \AF{lift′}
function itself turns any binary operation on array elements
into a binary operation on arrays that replicates scalars correctly.
Here we demonstrate the lifting \AF{\_+\_} for natural numbers.
\begin{code}[hide]%
\>[2]\AgdaFunction{s}\AgdaSpace{}%
\AgdaSymbol{:}\AgdaSpace{}%
\AgdaDatatype{Vec}\AgdaSpace{}%
\AgdaDatatype{ℕ}\AgdaSpace{}%
\AgdaNumber{3}\<%
\\
\>[2]\AgdaFunction{s}\AgdaSpace{}%
\AgdaSymbol{=}\AgdaSpace{}%
\AgdaNumber{1}\AgdaSpace{}%
\AgdaOperator{\AgdaInductiveConstructor{∷}}\AgdaSpace{}%
\AgdaNumber{2}\AgdaSpace{}%
\AgdaOperator{\AgdaInductiveConstructor{∷}}\AgdaSpace{}%
\AgdaNumber{3}\AgdaSpace{}%
\AgdaOperator{\AgdaInductiveConstructor{∷}}\AgdaSpace{}%
\AgdaInductiveConstructor{[]}\<%
\end{code}

\begin{mathpar}
\codeblock{
\begin{code}%
\>[2]\AgdaOperator{\AgdaFunction{\AgdaUnderscore{}+\AgdaUnderscore{}}}\AgdaSpace{}%
\AgdaSymbol{=}\AgdaSpace{}%
\AgdaFunction{lift′}\AgdaSpace{}%
\AgdaOperator{\AgdaPrimitive{ℕ.\AgdaUnderscore{}+\AgdaUnderscore{}}}\<%
\\
\>[2]\AgdaFunction{a}\AgdaSpace{}%
\AgdaSymbol{:}\AgdaSpace{}%
\AgdaRecord{Ar}\AgdaSpace{}%
\AgdaDatatype{ℕ}\AgdaSpace{}%
\AgdaNumber{3}\AgdaSpace{}%
\AgdaFunction{s}\<%
\\
\>[2]\AgdaFunction{z}\AgdaSpace{}%
\AgdaSymbol{:}\AgdaSpace{}%
\AgdaRecord{Ar}\AgdaSpace{}%
\AgdaDatatype{ℕ}\AgdaSpace{}%
\AgdaNumber{0}\AgdaSpace{}%
\AgdaInductiveConstructor{[]}\<%
\end{code}
}
\and
\codeblock{
\begin{code}%
\>[2]\AgdaFunction{ex₁}\AgdaSpace{}%
\AgdaFunction{ex₂}\AgdaSpace{}%
\AgdaSymbol{:}\AgdaSpace{}%
\AgdaRecord{Ar}\AgdaSpace{}%
\AgdaDatatype{ℕ}\AgdaSpace{}%
\AgdaNumber{3}\AgdaSpace{}%
\AgdaFunction{s}\<%
\\
\>[2]\AgdaFunction{ex₁}\AgdaSpace{}%
\AgdaSymbol{=}\AgdaSpace{}%
\AgdaFunction{a}\AgdaSpace{}%
\AgdaOperator{\AgdaFunction{+}}\AgdaSpace{}%
\AgdaFunction{z}\<%
\\
\>[2]\AgdaFunction{ex₂}\AgdaSpace{}%
\AgdaSymbol{=}\AgdaSpace{}%
\AgdaFunction{a}\AgdaSpace{}%
\AgdaOperator{\AgdaFunction{+}}\AgdaSpace{}%
\AgdaFunction{a}\<%
\end{code}
}
\and
\codeblock{
\begin{code}%
\>[2]\AgdaFunction{ex₃}\AgdaSpace{}%
\AgdaSymbol{:}\AgdaSpace{}%
\AgdaRecord{Ar}\AgdaSpace{}%
\AgdaDatatype{ℕ}\AgdaSpace{}%
\AgdaNumber{0}\AgdaSpace{}%
\AgdaInductiveConstructor{[]}\<%
\\
\>[2]\AgdaFunction{ex₃}\AgdaSpace{}%
\AgdaSymbol{=}\AgdaSpace{}%
\AgdaFunction{z}\AgdaSpace{}%
\AgdaOperator{\AgdaFunction{+}}\AgdaSpace{}%
\AgdaFunction{z}\<%
\end{code}
}
\and
\codeblock{
\begin{code}%
\>[2]\AgdaFunction{ex₄}%
\>[449I]\AgdaFunction{ex₅}\AgdaSpace{}%
\AgdaFunction{ex₆}\AgdaSpace{}%
\AgdaSymbol{:}\AgdaSpace{}%
\AgdaSymbol{∀}\AgdaSpace{}%
\AgdaSymbol{\{}\AgdaBound{n}\AgdaSpace{}%
\AgdaBound{s}\AgdaSymbol{\}}\<%
\\
\>[.][@{}l@{}]\<[449I]%
\>[6]\AgdaSymbol{→}\AgdaSpace{}%
\AgdaRecord{Ar}\AgdaSpace{}%
\AgdaDatatype{ℕ}\AgdaSpace{}%
\AgdaBound{n}\AgdaSpace{}%
\AgdaBound{s}\<%
\\
\>[6]\AgdaSymbol{→}\AgdaSpace{}%
\AgdaRecord{Ar}\AgdaSpace{}%
\AgdaDatatype{ℕ}\AgdaSpace{}%
\AgdaBound{n}\AgdaSpace{}%
\AgdaBound{s}\<%
\end{code}
}
\and
\codeblock{
\begin{code}%
\>[2]\AgdaFunction{ex₄}\AgdaSpace{}%
\AgdaBound{x}\AgdaSpace{}%
\AgdaSymbol{=}\AgdaSpace{}%
\AgdaBound{x}\AgdaSpace{}%
\AgdaOperator{\AgdaFunction{+}}\AgdaSpace{}%
\AgdaBound{x}\<%
\\
\>[2]\AgdaFunction{ex₅}\AgdaSpace{}%
\AgdaBound{x}\AgdaSpace{}%
\AgdaSymbol{=}\AgdaSpace{}%
\AgdaBound{x}\AgdaSpace{}%
\AgdaOperator{\AgdaFunction{+}}\AgdaSpace{}%
\AgdaFunction{z}\<%
\\
\>[2]\AgdaFunction{ex₆}\AgdaSpace{}%
\AgdaBound{x}\AgdaSpace{}%
\AgdaSymbol{=}\AgdaSpace{}%
\AgdaFunction{z}\AgdaSpace{}%
\AgdaOperator{\AgdaFunction{+}}\AgdaSpace{}%
\AgdaBound{x}\<%
\end{code}
}
\end{mathpar}
\begin{code}[hide]%
\>[2]\AgdaFunction{a}\AgdaSpace{}%
\AgdaSymbol{=}\AgdaSpace{}%
\AgdaInductiveConstructor{imap}\AgdaSpace{}%
\AgdaSymbol{λ}\AgdaSpace{}%
\AgdaBound{iv}\AgdaSpace{}%
\AgdaSymbol{→}\AgdaSpace{}%
\AgdaNumber{10}\<%
\\
\>[2]\AgdaFunction{z}\AgdaSpace{}%
\AgdaSymbol{=}\AgdaSpace{}%
\AgdaInductiveConstructor{imap}\AgdaSpace{}%
\AgdaSymbol{λ}\AgdaSpace{}%
\AgdaBound{iv}\AgdaSpace{}%
\AgdaSymbol{→}\AgdaSpace{}%
\AgdaNumber{20}\<%
\end{code}
In this example, \AF{a} is a 3-d array, and \AF{z} is a
scalar.  The lifted addition on arrays admits all the
desired variants.  The last three examples on the right show that it
still works for the cases when the rank is not known statically.

\begin{code}[hide]%
\>[2]\AgdaKeyword{data}\AgdaSpace{}%
\AgdaDatatype{SVup}\AgdaSpace{}%
\AgdaSymbol{(}\AgdaBound{X}\AgdaSpace{}%
\AgdaSymbol{:}\AgdaSpace{}%
\AgdaPrimitive{Set}\AgdaSymbol{)}\AgdaSpace{}%
\AgdaSymbol{:}\AgdaSpace{}%
\AgdaPrimitive{Set}\AgdaSpace{}%
\AgdaSymbol{→}\AgdaSpace{}%
\AgdaSymbol{(}\AgdaBound{d}\AgdaSpace{}%
\AgdaSymbol{:}\AgdaSpace{}%
\AgdaDatatype{ℕ}\AgdaSymbol{)}\AgdaSpace{}%
\AgdaSymbol{→}\AgdaSpace{}%
\AgdaSymbol{(}\AgdaBound{sh}\AgdaSpace{}%
\AgdaSymbol{:}\AgdaSpace{}%
\AgdaDatatype{Vec}\AgdaSpace{}%
\AgdaDatatype{ℕ}\AgdaSpace{}%
\AgdaBound{d}\AgdaSymbol{)}\AgdaSpace{}%
\AgdaSymbol{→}\AgdaSpace{}%
\AgdaPrimitive{Set}\AgdaSpace{}%
\AgdaKeyword{where}\<%
\\
\>[2][@{}l@{\AgdaIndent{0}}]%
\>[4]\AgdaKeyword{instance}\<%
\\
\>[4][@{}l@{\AgdaIndent{0}}]%
\>[6]\AgdaInductiveConstructor{scal}\AgdaSpace{}%
\AgdaSymbol{:}\AgdaSpace{}%
\AgdaDatatype{SVup}\AgdaSpace{}%
\AgdaBound{X}\AgdaSpace{}%
\AgdaBound{X}\AgdaSpace{}%
\AgdaNumber{0}\AgdaSpace{}%
\AgdaInductiveConstructor{[]}\<%
\\
\>[6]\AgdaInductiveConstructor{vect}\AgdaSpace{}%
\AgdaSymbol{:}\AgdaSpace{}%
\AgdaSymbol{∀}\AgdaSpace{}%
\AgdaSymbol{\{}\AgdaBound{n}\AgdaSymbol{\}}\AgdaSpace{}%
\AgdaSymbol{→}\AgdaSpace{}%
\AgdaDatatype{SVup}\AgdaSpace{}%
\AgdaBound{X}\AgdaSpace{}%
\AgdaSymbol{(}\AgdaDatatype{Vec}\AgdaSpace{}%
\AgdaBound{X}\AgdaSpace{}%
\AgdaBound{n}\AgdaSymbol{)}\AgdaSpace{}%
\AgdaNumber{1}\AgdaSpace{}%
\AgdaSymbol{(}\AgdaBound{n}\AgdaSpace{}%
\AgdaOperator{\AgdaInductiveConstructor{∷}}\AgdaSpace{}%
\AgdaInductiveConstructor{[]}\AgdaSymbol{)}\<%
\\
\>[6]\AgdaInductiveConstructor{arry}\AgdaSpace{}%
\AgdaSymbol{:}\AgdaSpace{}%
\AgdaSymbol{∀}\AgdaSpace{}%
\AgdaSymbol{\{}\AgdaBound{d}\AgdaSpace{}%
\AgdaBound{s}\AgdaSymbol{\}}\AgdaSpace{}%
\AgdaSymbol{→}\AgdaSpace{}%
\AgdaDatatype{SVup}\AgdaSpace{}%
\AgdaBound{X}\AgdaSpace{}%
\AgdaSymbol{(}\AgdaRecord{Ar}\AgdaSpace{}%
\AgdaBound{X}\AgdaSpace{}%
\AgdaBound{d}\AgdaSpace{}%
\AgdaBound{s}\AgdaSymbol{)}\AgdaSpace{}%
\AgdaBound{d}\AgdaSpace{}%
\AgdaBound{s}\<%
\\
\\[\AgdaEmptyExtraSkip]%
\>[2]\AgdaFunction{cst}\AgdaSpace{}%
\AgdaSymbol{:}\AgdaSpace{}%
\AgdaSymbol{∀}\AgdaSpace{}%
\AgdaSymbol{\{}\AgdaBound{a}\AgdaSymbol{\}\{}\AgdaBound{X}\AgdaSpace{}%
\AgdaSymbol{:}\AgdaSpace{}%
\AgdaPrimitive{Set}\AgdaSpace{}%
\AgdaBound{a}\AgdaSymbol{\}\{}\AgdaBound{d}\AgdaSpace{}%
\AgdaBound{s}\AgdaSymbol{\}}\AgdaSpace{}%
\AgdaSymbol{→}\AgdaSpace{}%
\AgdaBound{X}\AgdaSpace{}%
\AgdaSymbol{→}\AgdaSpace{}%
\AgdaRecord{Ar}\AgdaSpace{}%
\AgdaBound{X}\AgdaSpace{}%
\AgdaBound{d}\AgdaSpace{}%
\AgdaBound{s}\<%
\\
\>[2]\AgdaFunction{cst}\AgdaSpace{}%
\AgdaBound{x}\AgdaSpace{}%
\AgdaSymbol{=}\AgdaSpace{}%
\AgdaInductiveConstructor{imap}\AgdaSpace{}%
\AgdaSymbol{λ}\AgdaSpace{}%
\AgdaBound{\AgdaUnderscore{}}\AgdaSpace{}%
\AgdaSymbol{→}\AgdaSpace{}%
\AgdaBound{x}\<%
\\
\\[\AgdaEmptyExtraSkip]%
\>[2]\AgdaKeyword{infixr}\AgdaSpace{}%
\AgdaNumber{30}\AgdaSpace{}%
\AgdaOperator{\AgdaFunction{▴\AgdaUnderscore{}}}\<%
\\
\>[2]\AgdaOperator{\AgdaFunction{▴\AgdaUnderscore{}}}\AgdaSpace{}%
\AgdaSymbol{:}\AgdaSpace{}%
\AgdaSymbol{∀}\AgdaSpace{}%
\AgdaSymbol{\{}\AgdaBound{X}\AgdaSpace{}%
\AgdaBound{A}\AgdaSpace{}%
\AgdaBound{d}\AgdaSpace{}%
\AgdaBound{s}\AgdaSymbol{\}\{\{}\AgdaBound{t}\AgdaSpace{}%
\AgdaSymbol{:}\AgdaSpace{}%
\AgdaDatatype{SVup}\AgdaSpace{}%
\AgdaBound{X}\AgdaSpace{}%
\AgdaBound{A}\AgdaSpace{}%
\AgdaBound{d}\AgdaSpace{}%
\AgdaBound{s}\AgdaSymbol{\}\}}\AgdaSpace{}%
\AgdaSymbol{→}\AgdaSpace{}%
\AgdaBound{A}\AgdaSpace{}%
\AgdaSymbol{→}\AgdaSpace{}%
\AgdaRecord{Ar}\AgdaSpace{}%
\AgdaBound{X}\AgdaSpace{}%
\AgdaBound{d}\AgdaSpace{}%
\AgdaBound{s}\<%
\\
\>[2]\AgdaOperator{\AgdaFunction{▴\AgdaUnderscore{}}}\AgdaSpace{}%
\AgdaSymbol{\{\{}\AgdaSpace{}%
\AgdaArgument{t}\AgdaSpace{}%
\AgdaSymbol{=}\AgdaSpace{}%
\AgdaInductiveConstructor{scal}\AgdaSpace{}%
\AgdaSymbol{\}\}}\AgdaSpace{}%
\AgdaBound{a}\AgdaSpace{}%
\AgdaSymbol{=}\AgdaSpace{}%
\AgdaFunction{cst}\AgdaSpace{}%
\AgdaBound{a}\AgdaSpace{}%
\AgdaComment{--imap\ λ\ \AgdaUnderscore{}\ →\ a}\<%
\\
\>[2]\AgdaOperator{\AgdaFunction{▴\AgdaUnderscore{}}}\AgdaSpace{}%
\AgdaSymbol{\{\{}\AgdaSpace{}%
\AgdaArgument{t}\AgdaSpace{}%
\AgdaSymbol{=}\AgdaSpace{}%
\AgdaInductiveConstructor{vect}\AgdaSpace{}%
\AgdaSymbol{\}\}}\AgdaSpace{}%
\AgdaBound{a}\AgdaSpace{}%
\AgdaSymbol{=}\AgdaSpace{}%
\AgdaInductiveConstructor{imap}\AgdaSpace{}%
\AgdaSymbol{λ}\AgdaSpace{}%
\AgdaKeyword{where}\AgdaSpace{}%
\AgdaSymbol{(}\AgdaBound{i}\AgdaSpace{}%
\AgdaOperator{\AgdaInductiveConstructor{∷}}\AgdaSpace{}%
\AgdaInductiveConstructor{[]}\AgdaSymbol{)}\AgdaSpace{}%
\AgdaSymbol{→}\AgdaSpace{}%
\AgdaFunction{V.lookup}\AgdaSpace{}%
\AgdaBound{a}\AgdaSpace{}%
\AgdaBound{i}\<%
\\
\>[2]\AgdaOperator{\AgdaFunction{▴\AgdaUnderscore{}}}\AgdaSpace{}%
\AgdaSymbol{\{\{}\AgdaSpace{}%
\AgdaArgument{t}\AgdaSpace{}%
\AgdaSymbol{=}\AgdaSpace{}%
\AgdaInductiveConstructor{arry}\AgdaSpace{}%
\AgdaSymbol{\}\}}\AgdaSpace{}%
\AgdaBound{a}\AgdaSpace{}%
\AgdaSymbol{=}\AgdaSpace{}%
\AgdaBound{a}\<%
\\
\\[\AgdaEmptyExtraSkip]%
\>[2]\AgdaKeyword{infixr}\AgdaSpace{}%
\AgdaNumber{30}\AgdaSpace{}%
\AgdaOperator{\AgdaFunction{▾\AgdaUnderscore{}}}\<%
\\
\>[2]\AgdaOperator{\AgdaFunction{▾\AgdaUnderscore{}}}\AgdaSpace{}%
\AgdaSymbol{:}\AgdaSpace{}%
\AgdaSymbol{∀}\AgdaSpace{}%
\AgdaSymbol{\{}\AgdaBound{X}\AgdaSpace{}%
\AgdaBound{A}\AgdaSpace{}%
\AgdaBound{d}\AgdaSpace{}%
\AgdaBound{s}\AgdaSymbol{\}\{\{}\AgdaBound{t}\AgdaSpace{}%
\AgdaSymbol{:}\AgdaSpace{}%
\AgdaDatatype{SVup}\AgdaSpace{}%
\AgdaBound{X}\AgdaSpace{}%
\AgdaBound{A}\AgdaSpace{}%
\AgdaBound{d}\AgdaSpace{}%
\AgdaBound{s}\AgdaSymbol{\}\}}\AgdaSpace{}%
\AgdaSymbol{→}\AgdaSpace{}%
\AgdaRecord{Ar}\AgdaSpace{}%
\AgdaBound{X}\AgdaSpace{}%
\AgdaBound{d}\AgdaSpace{}%
\AgdaBound{s}\AgdaSpace{}%
\AgdaSymbol{→}\AgdaSpace{}%
\AgdaBound{A}\<%
\\
\>[2]\AgdaOperator{\AgdaFunction{▾\AgdaUnderscore{}}}\AgdaSpace{}%
\AgdaSymbol{\{\{}\AgdaSpace{}%
\AgdaArgument{t}\AgdaSpace{}%
\AgdaSymbol{=}\AgdaSpace{}%
\AgdaInductiveConstructor{scal}\AgdaSpace{}%
\AgdaSymbol{\}\}}\AgdaSpace{}%
\AgdaSymbol{(}\AgdaInductiveConstructor{imap}\AgdaSpace{}%
\AgdaBound{a}\AgdaSymbol{)}\AgdaSpace{}%
\AgdaSymbol{=}\AgdaSpace{}%
\AgdaBound{a}\AgdaSpace{}%
\AgdaInductiveConstructor{[]}\<%
\\
\>[2]\AgdaOperator{\AgdaFunction{▾\AgdaUnderscore{}}}\AgdaSpace{}%
\AgdaSymbol{\{\{}\AgdaSpace{}%
\AgdaArgument{t}\AgdaSpace{}%
\AgdaSymbol{=}\AgdaSpace{}%
\AgdaInductiveConstructor{vect}\AgdaSpace{}%
\AgdaSymbol{\}\}}\AgdaSpace{}%
\AgdaSymbol{(}\AgdaInductiveConstructor{imap}\AgdaSpace{}%
\AgdaBound{a}\AgdaSymbol{)}\AgdaSpace{}%
\AgdaSymbol{=}\AgdaSpace{}%
\AgdaFunction{V.tabulate}\AgdaSpace{}%
\AgdaSymbol{λ}\AgdaSpace{}%
\AgdaBound{i}\AgdaSpace{}%
\AgdaSymbol{→}\AgdaSpace{}%
\AgdaBound{a}\AgdaSpace{}%
\AgdaOperator{\AgdaFunction{\$}}\AgdaSpace{}%
\AgdaBound{i}\AgdaSpace{}%
\AgdaOperator{\AgdaInductiveConstructor{∷}}\AgdaSpace{}%
\AgdaInductiveConstructor{[]}\<%
\\
\>[2]\AgdaOperator{\AgdaFunction{▾\AgdaUnderscore{}}}\AgdaSpace{}%
\AgdaSymbol{\{\{}\AgdaSpace{}%
\AgdaArgument{t}\AgdaSpace{}%
\AgdaSymbol{=}\AgdaSpace{}%
\AgdaInductiveConstructor{arry}\AgdaSpace{}%
\AgdaSymbol{\}\}}\AgdaSpace{}%
\AgdaBound{a}\AgdaSpace{}%
\AgdaSymbol{=}\AgdaSpace{}%
\AgdaBound{a}\<%
\\
\\[\AgdaEmptyExtraSkip]%
\>[2]\AgdaKeyword{data}\AgdaSpace{}%
\AgdaDatatype{DyScalVec}\AgdaSpace{}%
\AgdaSymbol{(}\AgdaBound{X}\AgdaSpace{}%
\AgdaSymbol{:}\AgdaSpace{}%
\AgdaPrimitive{Set}\AgdaSymbol{)}\AgdaSpace{}%
\AgdaSymbol{:}\AgdaSpace{}%
\AgdaPrimitive{Set}\AgdaSpace{}%
\AgdaSymbol{→}\AgdaSpace{}%
\AgdaPrimitive{Set}\AgdaSpace{}%
\AgdaSymbol{→}\AgdaSpace{}%
\AgdaSymbol{(}\AgdaBound{d}\AgdaSpace{}%
\AgdaSymbol{:}\AgdaSpace{}%
\AgdaDatatype{ℕ}\AgdaSymbol{)}\AgdaSpace{}%
\AgdaSymbol{→}\AgdaSpace{}%
\AgdaSymbol{(}\AgdaBound{sh}\AgdaSpace{}%
\AgdaSymbol{:}\AgdaSpace{}%
\AgdaDatatype{Vec}\AgdaSpace{}%
\AgdaDatatype{ℕ}\AgdaSpace{}%
\AgdaBound{d}\AgdaSymbol{)}\AgdaSpace{}%
\AgdaSymbol{→}\AgdaSpace{}%
\AgdaPrimitive{Set}\AgdaSpace{}%
\AgdaKeyword{where}\<%
\\
\>[2][@{}l@{\AgdaIndent{0}}]%
\>[4]\AgdaKeyword{instance}\<%
\\
\>[4][@{}l@{\AgdaIndent{0}}]%
\>[6]\AgdaInductiveConstructor{s-s}\AgdaSpace{}%
\AgdaSymbol{:}%
\>[22]\AgdaDatatype{DyScalVec}\AgdaSpace{}%
\AgdaBound{X}\AgdaSpace{}%
\AgdaBound{X}\AgdaSpace{}%
\AgdaBound{X}\AgdaSpace{}%
\AgdaNumber{0}\AgdaSpace{}%
\AgdaInductiveConstructor{[]}\<%
\\
\>[6]\AgdaInductiveConstructor{s-v}\AgdaSpace{}%
\AgdaSymbol{:}\AgdaSpace{}%
\AgdaSymbol{∀}\AgdaSpace{}%
\AgdaSymbol{\{}\AgdaBound{n}\AgdaSymbol{\}}\AgdaSpace{}%
\AgdaSymbol{→}%
\>[22]\AgdaDatatype{DyScalVec}\AgdaSpace{}%
\AgdaBound{X}\AgdaSpace{}%
\AgdaBound{X}\AgdaSpace{}%
\AgdaSymbol{(}\AgdaDatatype{Vec}\AgdaSpace{}%
\AgdaBound{X}\AgdaSpace{}%
\AgdaBound{n}\AgdaSymbol{)}\AgdaSpace{}%
\AgdaNumber{1}\AgdaSpace{}%
\AgdaSymbol{(}\AgdaBound{n}\AgdaSpace{}%
\AgdaOperator{\AgdaInductiveConstructor{∷}}\AgdaSpace{}%
\AgdaInductiveConstructor{[]}\AgdaSymbol{)}\<%
\\
\>[6]\AgdaInductiveConstructor{s-a}\AgdaSpace{}%
\AgdaSymbol{:}\AgdaSpace{}%
\AgdaSymbol{∀}\AgdaSpace{}%
\AgdaSymbol{\{}\AgdaBound{d}\AgdaSpace{}%
\AgdaBound{s}\AgdaSymbol{\}}\AgdaSpace{}%
\AgdaSymbol{→}\AgdaSpace{}%
\AgdaDatatype{DyScalVec}\AgdaSpace{}%
\AgdaBound{X}\AgdaSpace{}%
\AgdaBound{X}\AgdaSpace{}%
\AgdaSymbol{(}\AgdaRecord{Ar}\AgdaSpace{}%
\AgdaBound{X}\AgdaSpace{}%
\AgdaBound{d}\AgdaSpace{}%
\AgdaBound{s}\AgdaSymbol{)}\AgdaSpace{}%
\AgdaBound{d}\AgdaSpace{}%
\AgdaBound{s}\<%
\\
\>[6]\AgdaInductiveConstructor{v-s}\AgdaSpace{}%
\AgdaSymbol{:}\AgdaSpace{}%
\AgdaSymbol{∀}\AgdaSpace{}%
\AgdaSymbol{\{}\AgdaBound{n}\AgdaSymbol{\}}\AgdaSpace{}%
\AgdaSymbol{→}%
\>[22]\AgdaDatatype{DyScalVec}\AgdaSpace{}%
\AgdaBound{X}\AgdaSpace{}%
\AgdaSymbol{(}\AgdaDatatype{Vec}\AgdaSpace{}%
\AgdaBound{X}\AgdaSpace{}%
\AgdaBound{n}\AgdaSymbol{)}\AgdaSpace{}%
\AgdaBound{X}\AgdaSpace{}%
\AgdaNumber{1}\AgdaSpace{}%
\AgdaSymbol{(}\AgdaBound{n}\AgdaSpace{}%
\AgdaOperator{\AgdaInductiveConstructor{∷}}\AgdaSpace{}%
\AgdaInductiveConstructor{[]}\AgdaSymbol{)}\<%
\\
\>[6]\AgdaInductiveConstructor{v-v}\AgdaSpace{}%
\AgdaSymbol{:}\AgdaSpace{}%
\AgdaSymbol{∀}\AgdaSpace{}%
\AgdaSymbol{\{}\AgdaBound{n}\AgdaSymbol{\}}\AgdaSpace{}%
\AgdaSymbol{→}%
\>[22]\AgdaDatatype{DyScalVec}\AgdaSpace{}%
\AgdaBound{X}\AgdaSpace{}%
\AgdaSymbol{(}\AgdaDatatype{Vec}\AgdaSpace{}%
\AgdaBound{X}\AgdaSpace{}%
\AgdaBound{n}\AgdaSymbol{)}\AgdaSpace{}%
\AgdaSymbol{(}\AgdaDatatype{Vec}\AgdaSpace{}%
\AgdaBound{X}\AgdaSpace{}%
\AgdaBound{n}\AgdaSymbol{)}\AgdaSpace{}%
\AgdaNumber{1}\AgdaSpace{}%
\AgdaSymbol{(}\AgdaBound{n}\AgdaSpace{}%
\AgdaOperator{\AgdaInductiveConstructor{∷}}\AgdaSpace{}%
\AgdaInductiveConstructor{[]}\AgdaSymbol{)}\<%
\\
\>[6]\AgdaInductiveConstructor{a-s}\AgdaSpace{}%
\AgdaSymbol{:}\AgdaSpace{}%
\AgdaSymbol{∀}\AgdaSpace{}%
\AgdaSymbol{\{}\AgdaBound{d}\AgdaSpace{}%
\AgdaBound{s}\AgdaSymbol{\}}\AgdaSpace{}%
\AgdaSymbol{→}\AgdaSpace{}%
\AgdaDatatype{DyScalVec}\AgdaSpace{}%
\AgdaBound{X}\AgdaSpace{}%
\AgdaSymbol{(}\AgdaRecord{Ar}\AgdaSpace{}%
\AgdaBound{X}\AgdaSpace{}%
\AgdaBound{d}\AgdaSpace{}%
\AgdaBound{s}\AgdaSymbol{)}\AgdaSpace{}%
\AgdaBound{X}\AgdaSpace{}%
\AgdaBound{d}\AgdaSpace{}%
\AgdaBound{s}\<%
\\
\>[6]\AgdaInductiveConstructor{a-a}\AgdaSpace{}%
\AgdaSymbol{:}\AgdaSpace{}%
\AgdaSymbol{∀}\AgdaSpace{}%
\AgdaSymbol{\{}\AgdaBound{m}\AgdaSpace{}%
\AgdaBound{n}\AgdaSpace{}%
\AgdaBound{sx}\AgdaSpace{}%
\AgdaBound{sy}\AgdaSymbol{\}\{@(}\AgdaKeyword{tactic}\AgdaSpace{}%
\AgdaFunction{dy-args-ok?}\AgdaSymbol{)}\AgdaSpace{}%
\AgdaBound{args}\AgdaSpace{}%
\AgdaSymbol{:}\<%
\\
\>[6][@{}l@{\AgdaIndent{0}}]%
\>[8]\AgdaDatatype{dy-args}\AgdaSpace{}%
\AgdaBound{m}\AgdaSpace{}%
\AgdaBound{n}\AgdaSpace{}%
\AgdaBound{sx}\AgdaSpace{}%
\AgdaBound{sy}\AgdaSymbol{\}}\AgdaSpace{}%
\AgdaSymbol{→}\AgdaSpace{}%
\AgdaDatatype{DyScalVec}\AgdaSpace{}%
\AgdaBound{X}\AgdaSpace{}%
\AgdaSymbol{(}\AgdaRecord{Ar}\AgdaSpace{}%
\AgdaBound{X}\AgdaSpace{}%
\AgdaBound{m}\AgdaSpace{}%
\AgdaBound{sx}\AgdaSymbol{)}\AgdaSpace{}%
\AgdaSymbol{(}\AgdaRecord{Ar}\AgdaSpace{}%
\AgdaBound{X}\AgdaSpace{}%
\AgdaBound{n}\AgdaSpace{}%
\AgdaBound{sy}\AgdaSymbol{)}\AgdaSpace{}%
\AgdaSymbol{(}\AgdaFunction{dy-args-dim}\AgdaSpace{}%
\AgdaBound{args}\AgdaSymbol{)}\AgdaSpace{}%
\AgdaSymbol{(}\AgdaFunction{dy-args-shp}\AgdaSpace{}%
\AgdaBound{args}\AgdaSymbol{)}\<%
\\
\\[\AgdaEmptyExtraSkip]%
\>[2]\AgdaFunction{▴ₗ}\AgdaSpace{}%
\AgdaSymbol{:}\AgdaSpace{}%
\AgdaSymbol{∀}\AgdaSpace{}%
\AgdaSymbol{\{}\AgdaBound{X}\AgdaSpace{}%
\AgdaBound{A}\AgdaSpace{}%
\AgdaBound{B}\AgdaSpace{}%
\AgdaBound{d}\AgdaSpace{}%
\AgdaBound{s}\AgdaSymbol{\}}\AgdaSpace{}%
\AgdaSymbol{\{\{}\AgdaBound{t}\AgdaSpace{}%
\AgdaSymbol{:}\AgdaSpace{}%
\AgdaDatatype{DyScalVec}\AgdaSpace{}%
\AgdaBound{X}\AgdaSpace{}%
\AgdaBound{A}\AgdaSpace{}%
\AgdaBound{B}\AgdaSpace{}%
\AgdaBound{d}\AgdaSpace{}%
\AgdaBound{s}\AgdaSymbol{\}\}}\AgdaSpace{}%
\AgdaSymbol{→}\AgdaSpace{}%
\AgdaBound{A}\AgdaSpace{}%
\AgdaSymbol{→}\AgdaSpace{}%
\AgdaRecord{Ar}\AgdaSpace{}%
\AgdaBound{X}\AgdaSpace{}%
\AgdaBound{d}\AgdaSpace{}%
\AgdaBound{s}\<%
\\
\>[2]\AgdaFunction{▴ₗ}\AgdaSpace{}%
\AgdaSymbol{\{\{}\AgdaInductiveConstructor{s-s}\AgdaSymbol{\}\}}\AgdaSpace{}%
\AgdaBound{a}\AgdaSpace{}%
\AgdaSymbol{=}\AgdaSpace{}%
\AgdaFunction{cst}\AgdaSpace{}%
\AgdaBound{a}\<%
\\
\>[2]\AgdaFunction{▴ₗ}\AgdaSpace{}%
\AgdaSymbol{\{\{}\AgdaInductiveConstructor{s-v}\AgdaSymbol{\}\}}\AgdaSpace{}%
\AgdaBound{a}\AgdaSpace{}%
\AgdaSymbol{=}\AgdaSpace{}%
\AgdaFunction{cst}\AgdaSpace{}%
\AgdaBound{a}\<%
\\
\>[2]\AgdaFunction{▴ₗ}\AgdaSpace{}%
\AgdaSymbol{\{\{}\AgdaInductiveConstructor{s-a}\AgdaSymbol{\}\}}\AgdaSpace{}%
\AgdaBound{a}\AgdaSpace{}%
\AgdaSymbol{=}\AgdaSpace{}%
\AgdaFunction{cst}\AgdaSpace{}%
\AgdaBound{a}\<%
\\
\>[2]\AgdaFunction{▴ₗ}\AgdaSpace{}%
\AgdaSymbol{\{\{}\AgdaInductiveConstructor{v-s}\AgdaSymbol{\}\}}\AgdaSpace{}%
\AgdaBound{a}\AgdaSpace{}%
\AgdaSymbol{=}\AgdaSpace{}%
\AgdaOperator{\AgdaFunction{▴}}\AgdaSpace{}%
\AgdaBound{a}\<%
\\
\>[2]\AgdaFunction{▴ₗ}\AgdaSpace{}%
\AgdaSymbol{\{\{}\AgdaInductiveConstructor{v-v}\AgdaSymbol{\}\}}\AgdaSpace{}%
\AgdaBound{a}\AgdaSpace{}%
\AgdaSymbol{=}\AgdaSpace{}%
\AgdaOperator{\AgdaFunction{▴}}\AgdaSpace{}%
\AgdaBound{a}\<%
\\
\>[2]\AgdaFunction{▴ₗ}\AgdaSpace{}%
\AgdaSymbol{\{\{}\AgdaInductiveConstructor{a-s}\AgdaSymbol{\}\}}\AgdaSpace{}%
\AgdaBound{a}\AgdaSpace{}%
\AgdaSymbol{=}\AgdaSpace{}%
\AgdaBound{a}\<%
\\
\>[2]\AgdaFunction{▴ₗ}\AgdaSpace{}%
\AgdaSymbol{\{\{}\AgdaSpace{}%
\AgdaArgument{t}\AgdaSpace{}%
\AgdaSymbol{=}\AgdaSpace{}%
\AgdaInductiveConstructor{a-a}\AgdaSpace{}%
\AgdaSymbol{\{}\AgdaArgument{args}\AgdaSpace{}%
\AgdaSymbol{=}\AgdaSpace{}%
\AgdaInductiveConstructor{n-n}\AgdaSymbol{\}}\AgdaSpace{}%
\AgdaSymbol{\}\}}\AgdaSpace{}%
\AgdaBound{a}\AgdaSpace{}%
\AgdaSymbol{=}\AgdaSpace{}%
\AgdaBound{a}\<%
\\
\>[2]\AgdaFunction{▴ₗ}\AgdaSpace{}%
\AgdaSymbol{\{\{}\AgdaSpace{}%
\AgdaArgument{t}\AgdaSpace{}%
\AgdaSymbol{=}\AgdaSpace{}%
\AgdaInductiveConstructor{a-a}\AgdaSpace{}%
\AgdaSymbol{\{}\AgdaArgument{args}\AgdaSpace{}%
\AgdaSymbol{=}\AgdaSpace{}%
\AgdaInductiveConstructor{n-0}\AgdaSymbol{\}}\AgdaSpace{}%
\AgdaSymbol{\}\}}\AgdaSpace{}%
\AgdaBound{a}\AgdaSpace{}%
\AgdaSymbol{=}\AgdaSpace{}%
\AgdaBound{a}\<%
\\
\>[2]\AgdaFunction{▴ₗ}\AgdaSpace{}%
\AgdaSymbol{\{\{}\AgdaSpace{}%
\AgdaArgument{t}\AgdaSpace{}%
\AgdaSymbol{=}\AgdaSpace{}%
\AgdaInductiveConstructor{a-a}\AgdaSpace{}%
\AgdaSymbol{\{}\AgdaArgument{args}\AgdaSpace{}%
\AgdaSymbol{=}\AgdaSpace{}%
\AgdaInductiveConstructor{0-n}\AgdaSymbol{\}}\AgdaSpace{}%
\AgdaSymbol{\}\}}\AgdaSpace{}%
\AgdaBound{a}\AgdaSpace{}%
\AgdaSymbol{=}\AgdaSpace{}%
\AgdaFunction{cst}\AgdaSpace{}%
\AgdaSymbol{(}\AgdaField{Ar.sel}\AgdaSpace{}%
\AgdaBound{a}\AgdaSpace{}%
\AgdaInductiveConstructor{[]}\AgdaSymbol{)}\<%
\\
\\[\AgdaEmptyExtraSkip]%
\>[2]\AgdaFunction{▴ᵣ}\AgdaSpace{}%
\AgdaSymbol{:}\AgdaSpace{}%
\AgdaSymbol{∀}\AgdaSpace{}%
\AgdaSymbol{\{}\AgdaBound{X}\AgdaSpace{}%
\AgdaBound{A}\AgdaSpace{}%
\AgdaBound{B}\AgdaSpace{}%
\AgdaBound{d}\AgdaSpace{}%
\AgdaBound{s}\AgdaSymbol{\}}\AgdaSpace{}%
\AgdaSymbol{\{\{}\AgdaBound{t}\AgdaSpace{}%
\AgdaSymbol{:}\AgdaSpace{}%
\AgdaDatatype{DyScalVec}\AgdaSpace{}%
\AgdaBound{X}\AgdaSpace{}%
\AgdaBound{A}\AgdaSpace{}%
\AgdaBound{B}\AgdaSpace{}%
\AgdaBound{d}\AgdaSpace{}%
\AgdaBound{s}\AgdaSymbol{\}\}}\AgdaSpace{}%
\AgdaSymbol{→}\AgdaSpace{}%
\AgdaBound{B}\AgdaSpace{}%
\AgdaSymbol{→}\AgdaSpace{}%
\AgdaRecord{Ar}\AgdaSpace{}%
\AgdaBound{X}\AgdaSpace{}%
\AgdaBound{d}\AgdaSpace{}%
\AgdaBound{s}\<%
\\
\>[2]\AgdaFunction{▴ᵣ}\AgdaSpace{}%
\AgdaSymbol{\{\{}\AgdaInductiveConstructor{s-s}\AgdaSymbol{\}\}}\AgdaSpace{}%
\AgdaBound{b}\AgdaSpace{}%
\AgdaSymbol{=}\AgdaSpace{}%
\AgdaFunction{cst}\AgdaSpace{}%
\AgdaBound{b}\<%
\\
\>[2]\AgdaFunction{▴ᵣ}\AgdaSpace{}%
\AgdaSymbol{\{\{}\AgdaInductiveConstructor{s-v}\AgdaSymbol{\}\}}\AgdaSpace{}%
\AgdaBound{b}\AgdaSpace{}%
\AgdaSymbol{=}\AgdaSpace{}%
\AgdaOperator{\AgdaFunction{▴}}\AgdaSpace{}%
\AgdaBound{b}\<%
\\
\>[2]\AgdaFunction{▴ᵣ}\AgdaSpace{}%
\AgdaSymbol{\{\{}\AgdaInductiveConstructor{s-a}\AgdaSymbol{\}\}}\AgdaSpace{}%
\AgdaBound{b}\AgdaSpace{}%
\AgdaSymbol{=}\AgdaSpace{}%
\AgdaBound{b}\<%
\\
\>[2]\AgdaFunction{▴ᵣ}\AgdaSpace{}%
\AgdaSymbol{\{\{}\AgdaInductiveConstructor{v-s}\AgdaSymbol{\}\}}\AgdaSpace{}%
\AgdaBound{b}\AgdaSpace{}%
\AgdaSymbol{=}\AgdaSpace{}%
\AgdaFunction{cst}\AgdaSpace{}%
\AgdaBound{b}\<%
\\
\>[2]\AgdaFunction{▴ᵣ}\AgdaSpace{}%
\AgdaSymbol{\{\{}\AgdaInductiveConstructor{v-v}\AgdaSymbol{\}\}}\AgdaSpace{}%
\AgdaBound{b}\AgdaSpace{}%
\AgdaSymbol{=}\AgdaSpace{}%
\AgdaOperator{\AgdaFunction{▴}}\AgdaSpace{}%
\AgdaBound{b}\<%
\\
\>[2]\AgdaFunction{▴ᵣ}\AgdaSpace{}%
\AgdaSymbol{\{\{}\AgdaInductiveConstructor{a-s}\AgdaSymbol{\}\}}\AgdaSpace{}%
\AgdaBound{b}\AgdaSpace{}%
\AgdaSymbol{=}\AgdaSpace{}%
\AgdaFunction{cst}\AgdaSpace{}%
\AgdaBound{b}\<%
\\
\>[2]\AgdaFunction{▴ᵣ}\AgdaSpace{}%
\AgdaSymbol{\{\{}\AgdaSpace{}%
\AgdaArgument{t}\AgdaSpace{}%
\AgdaSymbol{=}\AgdaSpace{}%
\AgdaInductiveConstructor{a-a}\AgdaSpace{}%
\AgdaSymbol{\{}\AgdaArgument{args}\AgdaSpace{}%
\AgdaSymbol{=}\AgdaSpace{}%
\AgdaInductiveConstructor{n-n}\AgdaSymbol{\}}\AgdaSpace{}%
\AgdaSymbol{\}\}}\AgdaSpace{}%
\AgdaBound{b}\AgdaSpace{}%
\AgdaSymbol{=}\AgdaSpace{}%
\AgdaBound{b}\<%
\\
\>[2]\AgdaFunction{▴ᵣ}\AgdaSpace{}%
\AgdaSymbol{\{\{}\AgdaSpace{}%
\AgdaArgument{t}\AgdaSpace{}%
\AgdaSymbol{=}\AgdaSpace{}%
\AgdaInductiveConstructor{a-a}\AgdaSpace{}%
\AgdaSymbol{\{}\AgdaArgument{args}\AgdaSpace{}%
\AgdaSymbol{=}\AgdaSpace{}%
\AgdaInductiveConstructor{n-0}\AgdaSymbol{\}}\AgdaSpace{}%
\AgdaSymbol{\}\}}\AgdaSpace{}%
\AgdaBound{b}\AgdaSpace{}%
\AgdaSymbol{=}\AgdaSpace{}%
\AgdaFunction{cst}\AgdaSpace{}%
\AgdaSymbol{(}\AgdaField{Ar.sel}\AgdaSpace{}%
\AgdaBound{b}\AgdaSpace{}%
\AgdaInductiveConstructor{[]}\AgdaSymbol{)}\<%
\\
\>[2]\AgdaFunction{▴ᵣ}\AgdaSpace{}%
\AgdaSymbol{\{\{}\AgdaSpace{}%
\AgdaArgument{t}\AgdaSpace{}%
\AgdaSymbol{=}\AgdaSpace{}%
\AgdaInductiveConstructor{a-a}\AgdaSpace{}%
\AgdaSymbol{\{}\AgdaArgument{args}\AgdaSpace{}%
\AgdaSymbol{=}\AgdaSpace{}%
\AgdaInductiveConstructor{0-n}\AgdaSymbol{\}}\AgdaSpace{}%
\AgdaSymbol{\}\}}\AgdaSpace{}%
\AgdaBound{b}\AgdaSpace{}%
\AgdaSymbol{=}\AgdaSpace{}%
\AgdaBound{b}\<%
\\
\\[\AgdaEmptyExtraSkip]%
\>[2]\AgdaFunction{lift}\AgdaSpace{}%
\AgdaSymbol{:}\AgdaSpace{}%
\AgdaSymbol{∀}\AgdaSpace{}%
\AgdaSymbol{\{}\AgdaBound{X}\AgdaSpace{}%
\AgdaBound{A}\AgdaSpace{}%
\AgdaBound{B}\AgdaSpace{}%
\AgdaBound{d}\AgdaSpace{}%
\AgdaBound{s}\AgdaSymbol{\}\{\{}\AgdaBound{t}\AgdaSpace{}%
\AgdaSymbol{:}\AgdaSpace{}%
\AgdaDatatype{DyScalVec}\AgdaSpace{}%
\AgdaBound{X}\AgdaSpace{}%
\AgdaBound{A}\AgdaSpace{}%
\AgdaBound{B}\AgdaSpace{}%
\AgdaBound{d}\AgdaSpace{}%
\AgdaBound{s}\AgdaSymbol{\}\}}\AgdaSpace{}%
\AgdaSymbol{→}\AgdaSpace{}%
\AgdaBound{A}\AgdaSpace{}%
\AgdaSymbol{→}\AgdaSpace{}%
\AgdaBound{B}\AgdaSpace{}%
\AgdaSymbol{→}\AgdaSpace{}%
\AgdaSymbol{(}\AgdaOperator{\AgdaBound{\AgdaUnderscore{}⊕\AgdaUnderscore{}}}\AgdaSpace{}%
\AgdaSymbol{:}\AgdaSpace{}%
\AgdaBound{X}\AgdaSpace{}%
\AgdaSymbol{→}\AgdaSpace{}%
\AgdaBound{X}\AgdaSpace{}%
\AgdaSymbol{→}\AgdaSpace{}%
\AgdaBound{X}\AgdaSymbol{)}\AgdaSpace{}%
\AgdaSymbol{→}\AgdaSpace{}%
\AgdaRecord{Ar}\AgdaSpace{}%
\AgdaBound{X}\AgdaSpace{}%
\AgdaBound{d}\AgdaSpace{}%
\AgdaBound{s}\<%
\\
\>[2]\AgdaFunction{lift}\AgdaSpace{}%
\AgdaSymbol{\{\{}\AgdaSpace{}%
\AgdaBound{t}\AgdaSpace{}%
\AgdaSymbol{\}\}}\AgdaSpace{}%
\AgdaBound{a}\AgdaSpace{}%
\AgdaBound{b}\AgdaSpace{}%
\AgdaOperator{\AgdaBound{\AgdaUnderscore{}⊕\AgdaUnderscore{}}}\AgdaSpace{}%
\AgdaSymbol{=}\AgdaSpace{}%
\AgdaInductiveConstructor{imap}\AgdaSpace{}%
\AgdaSymbol{λ}\AgdaSpace{}%
\AgdaBound{iv}\AgdaSpace{}%
\AgdaSymbol{→}\AgdaSpace{}%
\AgdaField{Ar.sel}\AgdaSpace{}%
\AgdaSymbol{(}\AgdaFunction{▴ₗ}\AgdaSpace{}%
\AgdaBound{a}\AgdaSymbol{)}\AgdaSpace{}%
\AgdaBound{iv}\AgdaSpace{}%
\AgdaOperator{\AgdaBound{⊕}}\AgdaSpace{}%
\AgdaField{Ar.sel}\AgdaSpace{}%
\AgdaSymbol{(}\AgdaFunction{▴ᵣ}\AgdaSpace{}%
\AgdaBound{b}\AgdaSymbol{)}\AgdaSpace{}%
\AgdaBound{iv}\<%
\\
\>[0]\<%
\end{code}

We have only presented the overloading between the \AD{Ar}
types of different shapes.  This still does not solve the problem of
implicit casts from base types such as \AD{ℕ} and vectors into arrays.
However, this can be solved by defining regular instances.  In the
code accompanying this paper, we define a similar \AD{lift} function
that extends the domain of the lifted binary operation and accepts
base types, vectors and arrays, and their combinations.

\subsection{A Convolutional Neural Network} \label{sec:cnn}
As a practical application, we consider a convolutional
neural network for recognising hand-written digits, implemented in APL.
The reference implementation we start from~\cite{cnninapl} is
written entirely in APL without relying on any external libraries
or frameworks. The implementation is very concise --- apart from
built-in operators, it only defines 10 new functions,
each of which is a single line of APL code.
Translating these functions into our embedded array language
serves two purposes.  First, we stress-test
abstractions used in our embedding and the extractor capabilities.
Second, we verify that all the shapes and ranks match, the indexing
is in-bound, no division by zero occurs, and that the functions are
terminating.  As APL is dynamically typed, it is difficult to be
sure that no runtime errors will occur.  Embedding the code into
Agda essentially requires us to define a type system for the operators
in use and guarantee that they hold.
We consider three representative samples of our encoding and explain
the details.

\begin{code}[hide]%
\>[0]\AgdaKeyword{module}\AgdaSpace{}%
\AgdaModule{CNN}\AgdaSpace{}%
\AgdaKeyword{where}\<%
\\
\>[0][@{}l@{\AgdaIndent{0}}]%
\>[2]\AgdaKeyword{open}\AgdaSpace{}%
\AgdaKeyword{import}\AgdaSpace{}%
\AgdaModule{Array}\<%
\\
\>[2]\AgdaKeyword{open}\AgdaSpace{}%
\AgdaKeyword{import}\AgdaSpace{}%
\AgdaModule{APL2}\<%
\\
\>[2]\AgdaKeyword{open}\AgdaSpace{}%
\AgdaKeyword{import}\AgdaSpace{}%
\AgdaModule{Agda.Builtin.Float}\<%
\\
\>[2]\AgdaKeyword{open}\AgdaSpace{}%
\AgdaKeyword{import}\AgdaSpace{}%
\AgdaModule{Data.Product}\AgdaSpace{}%
\AgdaKeyword{hiding}\AgdaSpace{}%
\AgdaSymbol{(}\AgdaOperator{\AgdaFunction{\AgdaUnderscore{}×\AgdaUnderscore{}}}\AgdaSymbol{)}\<%
\\
\>[2]\AgdaKeyword{postulate}\<%
\\
\>[2][@{}l@{\AgdaIndent{0}}]%
\>[4]\AgdaPostulate{⋯}\AgdaSpace{}%
\AgdaSymbol{:}\AgdaSpace{}%
\AgdaSymbol{∀}\AgdaSpace{}%
\AgdaSymbol{\{}\AgdaBound{a}\AgdaSymbol{\}\{}\AgdaBound{A}\AgdaSpace{}%
\AgdaSymbol{:}\AgdaSpace{}%
\AgdaPrimitive{Set}\AgdaSpace{}%
\AgdaBound{a}\AgdaSymbol{\}}\AgdaSpace{}%
\AgdaSymbol{→}\AgdaSpace{}%
\AgdaBound{A}\<%
\\
\\[\AgdaEmptyExtraSkip]%
\>[2]\AgdaFunction{A<B⇒K<2⇒A*2+K<B*2}\AgdaSpace{}%
\AgdaSymbol{:}\AgdaSpace{}%
\AgdaSymbol{∀}\AgdaSpace{}%
\AgdaSymbol{\{}\AgdaBound{n}\AgdaSpace{}%
\AgdaBound{s}\AgdaSymbol{\}\{}\AgdaBound{a}\AgdaSpace{}%
\AgdaBound{b}\AgdaSpace{}%
\AgdaBound{k}\AgdaSpace{}%
\AgdaSymbol{:}\AgdaSpace{}%
\AgdaRecord{Ar}\AgdaSpace{}%
\AgdaDatatype{ℕ}\AgdaSpace{}%
\AgdaBound{n}\AgdaSpace{}%
\AgdaBound{s}\AgdaSymbol{\}}\AgdaSpace{}%
\AgdaSymbol{→}\AgdaSpace{}%
\AgdaBound{a}\AgdaSpace{}%
\AgdaOperator{\AgdaFunction{<a}}\AgdaSpace{}%
\AgdaBound{b}\AgdaSpace{}%
\AgdaSymbol{→}\AgdaSpace{}%
\AgdaBound{k}\AgdaSpace{}%
\AgdaOperator{\AgdaFunction{<a}}\AgdaSpace{}%
\AgdaSymbol{(}\AgdaFunction{cst}\AgdaSpace{}%
\AgdaNumber{2}\AgdaSymbol{)}\AgdaSpace{}%
\AgdaSymbol{→}\AgdaSpace{}%
\AgdaSymbol{((}\AgdaBound{a}\AgdaSpace{}%
\AgdaOperator{\AgdaFunction{×}}\AgdaSpace{}%
\AgdaNumber{2}\AgdaSymbol{)}\AgdaSpace{}%
\AgdaOperator{\AgdaFunction{+}}\AgdaSpace{}%
\AgdaBound{k}\AgdaSymbol{)}\AgdaSpace{}%
\AgdaOperator{\AgdaFunction{<a}}\AgdaSpace{}%
\AgdaSymbol{(}\AgdaBound{b}\AgdaSpace{}%
\AgdaOperator{\AgdaFunction{×}}\AgdaSpace{}%
\AgdaNumber{2}\AgdaSymbol{)}\<%
\\
\>[2]\AgdaFunction{A<B⇒K<2⇒A*2+K<B*2}\AgdaSpace{}%
\AgdaSymbol{=}\AgdaSpace{}%
\AgdaPostulate{⋯}\<%
\\
\\[\AgdaEmptyExtraSkip]%
\>[2]\AgdaOperator{\AgdaFunction{\AgdaUnderscore{}+ᵣ′\AgdaUnderscore{}}}\AgdaSpace{}%
\AgdaSymbol{=}\AgdaSpace{}%
\AgdaPrimitive{primFloatPlus}\<%
\\
\\[\AgdaEmptyExtraSkip]%
\>[2]\AgdaComment{--\ Fuck\ you,\ unicode\ symbols!}\<%
\\
\>[2]\AgdaKeyword{infixr}\AgdaSpace{}%
\AgdaNumber{20}\AgdaSpace{}%
\AgdaOperator{\AgdaFunction{\AgdaUnderscore{}¨\AgdaUnderscore{}}}\<%
\\
\>[2]\AgdaOperator{\AgdaFunction{\AgdaUnderscore{}¨\AgdaUnderscore{}}}\AgdaSpace{}%
\AgdaSymbol{=}\AgdaSpace{}%
\AgdaOperator{\AgdaFunction{\AgdaUnderscore{}̈\AgdaUnderscore{}}}\<%
\end{code}

\paragraph{Logistic function}
After the convolution and fully-connected layers in the CNN, the
activation function is applied to each of the results.  The activation
function in use is called the standard logistic function
\(
\frac{1}{1 - e^x}
\), and it is being applied to all the elements of the resulting
array.  Here is the implementation in APL and in our embedding:
\begin{mathpar}
\codeblock{\apl{logistic←\{÷1+*-⍵\}}}
\and\codeblock{\begin{code}%
\>[2]\AgdaFunction{logistic}\AgdaSpace{}%
\AgdaSymbol{:}\AgdaSpace{}%
\AgdaSymbol{∀}\AgdaSpace{}%
\AgdaSymbol{\{}\AgdaBound{n}\AgdaSpace{}%
\AgdaBound{s}\AgdaSymbol{\}}\AgdaSpace{}%
\AgdaSymbol{→}\AgdaSpace{}%
\AgdaRecord{Ar}\AgdaSpace{}%
\AgdaPostulate{Float}\AgdaSpace{}%
\AgdaBound{n}\AgdaSpace{}%
\AgdaBound{s}\AgdaSpace{}%
\AgdaSymbol{→}\AgdaSpace{}%
\AgdaRecord{Ar}\AgdaSpace{}%
\AgdaPostulate{Float}\AgdaSpace{}%
\AgdaBound{n}\AgdaSpace{}%
\AgdaBound{s}\<%
\\
\>[2]\AgdaFunction{logistic}\AgdaSpace{}%
\AgdaBound{ω}\AgdaSpace{}%
\AgdaSymbol{=}\AgdaSpace{}%
\AgdaOperator{\AgdaFunction{÷ᵣ}}\AgdaSpace{}%
\AgdaNumber{1.0}\AgdaSpace{}%
\AgdaOperator{\AgdaFunction{+ᵣ}}\AgdaSpace{}%
\AgdaOperator{\AgdaFunction{*ᵣ}}\AgdaSpace{}%
\AgdaOperator{\AgdaFunction{-ᵣ}}\AgdaSpace{}%
\AgdaBound{ω}\<%
\end{code}}
\end{mathpar}
As can be seen, the implementations are almost identical.
There are two important reasons: the ability to define the
precedence and the associativity of the operators; and the automatic
casts that we explained before.  All the operators in APL are
right-associative, which we implement in Agda using \AK{infixr} statements.
We distinguish the operations on base types by adding a postfix
to the name, so instead of \AF{\_+\_}, \AF{\_-\_}, \etc{}
we have \AF{\_+ᵣ\_}, \AF{\_-ᵣ\_} when we the arguments are arrays
of base type \AF{Float}.  If we read the body right to left, the
function negates (\AF{-ᵣ\_}) its argument, then it computes the
exponent (\AF{*ᵣ\_}) of that result, then it adds \AN{1.0} to all
the elements, and finally it takes a reciprocal (\AF{÷ᵣ\_}).  The
function is shape- and rank-polymorphic; it does not require additional
proofs and it normalises to a single \AC{imap}.

\paragraph{Mean Squared Error}
The nice behaviour of the above function is not really surprising since
it just maps scalar operations over individual array
elements.  However, this is a common pattern in array-based
applications.  Here is another example that is used to compute the mean
error which is a sum of squared elements divided by two:
\begin{mathpar}
\codeblock{\apl{meansqerr←\{÷∘2+/,(⍺-⍵)*2\}}} \and
\codeblock{\begin{code}%
\>[2]\AgdaFunction{meansqerr}\AgdaSpace{}%
\AgdaSymbol{:}\AgdaSpace{}%
\AgdaSymbol{∀}\AgdaSpace{}%
\AgdaSymbol{\{}\AgdaBound{n}\AgdaSpace{}%
\AgdaBound{s}\AgdaSymbol{\}}\AgdaSpace{}%
\AgdaSymbol{→}\AgdaSpace{}%
\AgdaRecord{Ar}\AgdaSpace{}%
\AgdaPostulate{Float}\AgdaSpace{}%
\AgdaBound{n}\AgdaSpace{}%
\AgdaBound{s}\AgdaSpace{}%
\AgdaSymbol{→}\AgdaSpace{}%
\AgdaRecord{Ar}\AgdaSpace{}%
\AgdaPostulate{Float}\AgdaSpace{}%
\AgdaBound{n}\AgdaSpace{}%
\AgdaBound{s}\AgdaSpace{}%
\AgdaSymbol{→}\AgdaSpace{}%
\AgdaFunction{Scal}\AgdaSpace{}%
\AgdaPostulate{Float}\<%
\\
\>[2]\AgdaFunction{meansqerr}\AgdaSpace{}%
\AgdaBound{α}\AgdaSpace{}%
\AgdaBound{ω}\AgdaSpace{}%
\AgdaSymbol{=}\AgdaSpace{}%
\AgdaOperator{\AgdaFunction{\AgdaUnderscore{}÷ᵣ}}\AgdaSpace{}%
\AgdaNumber{2.0}\AgdaSpace{}%
\AgdaOperator{\AgdaFunction{\$}}\AgdaSpace{}%
\AgdaOperator{\AgdaFunction{\AgdaUnderscore{}+ᵣ′\AgdaUnderscore{}}}\AgdaSpace{}%
\AgdaOperator{\AgdaFunction{/}}\AgdaSpace{}%
\AgdaOperator{\AgdaFunction{,}}\AgdaSpace{}%
\AgdaSymbol{(}\AgdaBound{α}\AgdaSpace{}%
\AgdaOperator{\AgdaFunction{-ᵣ}}\AgdaSpace{}%
\AgdaBound{ω}\AgdaSymbol{)}\AgdaSpace{}%
\AgdaOperator{\AgdaFunction{×ᵣ}}\AgdaSpace{}%
\AgdaSymbol{(}\AgdaBound{α}\AgdaSpace{}%
\AgdaOperator{\AgdaFunction{-ᵣ}}\AgdaSpace{}%
\AgdaBound{ω}\AgdaSymbol{)}\<%
\end{code}}
\end{mathpar}
In addition to element-wise mapping we have a reduction of
the elements --- the \AF{\_/\_} operator.  On the right hand side
it gets an array that is being reduced, and the left operator is
a binary function that performs the actual operation.  We have a
flattened (\AF{,\_}) square of differences on the right, and
addition on \AD{Float}s on the left.  We need to flatten the
array on the right because according to the APL semantics, \AF{\_/\_} reduces
over the last axis of the array.  Also, in contrast to reductions
found in many functional languages, APL does not require the default
element but deduces it from the operation
in use.  We have encoded the same behaviour using instance resolution.
However, we had to supply the addition on floats
\AF{\_+ᵣ′\_}, rather than our generalised addition on the arrays
and vectors of floats \AF{\_+ᵣ\_}, because otherwise
Agda fails to instantiate hidden arguments to \AF{\_+ᵣ\_}.  Finally,
partial application of division on the right \apl{÷∘2} is a built-in
feature of mixfix operators in Agda.

\paragraph{Back Average Pool}
The reverse average pooling function requires us to specify
a shape restriction: the shape of the result must be twice
as big  as the shape of the input array (in every dimension).
\begin{mathpar}
\codeblock{\apl{backavgpool←\{2⌿2/⍵÷4\}⍤2}} \and
\codeblock{\begin{code}%
\>[2]\AgdaFunction{backavgpool}\AgdaSpace{}%
\AgdaSymbol{:}\AgdaSpace{}%
\AgdaSymbol{∀}\AgdaSpace{}%
\AgdaSymbol{\{}\AgdaBound{s}\AgdaSymbol{\}}\AgdaSpace{}%
\AgdaSymbol{→}\AgdaSpace{}%
\AgdaRecord{Ar}\AgdaSpace{}%
\AgdaPostulate{Float}\AgdaSpace{}%
\AgdaNumber{2}\AgdaSpace{}%
\AgdaBound{s}\AgdaSpace{}%
\AgdaSymbol{→}\AgdaSpace{}%
\AgdaRecord{Ar}\AgdaSpace{}%
\AgdaPostulate{Float}\AgdaSpace{}%
\AgdaNumber{2}\AgdaSpace{}%
\AgdaOperator{\AgdaFunction{\$}}\AgdaSpace{}%
\AgdaOperator{\AgdaFunction{▾}}\AgdaSpace{}%
\AgdaSymbol{(}\AgdaNumber{2}\AgdaSpace{}%
\AgdaOperator{\AgdaFunction{×}}\AgdaSpace{}%
\AgdaBound{s}\AgdaSymbol{)}\<%
\\
\>[2]\AgdaFunction{backavgpool}\AgdaSpace{}%
\AgdaSymbol{\{}\AgdaArgument{s}\AgdaSpace{}%
\AgdaSymbol{=}\AgdaSpace{}%
\AgdaSymbol{\AgdaUnderscore{}}\AgdaSpace{}%
\AgdaOperator{\AgdaInductiveConstructor{∷}}\AgdaSpace{}%
\AgdaSymbol{\AgdaUnderscore{}}\AgdaSpace{}%
\AgdaOperator{\AgdaInductiveConstructor{∷}}\AgdaSpace{}%
\AgdaInductiveConstructor{[]}\AgdaSymbol{\}}\AgdaSpace{}%
\AgdaBound{ω}\AgdaSpace{}%
\AgdaSymbol{=}\AgdaSpace{}%
\AgdaNumber{2}\AgdaSpace{}%
\AgdaOperator{\AgdaFunction{⌿ᵣ}}\AgdaSpace{}%
\AgdaNumber{2}\AgdaSpace{}%
\AgdaOperator{\AgdaFunction{/ᵣ′}}\AgdaSpace{}%
\AgdaBound{ω}\AgdaSpace{}%
\AgdaOperator{\AgdaFunction{÷ᵣ}}\AgdaSpace{}%
\AgdaNumber{4.0}\<%
\\
\>[2][@{}l@{\AgdaIndent{0}}]%
\>[4]\AgdaKeyword{where}\<%
\\
\>[4][@{}l@{\AgdaIndent{0}}]%
\>[6]\AgdaKeyword{infixr}\AgdaSpace{}%
\AgdaNumber{20}\AgdaSpace{}%
\AgdaOperator{\AgdaFunction{\AgdaUnderscore{}/ᵣ′\AgdaUnderscore{}}}\<%
\\
\>[6]\AgdaOperator{\AgdaFunction{\AgdaUnderscore{}/ᵣ′\AgdaUnderscore{}}}\AgdaSpace{}%
\AgdaSymbol{=}\AgdaSpace{}%
\AgdaOperator{\AgdaFunction{\AgdaUnderscore{}/ᵣ\AgdaUnderscore{}}}\AgdaSpace{}%
\AgdaSymbol{\{}\AgdaArgument{s}\AgdaSpace{}%
\AgdaSymbol{=}\AgdaSpace{}%
\AgdaSymbol{\AgdaUnderscore{}}\AgdaSpace{}%
\AgdaOperator{\AgdaInductiveConstructor{∷}}\AgdaSpace{}%
\AgdaInductiveConstructor{[]}\AgdaSymbol{\}}\<%
\end{code}}
\end{mathpar}
We specify this relation using our lifted arithmetic operations:
\AN{2} \AF{×} \AB{s}, where the left argument is of type \AD{ℕ},
and the right argument is \AD{Vec} \AD{ℕ} \AN{2}.  The multiplication
returns a 1-dimensional array of type
\AD{Ar} \AD{ℕ} \AN{1} (\AN{2} \AF{∷} \AC{[]}), and we typecast it
back to \AD{Vec} using the \AF{▾\_} function.

The function itself divides all the array elements by \AN{4.0} and
replicates them two times across each row (\AF{\_/ᵣ\_}), and two
times across each column (\AF{\_⌿ᵣ\_}).  Note that we have to
help Agda by specifying that \AB{s} is guaranteed to be of length 2.
Also, similarly to before, we
need to supply a hidden argument to \AF{\_/ᵣ\_}.  Rather than
doing this inside the application chain, we used a \AK{where}
syntax to define a local variant of the row replicator \AF{\_/ᵣ′\_}.

\paragraph{Average Pooling}
Our final example is an average pooling function.  It takes a
two-dimensional array of floats as an argument, where each axis
is divisible by two.  It partitions the array into sub-arrays
of shape [2,2] and computes the average of each partition.
Here is the implementation:
\apl{avg ← \{ (+/÷≢),⍵\} } \\
\apl{avgpool ← \{ (x y) ← ⍴⍵ ⋄ avg⍤2 ⊢ 0 2 1 3⍉(x÷2) 2 (y÷2) 2⍴ ⍵ \}}\\
\begin{code}%
\>[2]\AgdaFunction{avgpool}\AgdaSpace{}%
\AgdaSymbol{:}\AgdaSpace{}%
\AgdaSymbol{∀}\AgdaSpace{}%
\AgdaSymbol{\{}\AgdaBound{s}\AgdaSymbol{\}}\AgdaSpace{}%
\AgdaSymbol{→}\AgdaSpace{}%
\AgdaRecord{Ar}\AgdaSpace{}%
\AgdaPostulate{Float}\AgdaSpace{}%
\AgdaNumber{2}\AgdaSpace{}%
\AgdaOperator{\AgdaFunction{\$}}\AgdaSpace{}%
\AgdaOperator{\AgdaFunction{▾}}\AgdaSpace{}%
\AgdaSymbol{(}\AgdaBound{s}\AgdaSpace{}%
\AgdaOperator{\AgdaFunction{×}}\AgdaSpace{}%
\AgdaNumber{2}\AgdaSymbol{)}\AgdaSpace{}%
\AgdaSymbol{→}\AgdaSpace{}%
\AgdaRecord{Ar}\AgdaSpace{}%
\AgdaPostulate{Float}\AgdaSpace{}%
\AgdaNumber{2}\AgdaSpace{}%
\AgdaBound{s}\<%
\\
\>[2]\AgdaFunction{avgpool}\AgdaSpace{}%
\AgdaSymbol{\{}\AgdaBound{s}\AgdaSymbol{\}}\AgdaSpace{}%
\AgdaSymbol{(}\AgdaInductiveConstructor{imap}\AgdaSpace{}%
\AgdaBound{p}\AgdaSymbol{)}\AgdaSpace{}%
\AgdaSymbol{=}\AgdaSpace{}%
\AgdaInductiveConstructor{imap}\AgdaSpace{}%
\AgdaOperator{\AgdaFunction{\$}}\AgdaSpace{}%
\AgdaSymbol{λ}\AgdaSpace{}%
\AgdaBound{iv}\AgdaSpace{}%
\AgdaSymbol{→}\<%
\\
\>[2][@{}l@{\AgdaIndent{0}}]%
\>[4]\AgdaKeyword{let}%
\>[1204I]\AgdaBound{ix}\AgdaSpace{}%
\AgdaOperator{\AgdaInductiveConstructor{,}}\AgdaSpace{}%
\AgdaBound{ix<s}%
\>[19]\AgdaSymbol{=}\AgdaSpace{}%
\AgdaFunction{ix→a}\AgdaSpace{}%
\AgdaBound{iv}\<%
\\
\>[.][@{}l@{}]\<[1204I]%
\>[8]\AgdaBound{q}%
\>[19]\AgdaSymbol{=}\AgdaSpace{}%
\AgdaSymbol{λ}\AgdaSpace{}%
\AgdaSymbol{(}\AgdaBound{i}\AgdaSpace{}%
\AgdaOperator{\AgdaInductiveConstructor{,}}\AgdaSpace{}%
\AgdaBound{pf}\AgdaSymbol{)}\AgdaSpace{}%
\AgdaSymbol{→}\AgdaSpace{}%
\AgdaBound{p}\AgdaSpace{}%
\AgdaOperator{\AgdaFunction{\$}}\AgdaSpace{}%
\AgdaFunction{a→ix}\AgdaSpace{}%
\AgdaSymbol{((}\AgdaBound{ix}\AgdaSpace{}%
\AgdaOperator{\AgdaFunction{×}}\AgdaSpace{}%
\AgdaNumber{2}\AgdaSymbol{)}\AgdaSpace{}%
\AgdaOperator{\AgdaFunction{+}}\AgdaSpace{}%
\AgdaBound{i}\AgdaSymbol{)}\AgdaSpace{}%
\AgdaSymbol{(}\AgdaBound{s}\AgdaSpace{}%
\AgdaOperator{\AgdaFunction{×}}\AgdaSpace{}%
\AgdaNumber{2}\AgdaSymbol{)}\AgdaSpace{}%
\AgdaSymbol{(}\AgdaFunction{A<B⇒K<2⇒A*2+K<B*2}\AgdaSpace{}%
\AgdaBound{ix<s}\AgdaSpace{}%
\AgdaBound{pf}\AgdaSymbol{)}\<%
\\
\>[8]\AgdaBound{[2,2]}%
\>[19]\AgdaSymbol{=}\AgdaSpace{}%
\AgdaFunction{cst}\AgdaSpace{}%
\AgdaSymbol{\{}\AgdaArgument{s}\AgdaSpace{}%
\AgdaSymbol{=}\AgdaSpace{}%
\AgdaNumber{2}\AgdaSpace{}%
\AgdaOperator{\AgdaInductiveConstructor{∷}}\AgdaSpace{}%
\AgdaInductiveConstructor{[]}\AgdaSymbol{\}}\AgdaSpace{}%
\AgdaNumber{2}\<%
\\
\>[4]\AgdaKeyword{in}\AgdaSpace{}%
\AgdaOperator{\AgdaFunction{▾}}\AgdaSpace{}%
\AgdaSymbol{(}\AgdaOperator{\AgdaFunction{\AgdaUnderscore{}÷ᵣ}}\AgdaSpace{}%
\AgdaNumber{4.0}\AgdaSpace{}%
\AgdaOperator{\AgdaFunction{\$}}\AgdaSpace{}%
\AgdaOperator{\AgdaFunction{\AgdaUnderscore{}+ᵣ′\AgdaUnderscore{}}}\AgdaSpace{}%
\AgdaOperator{\AgdaFunction{/}}\AgdaSpace{}%
\AgdaOperator{\AgdaFunction{,}}\AgdaSpace{}%
\AgdaBound{q}\AgdaSpace{}%
\AgdaOperator{\AgdaFunction{¨}}\AgdaSpace{}%
\AgdaOperator{\AgdaFunction{ι}}\AgdaSpace{}%
\AgdaBound{[2,2]}\AgdaSymbol{)}\<%
\end{code}
In this example, a direct
implementation that uses indexing is actually more straightforward than
one expressed in index-free style.  The result of average
pooling is given by the \AC{imap}. Reading the body of the \AC{imap} right to left,
we obtain an array of indices (\AF{ι\_})
into a two-dimensional array of shape [2,2].  Then for each element (\AF{\_¨\_})
in that array we apply the function \AF{f} bound above it.  Then
we sum the elements up and divide them by \AN{4.0}.  The indices
returned by (\AF{ι\_}) are dependent pairs where the first component
is a 1-dimensional array representing the value of the index, and the
second component is a proof that the index is strictly less than the
array shape (in our case [2,2]).  In \AB{q}, we pattern-match
on the pair, and we compute selection into the argument of \AF{avgpool}
at index $2iv+i$.  The final argument to \AB{q} is a proof that this
index is within the bounds of the array.

Here we consider the extraction of \AF{avgpool} into \sac{}, slightly reformatted
for better readability.
\begin{lstlisting}[mathescape=false]
float[.,.] avgpool(int[2] x_1, float[.,.] x_3) {
  float[.,.] __ret;
  s = x_1;
  assert (shape (x_1)[0] == 2);
  assert (take (2, shape (x_3))
           == cons ((x_1[0] $* 2), cons ((x_1[1+0] $* 2), empty ([]))));
  #define p(__x) (x_3)[__x]
  __ret = with {
    (.<= iv_1 <=.) {
       i = iv_1[0];
       j = iv_1[1+0];
    } : (   (p(cons(((i $* 2) $+ 0), cons(((j $* 2) $+ 0), [])))
         $+ (p(cons(((i $* 2) $+ 0), cons(((j $* 2) $+ 1), [])))
         $+ (p(cons(((i $* 2) $+ 1), cons(((j $* 2) $+ 0), [])))
         $+ (p(cons(((i $* 2) $+ 1), cons(((j $* 2) $+ 1), [])))
         $+ 0.0f))))
        $/ 4.0f);
  }: genarray (s, zero_float ([]));
  assert (take (2, shape (__ret)) == x_1);
  return __ret;
}
\end{lstlisting}
In the extracted code, all the
local definitions are inlined, as well as
all the compound array operations.  We are very close to the code
that a programmer could write.
The assertions at the top are deduced from the type signature:
the first argument must be a
two-element array, and the shape of the second argument is twice
the shape of the first argument.  We use arithmetic operations
prefixed with \$, to indicate that these are operations on scalars
(int and float) to help the compiler with instantiating overloadings.
Before returning, the function asserts that the shape
of the returned result must be equal to the first argument.
The body of the \texttt{with}-loop performs 4 selections
into the argument array and averages them.
Finally, since \sac{} is a first-order language but \AC{imap}
is a higher-order construct, the extractor has inserted a
macro \AB{p} to mimic the application of the pattern-matched
argument of \AF{imap} as a function.

\section{\label{sec:related}Related Work}

\paragraph{Metaprogramming} There is a large body of
work on metaprogramming facilities in various programming
languages.  \citet{refl-masses} track the origins
of metaprogramming to Smith's work on reflection in Lisp~\cite{refl-lisp}.
%
%
Some prominent metaprogramming systems include
MetaOcaml~\cite{metaocaml}, MetaML~\cite{metaml},
reFlect~\cite{DBLP:journals/jfp/GrundyMO06},
Template Haskell~\cite{sheard2002template},
Racket~\cite{plt-tr1}, and various other Lisp/Scheme dialects.
However, these systems typically do not support dependent
types, so they are not well suited for our goal of statically
enforcing correctness of embedded programs.

\paragraph{Embedding}
Defining deep embeddings with static guarantees are a common application
of dependent types~\cite{10.5555/647849.737066,CHAPMAN200921,
10.1007/978-3-540-74464-1_7,10.1145/3236785,10.1145/1863495.1863497}.
These embeddings usually also define semantics of the embedded
language and therefore allow us to reason about the correctness
of program transformations and optimisations.
While the fact that this is possible is impressive in theory, the resulting
encodings are often difficult to use in practice. In this paper
we instead aim for a more lightweight approach.

\citet{deepshallow} propose to solve this problem with
a combination of deep and shallow embeddings.  Their idea is to define
a small deep embedding and leverage type classes in Haskell to define
the rest of the language on top of that.  It would be interesting to see
whether such an approach scales to dependently-typed embedded
languages.


\paragraph{Extraction}
The Coq proof assistant is equipped with extraction
capabilities~\cite{10.1007/978-3-540-69407-6_39,10.1007/3-540-39185-1_12},
which extracts functional code from Coq proofs (or programs).  The
default target language is Ocaml, but a few other options were added
recently.
Likewise, Agda itself has a mechanism for defining custom backends, of
which the GHC backend is the most prominent.
Other proof assistants provide similar extraction tools as well.
The main difference from our approach in this paper
is that these extractors are written as plugins to the proof
assistant, while we implement our extractors directly in the proof
assistant itself.
While it would be possible to implement extractors presented in this
paper as Coq plugins or Agda backends, conceptually they are more
heavyweight.  Our extractors and programs can (in principle)
communicate with each other. In addition, as they are just Agda programs, they can
be reflected themselves and their structure can be leveraged.

\paragraph{Dependently typed metaprogramming}
Several dependently-typed languages are equipped with metaprogramming
capabilities: Idris \cite{idris-refl}, Lean \cite{lean-refl},
Coq \cite{metacoq}, and Agda \cite{agda-refl}.  All of these
implement a similar API as described in this paper.  This is
reassuring, as it means our proposed approach is immediately portable
into many other contexts.
\citet{10.1145/3371071} introduce the Turnstile+ framework for
building dependently typed embedded DSLs and very much
shares the ideas advocated in this paper, suggesting that
our approach could work there as well.
\citet{10.1145/3371076} use MetaCoq to formally verify
the core type system of Coq. This combines nicely with
our approach, as we could use the verified core language
as a basis to verify our custom extractors.
\citet{10.1145/3372885.3373829} use MetaCoq to implement
a DSL combining deep and shallow approaches, in a way that
is quite similar to our own. While they are able to formally
reason about preservation of semantics (which we can't do yet), it is unclear
whether their approach scales to dependently-typed embedded languages.

\paragraph{Arrays}
Using dependent types to verify properties of array programs
is not a novel idea.  For example, Qube~\cite{TROJAHNER2009643}
and Remora~\cite{10.1007/978-3-642-54833-8_3} are dependently
typed languages that are focused on array programming.  Both
of these focus on automating the type inference, which is a
big advantage for programmers.  However, one has to rely on
the capabilities of the inference engine, which may fail for
some complex examples.
In~\cite{10.1007/978-3-642-41582-1_11} authors use Agda as
a frontend for Accelerate, an array library in Haskell.
The motivation of the work is similar to ours, to provide static
guarantees about array computations.  As the target language of
this work is Haskell, and Agda provides a backend for it,
the integration happens smoothly without requiring
any extraction techniques.

\section{\label{sec:concl}Conclusions and Future Work}

In this paper we investigate the idea of developing
embedded programs hand-in-hand with custom
code generators for them. We solve the well-known
conundrum of choosing between deep and shallow
embedding by leveraging the power of
reflection.
This allows us to enjoying the benefits
of shallow embedding, while keeping full access to the internal
structure of the embedded programs.

We apply this idea in the context of dependently-typed
languages to create verified implementations that can
be used in the context of an existing programming language.
We embed the target language in a theorem-prover, using
dependent types to encode the properties of interest.
Using reflection we bring the verified implementation
back into the original language.

We have demonstrated the approach by implementing
three embedded languages and two extractors, using
Agda as our host. Along the way we made some
improvements to the reflection API of Agda, and in the end we
used our embedding to implement (and verify) a practical
application --- a convolutional neural network.

The main advantages of our approach are twofold.  First,
our solution is fine-grained --- we can chose what part
of the application to embed, and what constructs of the
host language to extract.  Secondly, our extractors
can use the full power of dependent types to guarantee
safety of our embedded programs.

Right now we cannot yet guarantee that the
extracted code preservers the semantics of the original
implementation. While we rarely see
fully-verified compiler backends in the real world,
our approach is very close to enabling this.
We would need a formal semantics of the reflected language
and the proof that reflected programs respect it.
While this is non-trivial, a system like Agda could do
this in principle.

There is a number of improvements that can be added to
Agda and our extractors to make the resulting code more
efficient.  Supporting \AK{let}s in the internal syntax
would help to preserve sharing.  Recognising irrelevance
annotations in the extractors would help to eliminate
unused function arguments.  Introducing proper language
primitives to specify what exactly is an embedding would
be helpful. And finally, having access to more of
Agda's internals such as case trees would help to
generate more performant code.

Overall, this work only scratches the surface of extraction-based
compilation.  We never considered alternative theories supported
by theorem provers, \eg{} cubical type theory in Agda; we did not
consider recursive metaprogramming; we did not consider integrating
optimisations of extracted programs other than what rewriting rules
are capable to do.  All of these offer exciting research opportunities
on the way to make verified software easily accessible.
To paraphrase Jim Morrison: ``Safety and no surprise, The End''.


\bibliography{paper}

\end{document}